# TopoLa: A Universal Framework to Enhance Cell Representations for Single-cell and Spatial Omics through Topology-encoded Latent Hyperbolic Geometry

Kai Zheng[1,#], Shaokai Wang[2,#], Yunpei Xu[1,#], Qiming Lei[3], Qichang Zhao[1], Xiao Liang[1], Qilong Feng[1], Yaohang Li[4], Min Li[1], Jinhui Xu[5,*], Jianxin Wang[1,*]

[1]School of Computer Science and Engineering, Central South University, Changsha, 410083, China

[2]Department of Mathematics, The Hong Kong University of Science and Technology, Hong Kong, China

[3]School of Physics, Central South University, Changsha, 410083, China

[4]Department of Computer Science, Old Dominion University, Norfolk, VA 23529, United States

[5]Department of Computer Science and Engineering, State University of New York at Buffalo, Buffalo, NY 14260, United States

*To whom correspondence should be addressed:

Jianxin Wang. Email: jxwang@mail.csu.edu.cn
Jinhui Xu. Email: jinhui@buffalo.edu

**Abstract.**

Recent advances in cellular research demonstrate that scRNA-seq characterizes cellular heterogeneity, while spatial transcriptomics reveals the spatial distribution of gene expression. Cell representation is the fundamental issue in the two fields. Here, we propose Topology-encoded Latent Hyperbolic Geometry (TopoLa), a computational framework enhancing cell representations by capturing fine-grained intercellular topological relationships. The framework introduces a new metric, TopoLa distance (TLd), which quantifies the geometric distance between cells within latent hyperbolic space, capturing the network's topological structure more effectively. With this framework, the cell representation can be enhanced considerably by performing convolution on its neighboring cells. Performance evaluation across seven biological tasks, including scRNA-seq data clustering and spatial transcriptomics domain identification, shows that TopoLa significantly improves the performance of several state-of-the-art models. These results underscore the generalizability and robustness of TopoLa, establishing it as a valuable tool for advancing both biological discovery and computational methodologies.

**Keywords:** biological entity relations, topology-encoded latent hyperbolic geometry, biological data, single-cell RNA sequencing, spatial transcriptome

## Introduction

Single-cell RNA sequencing (scRNA-seq) has revolutionized our understanding of cellular heterogeneity and diversity, facilitating the identification of distinct cell types and developmental trajectories[1]. While scRNA-seq provides high-resolution profiles of individual cells, it lacks the spatial context crucial for understanding tissue organization and interactions, a gap that spatial transcriptomics (ST) fills by mapping gene expression to specific locations within tissues, offering an integrated view of tissue architecture and pathology[2]. The scRNA-seq and ST technologies significantly enhance our understanding of cellular functions and organization, propelling forward advancements in regenerative medicine and contemporary healthcare[3]. Despite their promise, both fields present critical biological challenges that must be addressed.

In many important applications of scRNA-seq and ST technologies, one key challenge is effectively representing cells. For scRNA-seq, these applications include the clustering of scRNA-seq data, integration of multi-batch and multi-omic data, the inference of rare cell populations, and so on. Similar tasks exist for ST, such as spatially informed clustering, multi-batch integration, and integrative analysis of multi-modal scRNA-seq and ST data.

In scRNA-seq, clustering is a fundamental step in scRNA-seq data analysis, as it involves data-driven, unbiased grouping of cells based on their representations generated from gene expression profiles to reveal significant biological insights. Existing methods can generally be described based on the characteristics of the obtained cell representations. For instance, SIMLR[4] utilizes multiple kernels to derive a robust cell-cell similarity metric between cells, offering a refined view of the data's underlying structure. scGNN[5] integrates three iterative multi-modal autoencoders to represent relationships among cells through topological abstraction based on both gene expression and transcriptional regulation information. However, integrating multi-batch and multi-omic data presents challenges for effective clustering that could compromise the accuracy of cell representations and impact downstream analyses. Multi-batch integration needs to address technical variability across experiments, which can obscure biological signals and hinder analyses like clustering and annotation[6]. Multi-omic integration needs to combine data types such as transcriptomics and proteomics to generate comprehensive cell representations while preserving biological signals[7]. The emergence of pre-trained large language models has made it possible to address both issues simultaneously, providing a unified framework for integrating multi-batch and multi-omic data[8]. For instance, scGPT corrects batch effects by using domain adaptation via reverse back propagation and domain-specific normalization, ensuring biological signals are preserved during multi-batch integration[8]. For multi-omic integration, it incorporates modality-specific tokens and fine-tunes the model with masked gene and value prediction, enabling the integration of diverse data types such as RNA, ATAC, and protein. High-quality cell representations not only contribute to effective clustering analyses, but they are also crucial for identifying rare cell populations, which aids in understanding immune responses and disease progression. For example, surprisal component analysis (SCA)[9] uses the concept of "surprisal" to assign scores to transcripts based on their expression deviations, capturing unexpected signals that define rare cells and generating cell representations that reflect these subtle gene expression differences.

In ST, clustering is equally crucial as it integrates spatial location data with gene expression profiles to cell representations, identifying distinct cellular patterns and offering deeper insights into tissue organization and cellular interactions not captured by scRNA-seq. Methods like Giotto employ graph-based clustering algorithms[10], such as Louvain, to construct spatial neighborhood graphs that integrate these data, producing enriched cell-type representations. Similarly, GraphST combines graph neural networks with self-supervised contrastive learning to generate cell representations that account for both spatial proximity and gene expression profiles[11]. Another important task in ST is multi-batch integration, where technical variability across different experimental batches complicates data analysis. GraphST not only excels at clustering individual slices but also addresses the challenge of integrating serial tissue slices by implicitly correcting batch effects. It achieves this by leveraging spatial neighborhood information and contrastive learning, smoothing feature distributions across batches to generate consistent cell representations. Moreover, integrative analysis of scRNA-seq and ST marks a significant advancement in the field. This approach merges the high-resolution gene expression data provided by scRNA-seq with the spatial context offered by ST, offering more comprehensive cell representations. For instance, integrative and reference-informed tissue segmentation (IRIS) focuses on identifying spatial domains based on cell type composition and generates cell representations that reflect the proportion of different cell types[12].

As stated earlier, cell representations are crucial across various tasks, with numerous related models for constructing their representations. However, methods derived from different technologies and principles introduce biases and limitations, leading to difficulties in capturing detailed cellular information. Mitigating such biases in cell representation is a crucial and unresolved issue in the integrative analysis of scRNA-seq and ST data.

To address this challenge, we introduce Topology-encoded Latent Hyperbolic Geometry (TopoLa),

a novel framework designed to capture fine-grained intercellular relationships. Based on latent hyperbolic geometry, TopoLa models intercellular interactions in scRNA-seq and ST data through latent space embeddings. The framework contains a distance measure, known as TopoLa distance (TLd), which assesses the cell network geometric structural similarity between cells by counting the weighted number of even-hop paths that connect them (Methods section). Additionally, TopoLa includes a component, spatial convolution via topology-encoded latent hyperbolic geometry (TopoConv), which utilizes TLd to convolve neighboring cells especially those with similar topological structures, thereby enhancing cell representations.

The TopoLa framework demonstrates its transformative potential for assessing intercellular relationships through three key features. First, we show that the topological similarities between cells (nodes) can be encoded into a latent hyperbolic space, enabling more precise measurement of the geometric structure of cell networks. This conclusion is validated through proofs based on the principle of maximum entropy (**Supplementary text 2.1** and **Method**). Second, the TLd enables the determination of the positional distribution of cells in latent hyperbolic space. By convolving neighboring cells, it introduces the geometric structure of the cell network into cell representations (**Supplementary text 2.1**, **Theorem 2**), thereby correcting biases of cell representations. Furthermore, we provide the mathematical reasons behind TopoLa's enhanced cell representations (**Supplementary text 2.2**, **Theorem 3**, and **Theorem 4**). Third, since TopoConv enhances cell representations without altering their sizes, it can be integrated with existing methods.

We demonstrate TopoLa's universal improvements over multiple state-of-the-art models across seven critical biological tasks: clustering of scRNA-seq data, single-cell multi-batch integration, single-cell multi-omic integration, rare cell identification, spatially informed clustering of ST, vertical integration of multiple tissue slices of ST, and spatially informed clustering with the integration of scRNA-seq and ST. These results validate TopoLa's effectiveness as a universal solution for the clustering analysis of scRNA-seq and ST data.

## Results

### Overview of TopoLa

We introduce TopoLa, a latent hyperbolic geometry framework designed to enhance cell representations by capturing fine-grained intercellular relationships. The TopoLa framework captures the geometric structure of cells within a cell network and embeds them in a latent hyperbolic space. As shown in **Fig. 1**, the TopoLa framework consists of two components: TLd and TopoConv. Specifically, TLd represents the distance between cells in this latent hyperbolic space, reflecting their geometric similarities between them in the cell network. TopoConv enhances cell representations by convolving neighboring cells especially those with similar topological structures based on their positions within the latent hyperbolic space.

TopoLa can be applied to any form of cell network represented as a matrix. It first computes the singular value decomposition (SVD) of the cell network matrix $\boldsymbol{A}$ to obtain its singular values (**Fig. 1a**). These singular values are calculated through the equation $\sigma_i^* = \frac{\sigma_i^2}{\sigma_i^2 + \lambda}$, where $\lambda$ is a regularization parameter. TLd matrix $\boldsymbol{D}_{topo}$, which reflects the geometric relationships between cells and maps them into a latent hyperbolic space (**Fig. 1b**), is generated by multiplying the regularized singular values $\sigma_i^*$ with the left singular matrix. The TLd between cells represents the energy distance in a latent hyperbolic space. TLd is determined by both the energy (spatial distance) and the chemical potential (a function of cell degree in cell network). The physical significance of TLd is further detailed in the Methods section. After computing $\boldsymbol{D}_{topo}$, cells are convolved based on their positional relationships with other cells in the latent hyperbolic space. This convolution incorporates information from neighboring cells, mitigating measuring errors in scRNA-seq and ST sequencing technologies. Once TopoConv is applied to each cell, a new cell representation is obtained, referred to as the TopoLa network. The TopoLa framework is described in detail in the Methods section.

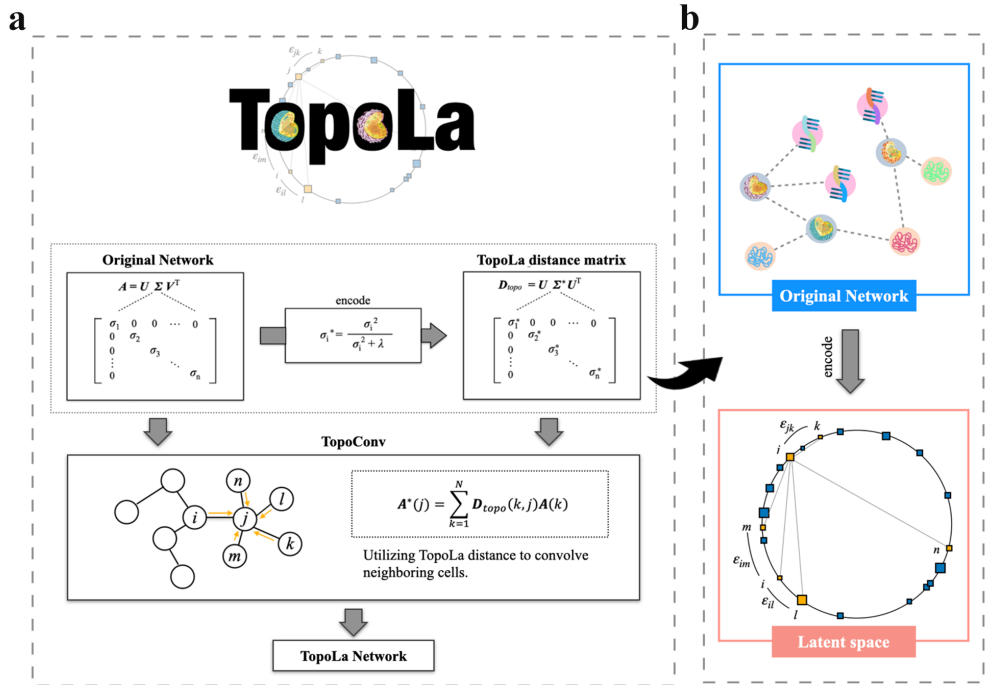

**Fig. 1 Overview of the Topology-encoded Latent Hyperbolic Geometry (TopoLa) framework. a** The original network matrix $\boldsymbol{A}$ is decomposed using singular value decomposition (SVD) into singular values and their corresponding singular vectors. The TLd matrix $\boldsymbol{D}_{topo}$ is obtained by transforming the singular values and multiplying them with one of their singular vectors. TopoConv is then applied to convolve neighboring cells, enhancing their representations to construct TopoLa network. **b** The original network, representing intercellular relationships, is encoded into latent hyperbolic space. In this space, the spatial relationships effectively capture the geometric structures of intercellular interactions.

**Enhancing single-cell RNA-seq clustering with TopoLa**

Single-cell RNA sequencing (scRNA-seq) has revolutionized our understanding of gene expression by enabling detailed profiling at both the global transcriptome and single-cell levels[13]. This powerful technology generates vast amounts of data, necessitating robust computational analyses to effectively unveil the underlying biological insights[14]. Among these analytical methods, clustering techniques have proven indispensable, particularly for cell clustering and cell type discovery[15]. Consequently, clustering and cell type classification are critical components of scRNA-seq data analysis.

SIMLR, a widely used method, employs multi-kernel learning for scRNA-seq data clustering[4]. This technique integrates multiple kernels to derive a robust similarity metric that optimally captures the structure of the data. Here, we introduce TopoLa to enhance the similarity metrics integrated by SIMLR. By viewing the similarity as a network of relationships between cells, TopoLa maps the cells into a latent hyperbolic space and uses TLd to measure the geometric relationships between them. Spatial convolution is then performed on neighboring cells to update the relationships within the original network, resulting in a new method referred to as SIMLR+TopoLa. The model schematic is provided in **Supplementary Fig. 1**.

To assess the effectiveness of SIMLR+TopoLa, we apply it to 30 scRNA-seq datasets from different species and tissues, sequenced using various platforms (Methods section). As shown in **Fig. 2a**, SIMLR+TopoLa outperforms SIMLR, with relative improvements of 5.8% in Adjusted Rand Index (ARI) (**Supplementary text 1.1**) and 2.4% in Normalized Mutual Information (NMI) (**Supplementary text 1.2**). These gains underscore TopoLa's enhanced ability to capture the underlying structure of scRNA-seq data, leading to more accurate clustering results. The overall improvement in ARI and NMI across all datasets is further summarized in the box plots in **Fig. 2b**.

In addition to SIMLR, TopoLa can enhance embedded cell representation from deep learning methods like scGNN[5], which models complex cell-to-cell relationships in scRNA-seq data using graph neural networks and multimodal autoencoders. Notably, the integration of TopoLa with scGNN (scGNN+TopoLa) further improves performance, achieving an ARI of 0.7784 and an NMI of 0.5487. The model schematic is provided in **Supplementary Fig. 2**. This represents increases of 2.6% in ARI and 6.1% in NMI compared to scGNN, highlighting the robustness of TopoLa's enhancement. These results show that TopoLa benefits not just similarity-based methods like SIMLR but also cell graph-based approaches such as scGNN. The improvements are also summarized using boxplots in **Fig. 2e**.

To further elucidate the role of TopoLa in refining cell type delineation, the cell representations on Puram and Baron_human3 datasets are visualized through t-SNE. The resulting plots (**Figs. 2c, f**) indicate that the addition of TopoLa improved the clustering of different cell types. In the Puram dataset, analysis with SIMLR and SIMLR+TopoLa reveals that T cells are initially embedded within clusters of "Macrophage", "Fibroblast", and "Endothelial" cells, making them challenging to distinguish. However, following the application of TopoLa, T cells emerge as spatially discrete populations, markedly improving their detectability and identification. A similar pattern is observed in the Baron_human1 dataset when analyzed using scGNN and scGNN+TopoLa. Initially, "ductal" cells are obscured by the surrounding "delta", "beta", "quiescent_stellate", and "gamma" cells, complicating their identification. The introduction of TopoLa enhances the spatial resolution of the "ductal" cells, allowing for their clear separation from neighboring populations. Additionally, visualizations for all four methods across all datasets are available in **Supplementary Figs. 3-18**.

These findings emphasize the usefulness of TopoLa in improving the precision and accuracy of cell type clustering in single-cell analyses.

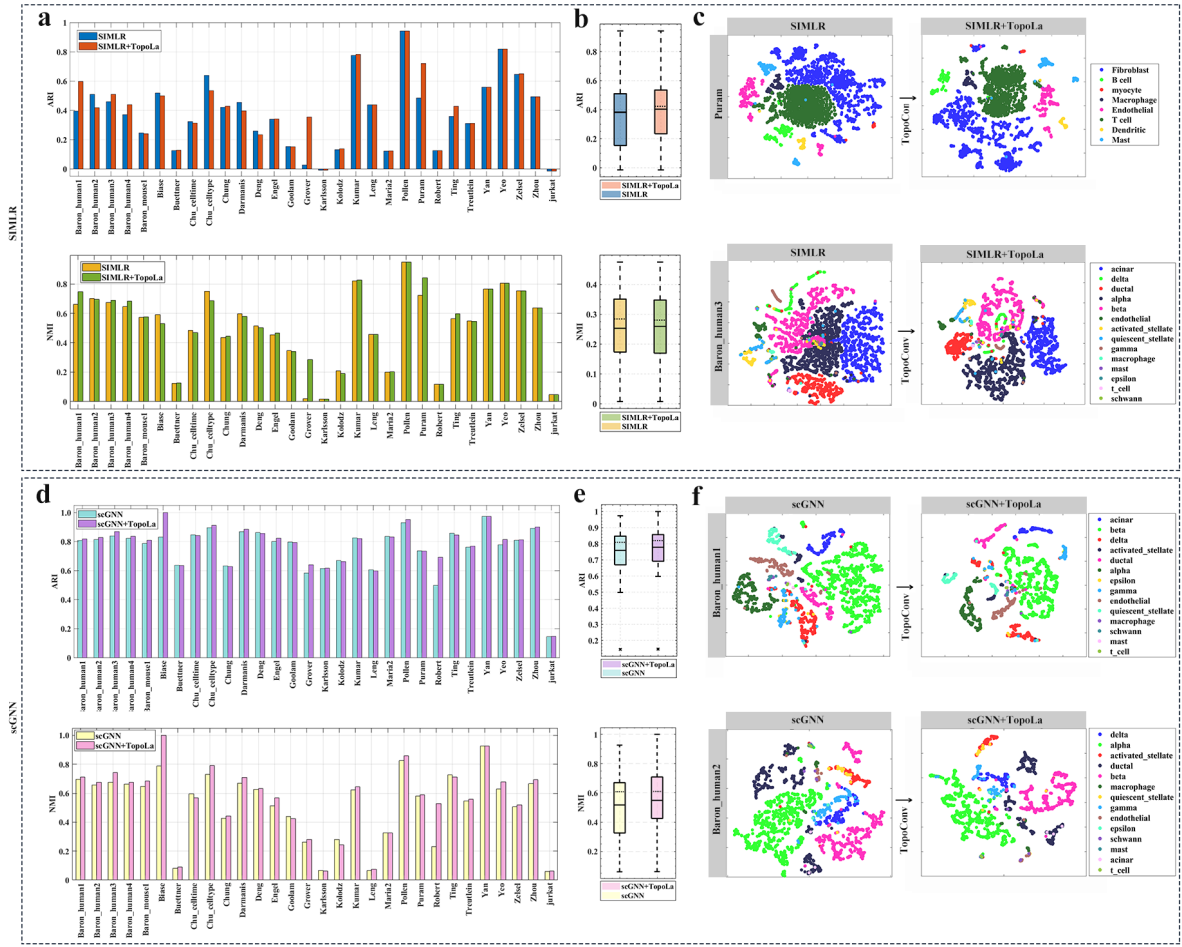

**Fig. 2 Enhanced cell type identification using TopoLa on single-cell RNA-seq data. a** ARI and NMI performance of SIMLR and SIMLR+TopoLa across 30 single-cell datasets. **b** Boxplots displaying the distribution of ARI and NMI values across all datasets for SIMLR and SIMLR+TopoLa, with the solid line representing the mean and the dashed line indicating the median. **c** t-SNE visualizations of cell type representations in the Baron_human1 and Baron_human2 datasets using SIMLR and SIMLR+TopoLa. **d** ARI and NMI performance of scGNN and scGNN+TopoLa across 30 single-cell datasets. **e** Boxplots displaying the distribution of ARI and NMI values across all datasets for SIMLR and scGNN+TopoLa, with the solid line representing the mean and the dashed line indicating the median. **f** t-SNE visualizations of cell type representations in the Baron_human1 and Baron_human2 datasets using scGNN and scGNN+TopoLa.

**Enhancing single-cell multi-batch integration with TopoLa**

Batch effects, which are technical variations introduced across different experimental batches, pose significant challenges in the analysis of scRNA-seq data[16]. These variations may arise from differences in reagent lots, instrument performance, or operator handling, leading to inconsistencies that are unrelated to biological conditions[17]. Such technical artifacts can obscure true biological signals, thereby complicating downstream analyses and data interpretation. Therefore, integrating scRNA-seq data across multiple batches is essential for mitigating these effects and enhancing the biological reliability of the findings. Robust batch correction methods are needed to accurately distinguish genuine biological variation from technical noise, enabling more precise cell-type classification and comparative analyses[18].

Generative pretrained models, such as scGPT, have recently become effective solutions for integrating multiple batches of data[8]. To address batch effects in integration tasks, scGPT leverages a shared set of gene tokens between the pretrained model and the current dataset, selects highly variable genes (HVGs) as input and normalizes expression values before model training. The model is initialized with pretrained weights and fine-tuned using objectives like cell contrastive learning and domain-specific normalization to explicitly correct batch effects. scGPT implements downstream clustering algorithms by converting model embeddings into cell graphs.

We refine the cell graph to enhance scGPT's performance on multi-batch integration. This refined approach, termed scGPT+TopoLa for multi-batch integration, incorporates spatial information by leveraging hyperbolic geometry, thereby improving the robustness and accuracy of cell-type identification. The model schematic is provided in **Supplementary Fig. 19**.

We applied the method on three different scRNA-seq datasets: PBMC 10k (two batches), perirhinal cortex (two batches), and a COVID-19 dataset (two batches). To quantitatively assess the enhancements introduced by scGPT+TopoLa, we employed the NMI and ARI metrics. These metrics assess the consistency between predicted and actual labels, as well as the accuracy of clustering in relation to the true labels.

The visualizations in **Figs. 3a-c** illustrate the comparative integration results for the PBMC 10k, Cortex, and COVID-19 datasets, highlighting the improvements introduced by scGPT+TopoLa. In the PBMC dataset, scGPT struggles to effectively distinguish between "CD14+ Monocytes," "Dendritic Cells", and "Other" cell types, which are clustered together in **Fig. 3a**. After integrating geometric structural information via TopoLa, these cell types become more clearly separated, indicating improved resolution in identifying subtle population differences. In the Cortex dataset, scGPT struggles to accurately cluster "astrocyte" cells, which are incorrectly split into two distinct groups. The introduction of TopoLa addresses this issue by promoting tighter clustering of "astrocytes", as seen in **Fig. 3b**. This improvement highlights the ability of scGPT+TopoLa to enhance cell-type specificity in complex and heterogeneous tissue samples. In the COVID-19 dataset, scGPT fails to adequately distinguish between "M2 Macrophages" and "Macrophages," leading to overlapping clusters (**Fig. 3c**). With the incorporation of TopoLa, these cell types are more distinctly separated, demonstrating the model's capacity to disentangle closely related but functionally distinct cell populations.

As shown in **Fig. 3d**, scGPT+TopoLa outperforms scGPT alone across all datasets. On average, scGPT+TopoLa improves the NMI by approximately 3.2% over scGPT. Although this enhancement is modest, it is consistent across different biological conditions, demonstrating the broad applicability of the method. More notably, the ARI exhibits an average improvement of 15.4%, with the most substantial gain observed in the Cortex dataset (26.8% increase), reflecting the model's robustness in resolving complex and heterogeneous cell populations. These improvements indicate scGPT+TopoLa's capability to achieve more accurate clustering and better separation of biologically distinct cell types while effectively mitigating batch effects.

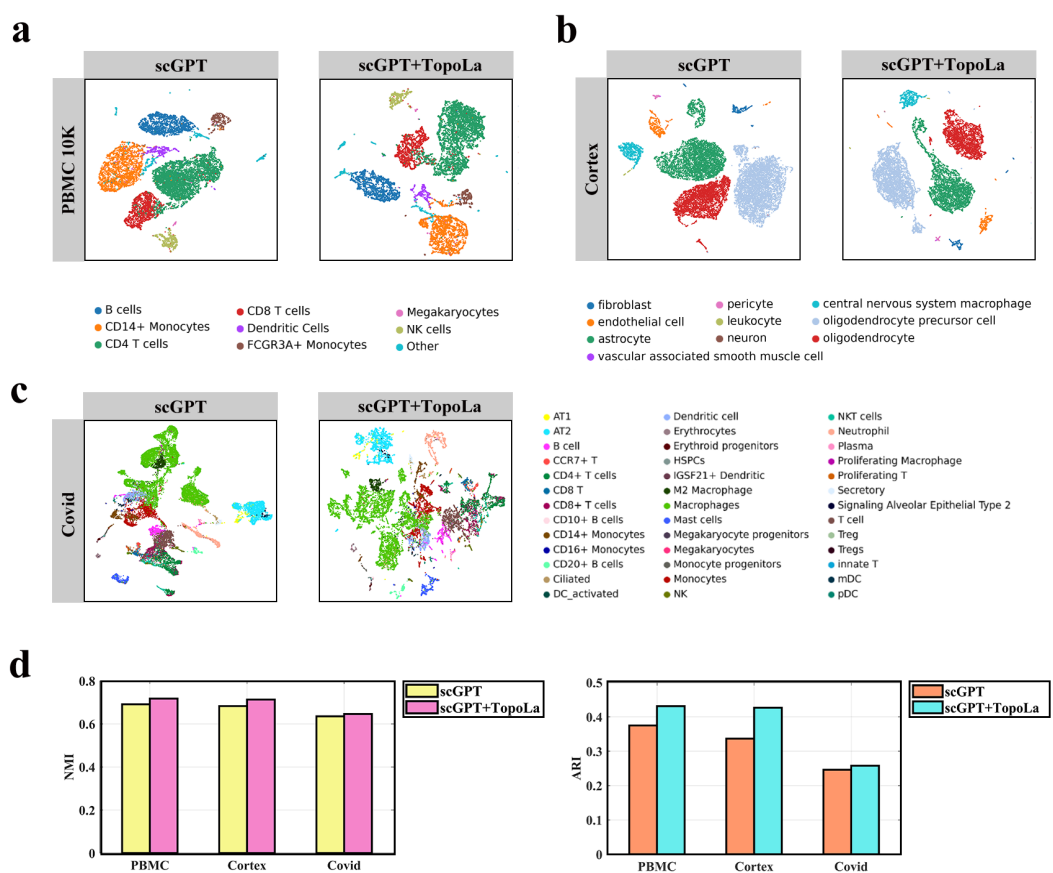

**Fig. 3 Enhanced multi-batch integration using TopoLa on single-cell RNA-seq data. a** UMAP visualizations of the PBMC 10k dataset comparing the performance of scGPT and scGPT+TopoLa, with cells labeled by their true cell types. **b** UMAP visualizations of the Cortex dataset comparing the performance of scGPT and scGPT+TopoLa, with cells labeled by their true cell types. **c** UMAP visualizations of the COVID-19 dataset comparing the performance of scGPT and scGPT+TopoLa, with cells labeled by their true cell types. **d** Quantitative comparisons of performance metrics (NMI, ARI) across the three datasets (PBMC 10k, Cortex, and COVID-19).

scGPT+TopoLa represents a significant advance in scRNA-seq data integration by effectively addressing both batch effects and the biological complexity inherent in multi-batch datasets. By fusing geometric structural information into the scGPT framework, this approach leads to more coherent clustering and enhanced cell-type resolution. The results demonstrate that scGPT+TopoLa enhances cluster separation and provides finer granularity in distinguishing biologically relevant subpopulations, particularly in cases where standard scGPT encounters limitations.

**Enhancing single-cell multi-omic integration with TopoLa**

Recent advancements in scRNA-seq have revolutionized the exploration of regulatory landscapes across multiple omics layers, such as chromatin accessibility (scATAC-seq), DNA methylation (snmC-seq, sci-MET), and the transcriptome (scRNA-seq), offering unprecedented opportunities to uncover regulatory mechanisms in diverse cell types[19]. Despite the emergence of simultaneous multi-omics assays, the unpaired nature of the resulting data necessitates efficient in silico integration[20]. This integration is challenging due to distinct feature spaces, such as accessible chromatin regions in scATAC-seq versus genes in scRNA-seq. Strategies like converting multimodal data into a common feature space can lead to information loss. Nonlinear manifold alignment methods, though promising, are limited to small datasets[21]. As data volumes grow to encompass millions of cells, scalable and accurate computational integration methods become essential[22]. Addressing these challenges will enable deeper insights into the cellular processes and provide a comprehensive understanding of the cellular functions and regulatory mechanisms.

For multi-omic integration, scGPT employs a unified generative pretraining workflow specifically tailored for non-sequential omics data. The integration process begins by retaining a common set of gene tokens between the pretrained model and the current dataset, followed by selecting a subset of HVGs for input. Expression values are preprocessed before model training to normalize data across different omics modalities. For datasets containing additional ATAC or protein data, scGPT inherits the trained gene embeddings for RNA data and trains additional token embeddings from scratch.

Recognizing that cell embeddings can be viewed as a network of relationships between cells and abstract features, we map these cells into a latent hyperbolic space and use TLd to measure geometric relationships. Subsequently, we perform spatial convolution on neighboring cells to update the relationships within the original network (cell embeddings generated by scGPT), thereby enhancing the robustness and accuracy of the integrated cell representations. The model schematic is provided in **Supplementary Fig. 20**.

Quantitative evaluations of clustering performance further substantiate these observations (**Supplementary Fig. 21a**). Notably, scGPT+TopoLa for multi-omic integration exhibits significant improvements across all key metrics compared to scGPT. Specifically, ARI shows a 45.7% increase, rising from 0.151 to 0.220, indicating enhanced clustering accuracy and agreement with ground truth labels. NMI is improved by 8.1%, from 0.541 to 0.585, reflecting better consistency between clusters and known cell type annotations. ASW (Average Silhouette Width), which measures cluster cohesion and separation, increases from 0.483 to 0.505, demonstrating more distinct and compact clustering (**Supplementary text 1.3**). **Supplementary Fig. 21b** displays UMAP visualizations of the BMMC dataset, comparing the clustering performance of scGPT and scGPT+ TopoLa, with cells labeled by their true cell types. The incorporation of TopoLa into scGPT results in more distinct and well-defined clusters. Notably, the scGPT+TopoLa achieves a clearer separation between "CD4+ T naive" and "CD8+ T naive" cell types, which are less distinctly differentiated in the clustering produced by the original scGPT model.

The integration of topological alignment in scGPT+TopoLa, as demonstrated by both visual (UMAP) and quantitative (clustering metrics) analyses, not only reinforces the model's robustness but also significantly enhances its capacity for multi-omics data integration. By effectively capturing the relationships between different omics layers, such as transcriptomics, chromatin accessibility, and proteomics, scGPT+TopoLa maintains the integrity of biological signals while harmonizing data from distinct feature spaces. This capacity is crucial for preserving cellular identity and regulatory dynamics during the integration process.

The results highlight how scGPT+TopoLa excels at managing the inherent challenges of integrating high-dimensional, non-linear datasets, where different omics modalities contribute unique and complementary insights. The improved alignment between omics layers ensures that key regulatory mechanisms are accurately captured and preserved, facilitating a more comprehensive understanding of complex cellular processes. This advancement establishes scGPT+TopoLa as a powerful and scalable tool for downstream analyses, particularly in the context of resolving cellular heterogeneity, reconstructing regulatory networks, and exploring cross-modality relationships in large-scale multi-omics studies.

**Enhancing rare cell identification in single-cell analyses with TopoLa**

The identification of rare cells using scRNA-seq represents a pivotal advancement in modern biology, enabling the detection of cell populations that, although small in number, have substantial biological relevance. These rare cells often constitute only a minor fraction of the total cellular milieu, but play critical roles in processes such as disease progression and immune responses[23]. In oncology, for example, rare subpopulations like cancer stem cells are implicated in tumor recurrence and drug resistance[24,25]. In immunology, scarce immune cell subsets may act as key mediators in pathogen response or contribute to the pathogenesis of autoimmune diseases[26]. Therefore, the precise identification of these cells not only enhances our understanding of complex biological systems, but also has profound implications for developing targeted therapeutic strategies[48]. Despite its transformative potential, scRNA-seq analysis faces considerable challenges in reliably identifying rare cells[27]. The vast dimensionality of scRNA-seq data, typically encompassing tens of thousands of gene expression features, further complicates efforts to isolate distinctive markers specific to rare cell types[28]. Addressing these obstacles requires the development of sophisticated analytical techniques capable of disentangling the complexities inherent in scRNA-seq data[25].

SCA is an information-theoretic dimensionality reduction technique designed to tackle challenges in scRNA-seq analysis[9]. Unlike traditional methods such as Principal Component Analysis (PCA) or t-SNE, which may overlook signals from rare or subtly defined cell populations, SCA leverages "surprisal scores" to highlight unexpected gene expression patterns, revealing biologically meaningful variations that might otherwise be masked by noise or dominant cell types. This approach is particularly valuable in fields like tumor immunology, where rare yet critical cell populations are key to understanding disease mechanisms. However, SCA can amplify noise and batch effects, especially in complex datasets with substantial technical variation, limiting its robustness.

To address this limitation, we introduce SCA enhanced with TopoLa (SCA+TopoLa). By mapping cell embeddings into a latent hyperbolic space and viewing the association of cells as a network of relationships between themselves and abstract features, we use TLd to measure geometric relationships. We then apply spatial convolution on neighboring cells to update these relationships within the original network, ultimately enhancing the robustness and accuracy of the integrated cell representations. The detailed computational workflow is illustrated in **Supplementary Fig. 22**.

To comprehensively evaluate the enhancements provided by SCA+TopoLa, we applied the method to 20 scRNA-seq datasets spanning diverse tissues and sequencing platforms. As shown in **Fig. 4a**, SCA+TopoLa mostly outperforms SCA in F1 scores (**Supplementary text 1.4**) across all datasets, with particularly significant improvements in challenging datasets like the hippocampus, where the F1 score nearly doubles. In complex environments such as tonsil and pancreas, SCA+TopoLa maintains a clear performance advantage, reflecting its robustness in diverse cellular contexts. As shown in **Fig. 4b**, on average, the F1 score increases from 0.3568 with SCA to 0.4571 with SCA+TopoLa, representing a 28.1% improvement.

Beyond the numerical improvements, **Figs. 4c–f** provide deeper insights into how SCA+TopoLa enhances cell-type separation and rare cell identification. In the Airway dataset, identifying rare cell populations, such as Goblet, Ionocyte, Neuroendocrine, and Tuft cells, serves as a crucial benchmark for model evaluation. As shown in **Figs. 4c-d**, both SCA and SCA+TopoLa models successfully detect these rare populations. However, a detailed examination reveals that the SCA model misclassifies a considerable number of non-rare cells, leading to a higher false positive rate. This misclassification results in significant overlap between rare and non-rare cell clusters, diminishing the clarity of rare cell signatures. In contrast, SCA+TopoLa remarkably improves precision, effectively isolating rare cell types with fewer false positives, demonstrating an enhanced capacity to capture subtle cellular distinctions within the complex and heterogeneous airway tissue environment.

This pattern is further observed in the hippocampus dataset (**Figs. 4e-f**), where identifying rare cell types like Ependymal and CA2 cells is critical. Although both models detect these populations, the SCA model again misclassifies a range of non-rare cells, erroneously including them within the rare cell cluster space and increasing the false positive rate. Conversely, SCA+TopoLa exhibits finer clustering of rare cells, with significantly fewer misclassifications. The improved granularity of SCA+TopoLa is particularly evident in its capacity to preserve the integrity of rare cell clusters, indicating that the integration of geometric structure offers a robust framework for differentiating rare from abundant cell types.

Visualizations for the other datasets are available in **Supplementary Figs. 23-27**. The comparative performance across both datasets reveals that while SCA shows some capacity for rare cell detection, it frequently conflates non-rare cells with rare categories, undermining classification accuracy. SCA+TopoLa mitigates this limitation by incorporating geometric structural features, leading to more distinct clustering and fewer false positives. These findings underscore the importance of leveraging topological approaches in single-cell analysis, especially in scenarios where accurate rare cell identification is essential.

Collectively, these results indicate that integrating geometric structural information via SCA+TopoLa not only enhances overall clustering performance but also significantly improves the detection and separation of rare cell types, leading to more biologically meaningful conclusions. The combination of improved quantitative metrics, clearer visualizations, and more precise rare cell identification demonstrates that SCA+TopoLa represents a substantial advancement over conventional SCA, particularly in datasets characterized by complex cellular compositions and subtle cell-type distinctions.

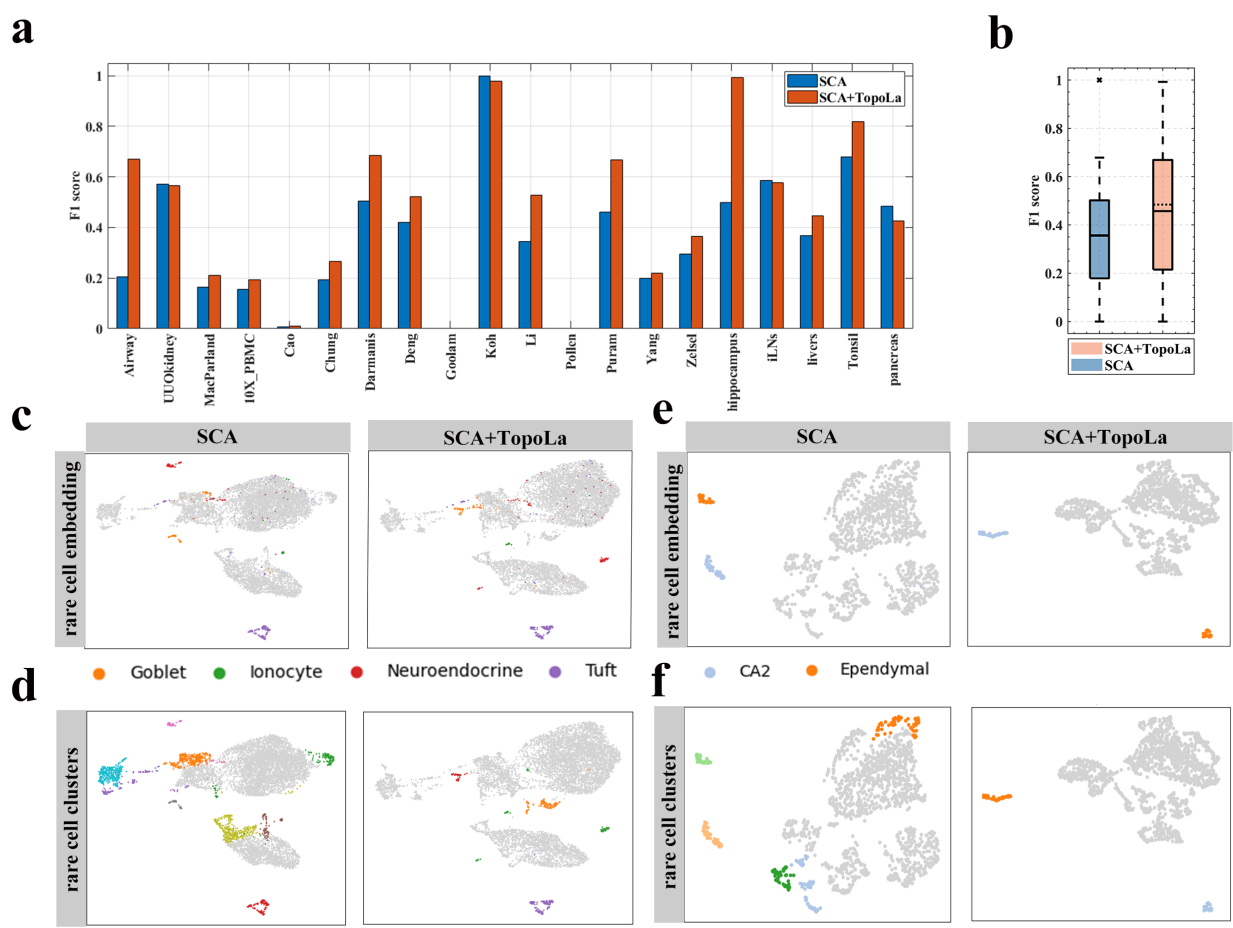

**Fig. 4 Improved rare cell identification with SCA+TopoLa. a** Comparison of F1 scores across 20 datasets between SCA and SCA+TopoLa. **b** Boxplot illustrates the distributions of F1 scores across all datasets. The solid line in each box represents the mean F1 scores, while the dashed line represents the median F1 scores. **c** UMAP visualizations of the rare cell embeddings obtained by SCA and SCA+TopoLa on the Airway dataset. **d** UMAP visualizations of rare cell clustering results obtained by SCA and SCA+TopoLa on the Airway dataset.

**e** UMAP visualizations of the rare cell embeddings obtained by SCA and SCA+TopoLa on the hippocampus dataset. **f** UMAP visualizations of rare cell clustering results obtained by SCA and SCA+TopoLa on the hippocampus dataset.

## TopoLa enhances spatially informed clustering in ST data

ST plays a pivotal role in revealing the spatial organization of tissues by clustering cells based on their spatially resolved gene expression profiles[2]. Clustering is essential for identifying spatial domains and uncovering cellular architectures within tissues, providing a comprehensive understanding of tissue organization and cell-cell interactions[29]. However, one of the key challenges in ST data analysis is the influence of noise, which can distort the accuracy of clustering and lead to the misidentification of spatial domains. Noise in ST datasets often arises from technical variations, such as sequencing errors or dropouts, as well as biological variability, making it difficult to capture the true spatial structure of tissues[30].

GraphST is a widely adopted method for clustering in ST. While effective, it falls short in capturing the intercellular geometric structure, which is vital for resolving fine-grained spatial domains with precision. To overcome this limitation, we present a novel model, GraphST+TopoLa, which integrates geometric structural information using TopoLa. By incorporating topological features, TopoLa enhances the clustering process, leading to a more faithful representation of spatial domains. The combination of GraphST's graph-based clustering with TopoLa's geometry-aware approach results in a more refined and accurate identification of spatial domains.

To validate the applicability of GraphST+TopoLa in ST data analysis, we applied it to the DLPFC dataset[31] and assessed the impact of network enhancement on the spatial domain analysis. The dataset comprises spatially resolved transcriptomic profiles from 12 slices, illustrating the four or six layers of the DLPFC and white matter (WM). We used the GraphST algorithm as the baseline method to evaluate the enhanced performance of GraphST+TopoLa. The detailed computational workflow of the GraphST+TopoLa model is illustrated in **Supplementary Fig. 28**.

The results underscore the capability of TopoLa to refine spatial domain identification within ST data. For instance, in slice 151509 (**Fig. 5a**), GraphST fails to distinguish between Layer 3 and Layer 4 of the DLPFC, resulting in significant overlap between these regions. Quantitative assessment using the ARI reflects this limitation, with a relatively low score of 0.3950. However, after the integration of TopoLa-enhanced embeddings, these two layers are clearly differentiated, resulting in a notable increase in the ARI to 0.5313, indicating a more accurate spatial delineation.

Moreover, TopoLa significantly improves the accuracy of cell type identification, particularly in complex tissue regions. In slice 151675 (**Fig. 5a**), GraphST incorrectly classifies approximately 25% of Layer 3 cells as WM, as reflected in the ARI score of 0.4492. The application of TopoLa corrects these errors, reassigning the misclassified cells to their correct layers and improving the ARI to 0.6792. This substantial enhancement demonstrates TopoLa's ability to resolve ambiguities in spatial domain identification that are challenging for traditional models. Clustering results produced by GraphST and GraphST+TopoLa on the other slices are shown in **Supplementary Figs. 29-31**.

GraphST recommends the use of mclust[32], a clustering method that requires prior knowledge of the number of cell types. To evaluate performance in scenarios where the number of cell types is unknown, we also apply the Louvain algorithm[33]. As shown in **Fig. 5b**, TopoLa enhances the performance of GraphST across both clustering methods. Specifically, when the number of label types is predefined,

TopoLa boosts GraphST's accuracy by an average of 5.2% across 12 slices, as measured by ARI. Remarkably, in cases where the number of label types is unknown, TopoLa yields an even greater improvement, with an average enhancement of 20.9% across the same slices. These consistent gains across different analytical frameworks highlight TopoLa's versatility and robustness in improving ST analyses.

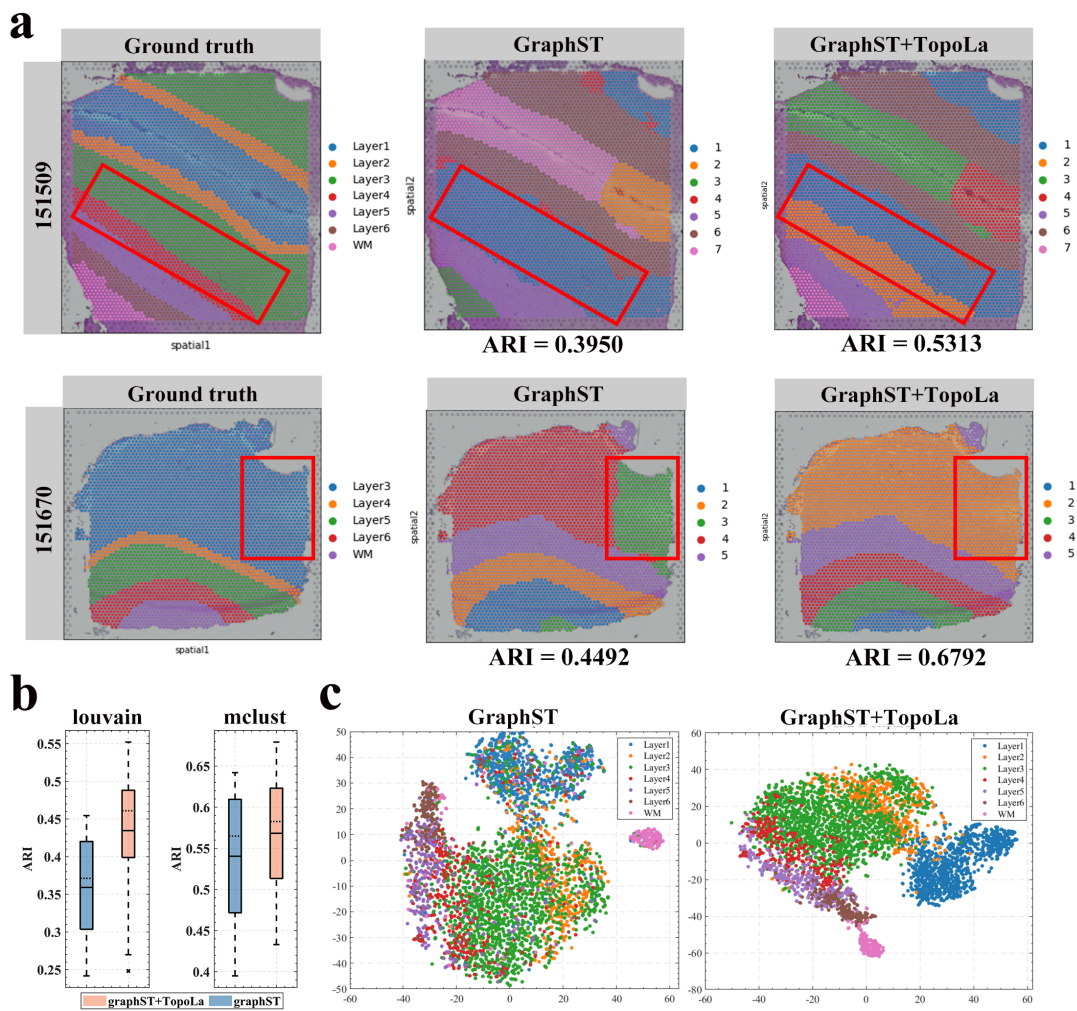

**Fig. 5 Enhanced spatial domain identification using ST data with TopoLa integration. a** Clustering results produced by GraphST and GraphST+TopoLa on slices 151509 and 151670 from the DLPFC dataset. **b** Boxplots displaying the performance of graphST and graphST+TopoLa measured in the format of ARI on the tissue slices using the Louvain and mclust clustering algorithms. The solid line in each box represents the mean ARI scores, while the dashed line represents the median ARI scores. **c** Visualizations of cell embeddings from slice 151509, comparing GraphST results before (left) and after (right) TopoLa enhancement. The embeddings are visualized with distinct colors representing different cell types, showcasing the improved dispersion of different cell types.

To visualize the structural improvements provided by TopoLa, we conduct a visualization analysis of the cell embeddings for slice 151509 (**Fig. 5c**). The resulting plots reveal that the TopoLa-enhanced embeddings exhibit a clearer hierarchical distribution compared to the original GraphST embeddings. This is particularly evident in the distinct separation between Layer 3 and Layer 4 cells, which is less significant in the non-enhanced embeddings. Additional visualizations for the remaining datasets are available in **Supplementary Figs. 32-34**. The clearer segregation of cell types in the TopoLa-enhanced plot directly correlates with the observed improvements in ARI scores, suggesting that the geometric structural information integrated by TopoLa plays a crucial role in refining cell type classification and spatial domain identification.

In conclusion, our analysis strongly supports the conclusion that TopoLa significantly enhances the performance of GraphST in both known and unknown cell type scenarios, making it a valuable tool for the accurate analysis of ST data.

**Enhancing batch effect correction in vertical integration of multiple tissue slices with TopoLa**

ST has emerged as a transformative technology for the comprehensive analysis of serial tissue slices, enabling high-resolution mapping of gene expression across multiple layers[34]. This capability to examine continuous tissue slices facilitates precise vertical integration, which is essential for reconstructing three-dimensional tissue architectures and deciphering the complex spatial organization of cellular environments[35]. Vertical axis reconstruction offers critical insights into the dynamic interplay between cell types and the spatial relationships across distinct tissue regions[36].

However, batch effects remain a major obstacle to the reliable integration of such spatial data. These effects arise from technical variations in sample processing, sequencing protocols, and other experimental factors, introducing non-biological noise that can mask genuine biological signals[37]. The resulting inconsistencies compromise the reproducibility of spatial analyses and may lead to biased conclusions, ultimately hindering the efficacy of vertical integration in accurately reconstructing tissue architecture[38]. Addressing these batch effects is therefore crucial for unlocking the full potential of ST in the layered analysis of tissue structure.

GraphST uses spatial information to correct batch effects, which makes it better than non-spatial methods. It achieves this by mapping data from different batches into a common space, enabling the integration of cell representations from different batches into a unified cell network. This helps establish geometric relationships across batches. However, GraphST does not fully consider the geometric relationships between cells from different batches, potentially leading to suboptimal batch alignment during integration.

To address this limitation, we introduce TopoLa, capturing inter-batch geometric relationships, to GraphST for enhancing integration accuracy. By incorporating geometric structure information, the combined model, GraphST+TopoLa, offers a more refined approach to batch correction, enabling consistent alignment across tissue slices. The detailed computational workflow of the GraphST+TopoLa model for multi-batch integrationis illustrated in **Supplementary Fig. 35**.

In **Supplementary Fig. 36a**, we first illustrate the alignment of serial slices from mouse breast cancer samples using the PASTE algorithm. The uncorrected UMAP embeddings of two consecutive slices (**Supplementary Fig. 36b**) show pronounced batch effects, evident in the distinct separation between batches. After applying batch correction (**Supplementary Fig. 36c**), we compare the performance of GraphST and GraphST+TopoLa. The first column, colored by batch, reveals that GraphST+TopoLa achieves a more homogeneous mixing of cells across batches, indicating enhanced batch effect correction. The second column, colored by true cell types, demonstrates improved clustering accuracy with GraphST+TopoLa.

To quantify these improvements, we apply the integration Local Inverse Simpson's Index (iLISI) and ASW metrics (**Supplementary Fig. 36d**). The iLISI score (**Supplementary text 1.5**), which measures the degree of batch mixing, increases from 0.8967 with GraphST to 0.9029 with GraphST+TopoLa. Similarly, the ASW score (**Supplementary text 1.3**), which evaluates clustering accuracy by balancing separation between clusters and cohesion within clusters, improves from 0.0162 with GraphST to 0.0259 with GraphST+TopoLa (an increase of approximately 60%). This substantial enhancement indicates that GraphST+TopoLa significantly improves the preservation of spatial boundaries, reducing artificial blending while maintaining distinct cell type identities, which is crucial for accurate spatial reconstruction and biological interpretation.

These results demonstrate that introducing TopoLa significantly enhances alignment accuracy and spatial structure preservation by integrating geometric relationships across batches. GraphST+TopoLa effectively addresses limitations in existing approaches, offering a more reliable solution for mitigating batch effects, and enabling consistent reconstructions of tissue architecture across serial slices. While GraphST provides a strong foundation for integrating ST data, the incorporation of TopoLa represents a critical improvement, offering a more robust framework for analyzing complex spatial relationships within tissue environments. By preserving both alignment and spatial integrity, GraphST+TopoLa facilitates deeper insights into the architecture of tissues across multiple layers.

## TopoLa advances integrative analyses of scRNA-seq and ST data

scRNA-seq has significantly advanced our understanding of cellular heterogeneity, uncovering novel cell types and states across various contexts. However, traditional scRNA-seq analysis, which clusters cells based on their gene expression profiles, often struggles with samples containing cells in transitional states, such as those found in tumors or during development[39]. This limitation can lead to unreliable clustering and hinder pathway analyses.

Integrating scRNA-seq with ST data offers a powerful solution to these challenges. ST technologies preserve spatial context while providing transcriptome-wide gene expression profiles, allowing researchers to map gene expression to specific tissue locations[40]. Spatially informed clustering combines spatial information with gene expression data to identify distinct cell or domain clusters within tissues[12]. This approach facilitates the characterization of how cells interact with their spatial neighbors, the detection of genes with spatially varying expression, and the uncovering of spatial trajectories or RNA velocity patterns[41]. By using spatial information, researchers can gain deeper insights into the functional organization of tissues, improving the accuracy of cell-type classification and advancing the study of complex biological processes.

IRIS is a computational method designed for detecting spatial domains in ST studies[12]. It directly models the cell type compositional heterogeneity across spatial locations, segmenting the tissue into biologically relevant domains characterized by distinct cell type compositions. To perform spatial clustering, IRIS integrates scRNA-seq data to characterize cell type compositions in ST tissues, accounting for compositional similarities within and across tissue slices. Focusing on cell type composition rather than transcriptomic heterogeneity, IRIS simplifies the detection task and enhances computational efficiency, making it feasible for large-scale ST datasets. In our study, all cells from the four slices of the same donor obtained through IRIS, along with their cell type proportions, are considered as a network of cellular relationships. This network is then mapped into a latent hyperbolic space using TopoLa. Spatial convolution is performed on neighboring cells based on their spatial positions to update cell representations accordingly. We refer to this new method as IRIS+TopoLa. The detailed computational workflow is illustrated in **Supplementary Fig. 37**.

To validate the applicability of IRIS+TopoLa for spatially informed clustering with the integration of scRNA-seq and ST data, we applied it to several ST datasets. We first examined the gold-standard human DLPFC dataset[31], acquired using 10x Visium. The dataset includes 12 annotated brain tissue slices derived from three distinct donors. These slices include seven spatial domains, specifically six cortical layers as well as the white matter. Additionally, scRNA-seq data from post-mortem brain tissue, sequenced using 10x Chromium and comprising 44 cell types, was utilized as a reference.

The results demonstrate that incorporating TopoLa significantly enhances spatially informed clustering using IRIS, particularly in the human DLPFC. In **Fig. 6a** (top row), the visualization of slice 151670 clearly illustrates this improvement. Specifically, while IRIS (middle panel, ARI = 0.4116) struggles to differentiate between layers 5 and 6, leading to substantial overlap and misclassification, the embeddings enhanced by IRIS+TopoLa (right panel, ARI = 0.7126) achieve a distinct and accurate separation of these layers. The visualization of the results on the other 10 slices is provided in

**Supplementary Figs. 38-40**. These improvements demonstrate TopoLa's ability to refine the spatial structure, ensuring that the cellular architecture is more faithfully represented.

Similarly, in **Fig. 6a** (bottom row), the visualization of slice 151671 underscores another critical enhancement. Here, IRIS (middle panel, ARI = 0.5308) misclassifies a significant portion of layer 3 cells as layer 4, reflecting its limitations in distinguishing closely related cellular populations. Upon applying IRIS+TopoLa (right panel, ARI = 0.7794), these classification errors are substantially reduced, with layer 3 cells now correctly identified. This correction highlights TopoLa's ability to improve the resolution of clusters, particularly in cases where cellular boundaries are ambiguous or closely aligned.

In **Fig. 6b**, we present performance metrics across 12 different cortical slices, providing a comprehensive view of TopoLa's impact on clustering accuracy. The ARI scores exhibit a marked improvement with the integration of TopoLa, outperforming IRIS with an average enhancement of 11.7%. This increase underscores the efficacy of TopoLa in accurately capturing cellular relationships and enhancing clustering precision. The NMI scores reflect a similar trend, with TopoLa leading to more coherent and reliable clustering results, as evidenced by the higher median values in the box plots.

**Fig. 6c** provides a detailed visualization of slice 151670. The use of IRIS results in a scattered clustering of Layer 3 cells, with noticeable mixing with cells from other layers, which could lead to erroneous biological interpretations. However, with TopoLa enhancement, the Layer 3 cells are clearly delineated and no longer mixed with other cell types, reinforcing the accuracy and precision of the clustering. Visualization of cell embeddings from the other 11 slices is provided in **Supplementary**

**Figs. 41-43.** This precise delineation is crucial for downstream analyses, where accurate identification of cell types impacts the understanding of cellular functions and interactions.

In summary, these findings robustly confirm that TopoLa significantly enhances the clustering accuracy and robustness of IRIS. By addressing both the spatial structure and the inherent variability within datasets, TopoLa emerges as a critical advancement for ST, enabling more accurate and reliable analysis of complex tissue architectures. The incorporation of TopoLa into integrative analyses of scRNA-seq and ST data facilitates deeper insights into tissue organization and cellular dynamics, advancing the field toward more comprehensive and precise biological interpretations.

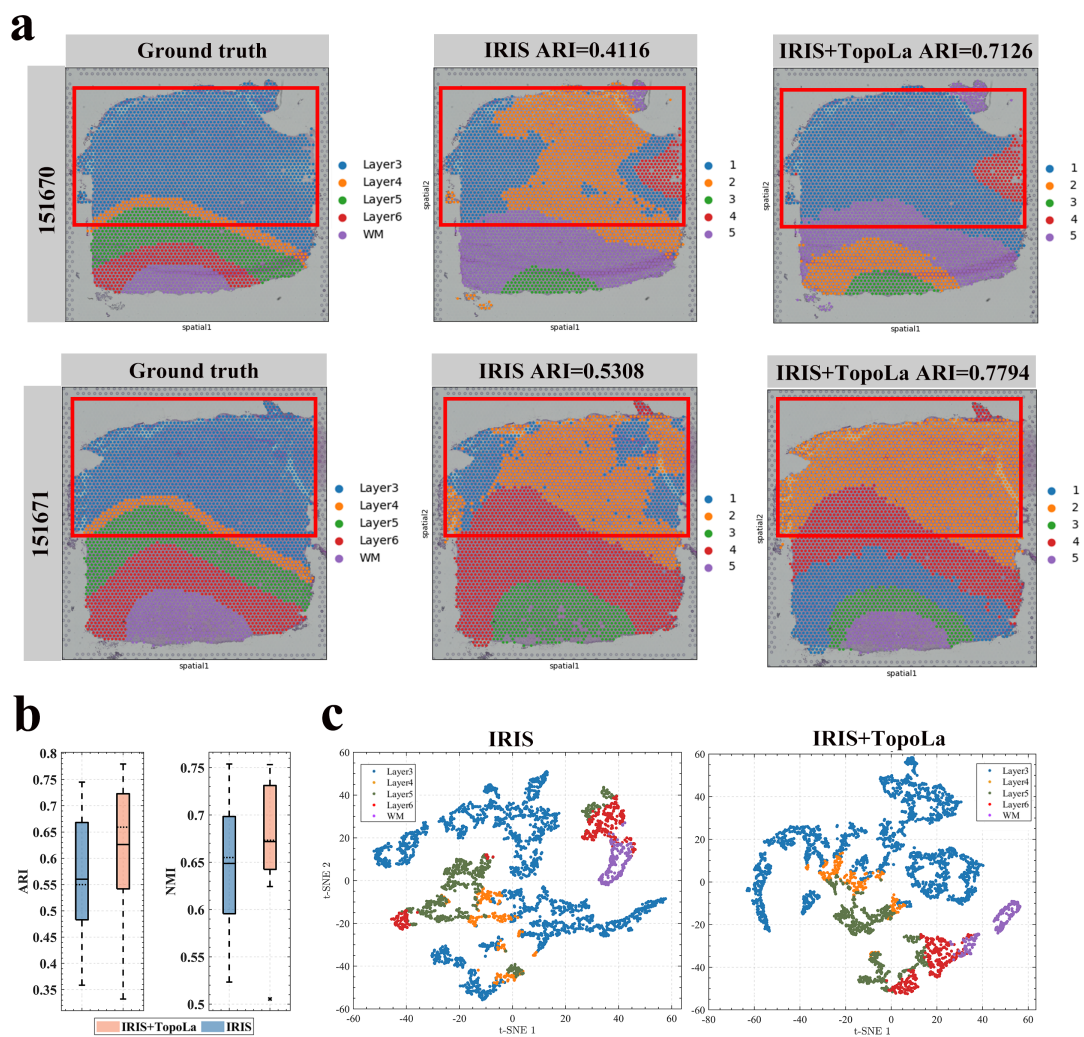

**Fig. 6 Identification of spatial domain using IRIS and IRIS+TopoLa on ST and scRNA-seq data. a**

Clustering results using IRIS and IRIS+TopoLa on slices 151670 and 151671 from the DLPFC dataset. **b** Comparative performance evaluation of spatial domain identification using ARI and NMI across 12 slices. The solid line in each box represents the mean ARI and NMI values, while the dashed line represents the median ARI and NMI values. **c** Visualizations of cell embeddings generated by IRIS and IRIS+TopoLa. The left panel shows the initial spatial distribution of cell types in slice 151670, with each type represented by a distinct color. The right panel illustrates the refined distribution after TopoLa enhancement.

## Discussion

We introduce TopoLa, a novel latent hyperbolic geometry framework that enhances cell representations by capturing fine-grained intercellular relationships. The framework includes two key components: TLd and TopoConv. Specifically, TopoLa maps cells into a latent hyperbolic space based on their geometric structural feature within the cellular network. TLd quantifies the distances between cells in this hyperbolic space. To the best of our knowledge, TopoLa is the first generalizable framework capable of enhancing cell representations to facilitate across various types of cellular studies. We further validate TopoLa's effectiveness by demonstrating the universal performance improvements of TopoLa over multiple state-of-the-art models across seven key biological tasks (**Supplementary Table 1**). Notably, TopoLa achieves significant enhancements in single-cell multi-batch integration (15.4%), single-cell multi-omic integration (45.7%), and spatially informed clustering with the integration of scRNA-seq and ST (11.7%).

In addition to its applications in specific biological tasks, we have also examined the physical significance of TLd and the mathematical principles behind TopoConv. For TLd, we conducted an in-depth analysis of its physical meaning and its role in accurately measuring geometric similarities between cells in cellular networks. Guided by the principle of maximum entropy, TopoLa achieves greater precision compared to existing energy-distance-based latent hyperbolic geometry frameworks (**Supplementary text 3, Supplementary Fig. 44**). The traditional energy distance measure primarily focus on local node connectivity, often overlooking the crucial impact of global connectivity and topological structure on latent space embeddings. In contrast, TLd incorporates global connectivity, making it more precise in capturing the geometric structure of networks. By analyzing physical properties such as paths and topologies, TLd provides valuable insights into intercellular geometric relationships, framed through the perspectives of graph theory and network geometry. For TopoConv, we further explore its mathematical foundations and the mechanisms by which it enhances cell representations. We use matrix perturbation theory and singular value gap analysis to justify the mathematical reasons underlying TopoConv's ability to improve cell representations, we apply matrix perturbation theory and singular value gap analysis. Our results show that convolving neighboring cells especially those with similar topological structures in latent space increases the singular value gap, effectively reducing noise interference in singular subspaces. This mathematical analysis not only explains how TopoConv enhances cell representations but also establishes a new theoretical foundation for future research.

The essence of this study lies in the introduction of a novel network geometry framework, TopoLa, designed to capture the geometric relationships between cells within cell networks. We assessed the applicability of TopoLa for scRNA-seq and ST analyses across seven key biological tasks. Beyond these applications, TopoLa holds potential for other cell representation-based tasks, such as cell-cell communication, gene regulatory network inference, and spatial multi-omics reconstruction.

As TopoLa is a general network geometry framework, it is not limited to capturing the geometric structure of cell representations alone but can be extended to networks involving general entities and their associated applications. For instance, in link prediction, which involves predicting future connections based on existing ones, integrating network geometric features through TopoLa could enhance the performance of current link prediction methods. Moreover, deep learning approaches generate entity embeddings, such as "word embeddings" and "graph embeddings". By introducing the network geometry of TopoLa into these embeddings, it may be possible to improve the performance of various downstream tasks. In addition to enhancing representations, TopoLa offers a new metric, TLd, for evaluating relationships between entities. TLd could complement existing metrics for analyzing and assessing relationships, potentially benefiting models that involve contrastive learning and knowledge distillation, and other related models.

The TopoLa framework offers a versatile solution for working with arbitrary matrices, enhancing representations and networks by integrating geometric structural features. This flexibility facilitates seamless integration into various approaches that employ networks or representations. By embedding geometric structures, TopoLa provides a novel way for improving the performance of diverse techniques in tasks related to representational analysis.

In conclusion, TopoLa represents a foundational tool in cell analysis, offering a novel perspective on intercellular geometric relationships. We anticipate that TopoLa will provide fresh insights into cell science. The methodology behind TopoLa has the potential for a wide spectrum of applications across bioinformatics, deep learning, social networks, cybersecurity, and other fields involving matrix analysis.

## Methods

Building on accurately encoding the geometric structure of cell networks, TopoLa offers two components: TLd and TopoConv. TLd represents the distance between cells in latent hyperbolic space, while TopoConv leverages the spatial relationships of cells in this space to enhance cell representations. In the following sections, we outline the theoretical basis of this framework and the key innovations in TLd and TopoConv that drive its enhanced performance, and a description of the datasets used for validation.

### TopoLa framework and TopoLa distance

TopoLa is a network geometry framework designed to model the geometrical structure of cell networks. However, the existing energy distance-based network geometry framework (latent hyperbolic geometry), face two key challenges: (1) Its application to real-world problems remains largely unexplored, and (2) It requires sufficiently large networks to accurately model geometry through the statistics of 2-hop paths. To address these challenges, we refine existing latent hyperbolic geometry frameworks by incorporating the principle of maximum entropy.

Previous studies have shown that when the logarithm of thermodynamic activity is high, the expected number of common neighbors closely approximates the complement of the energy distance[42]. We demonstrate that, under the same network, applying weighted statistics to 2-hop paths yields more accurate measurements of energy distance compared to unweighted statistics (**Supplementary text 2.1**, **Theorem 1**). The energy distance in latent space represents the ratio between energy (spatial distance) and chemical potential (a function of the expected degree)[42,43]. Consequently, we incorporate node (cell) degree and global connectivity as weighting factors into the computation of energy distance, and propose the TLd matrix $\boldsymbol{D}_{topo}$ satisfying the following formula:

$$\boldsymbol{D}_{topo} + \boldsymbol{D}_{topo}^2 + \boldsymbol{D}_{topo}^3 + \boldsymbol{D}_{topo}^4 + \cdots = \frac{\boldsymbol{A}\boldsymbol{A}^T}{\lambda} \tag{1}$$

where $\boldsymbol{A}$ is the adjacency matrix of the original network, $\lambda$ is a coefficient modulating the influence of degree. $\boldsymbol{D}_{topo}^n$ is the matrix power of $\boldsymbol{D}_{topo}$, which represents the $n$-hop path matrix of $\boldsymbol{D}_{topo}$ and $\boldsymbol{D}_{topo}^n(i,j)$ is the weighted sums of all $n$-hop paths between $i$ and $j$ in $\boldsymbol{D}_{topo}$. $\boldsymbol{A}\boldsymbol{A}^T(i,j)$ is the weighted sums of all 2-hop paths between $i$ and $j$ in $\boldsymbol{A}$. Aggregating all hop paths not only introduces both global connectivity and degree information (see section-"Physical significance of the TopoLa distance") but also facilitates the derivation of a closed-form solution. This improves the importance of 2-hop paths for low-degree nodes. Through the Infinite Neumann series, $\boldsymbol{D}_{topo}$ can be solved as:

$$\boldsymbol{D}_{topo}\big(\boldsymbol{I} - \boldsymbol{D}_{topo}\big)^{-1} = \frac{\boldsymbol{A}\boldsymbol{A}^T}{\lambda} \tag{2}$$

$$\boldsymbol{D}_{topo} = \boldsymbol{A}\boldsymbol{A}^T(\lambda\boldsymbol{I} + \boldsymbol{A}\boldsymbol{A}^T)^{-1} \tag{3}$$

where $\boldsymbol{I}$ is the identity matrix. Therefore, we propose the TopoLa framework, which consists of latent space embedding, energy distance measure, and spatial specificity (such as the triangle inequality). Eq. 3 enables the derivation of a closed-form solution for $\boldsymbol{D}_{topo}$, which allows the calculation of its exact value (The process for singular value calculation is depicted in **Fig. 1a**). Therefore, the topologically encoded latent hyperbolic geometry can offer a novel perspective for theoretical analysis in these domains. The TLd between node $i$ and node $j$ in the network is expressed as:

$$d_{topo}(i,j) = \frac{1}{\lambda}|2-\text{hop}| - \frac{1}{\lambda^2}|4-\text{hop}| + \frac{1}{\lambda^3}|6-\text{hop}| - \cdots \tag{4}$$

Our findings reveal that $d_{topo}$ captures global connectivity information, which can be used to measure topological similarity between two nodes, a concept that we define as 'global connectivity-based topological similarity' (see section-"Physical significance of the TopoLa distance"). This establishes a clear correlation between the topological structure of nodes and their spatial positioning in latent space, notably clustering nodes with similar topology (**Supplementary text 2.1**, **Theorem 2**).

In summary, $\boldsymbol{D}_{topo}$ has the following two unique properties which are useful in a variety of applications:

1. $\boldsymbol{D}_{topo}$ encodes the precise information of nodes' global connectivity, topological structure, and degree (see section-"Physical significance of the TopoLa distance").

2. $\boldsymbol{D}_{topo}$ enables the utilization of the chemical potential across various networks for energy distance measure, instead of employing a uniform metric scale for all (see section-"Physical significance of the TopoLa distance").

Through **Theorem 2**, we demonstrate that $\frac{||\boldsymbol{D}_{topo}(i)-\boldsymbol{D}_{topo}(j)||_F^2}{||\boldsymbol{y}||_F^2} \leq ||\boldsymbol{A}(i)-\boldsymbol{A}(j)||_F^2$, which further proves that $\boldsymbol{D}_{topo}$ reflects the nodes' topological structure (**Supplementary text 2.2**), where $\boldsymbol{D}_{topo}(i)$ is the $i$-th row of matrix $\boldsymbol{D}_{topo}$, denoting the set of distances between node $i$ and the other nodes, $\boldsymbol{y}$ represents any row within matrix $\boldsymbol{A}$ and $\boldsymbol{A}(i)$ is the $i$-th row of matrix $\boldsymbol{A}$, denoting the topological structure of node $i$, and $||.||_F$ denotes the Frobenius norm.

Cell representations as matrices can be interpreted as networks, enabling the geometric structure of cell networks to reveal fine-grained intercellular relationships. TopoLa utilizes these representations to model the geometric structure of the cell network and computes the geometric structural similarity between cells using the TLd. Additionally, TopoLa includes a specialized component called TopoConv, which enhances cellular representations by convolving neighboring cells.

**Physical significance of the TopoLa distance**

Our way to exploring the physical significance of TopoLa distance is to differentiate various types of paths connecting nodes $i$ and $j$ in original network (cell representations). By counting the number of each type of paths, it allows us to take into consideration not only local and global connectivity, but also node degrees in an implicit manner. Particularly, we focus on the classification of $n$-hop paths connecting nodes $i$ and $j$. Due to the possible existence of loop or loops on these paths, we establish a classification system for all such $n$-hop paths, based on the length of a path actually traversed (i.e., the number of hops after removing all the loops on the path). Clearly, there are a total of ($n$-1) types. To facilitate a clear differentiation, we represent these path types with polygons. For example, a path that actually traverses 2-hop is defined as a $P_2$ path, which can be viewed as a triangle if adding a direct edge back from node $j$ to node $i$. Similarly, a path that actually traverses 3 hops can be defined as $P_3$ (represented as a quadrangle), and a path that actually traverses 4 hops can be defined as $P_4$ (represented as a pentagon; **Supplementary Fig. 45a**). The quantity of these types can be expressed as $|P_l| = loop(a_l)$, where $a_l$ is the set of loop-free $l$-hop paths connecting nodes $i$ and $j$ and $loop(\cdot)$ represents the number of different ways of adding loops to paths in $a_l$ to form $n$-hop paths between nodes $i$ and $j$. While there is a correlation between $|P_l|$ and $|a_l|$, accurately measuring this relationship remains challenging. Nevertheless, our observations suggest that a proportional relationship exists between the quantity of $P_l$, characterized by solely loops between node $i$ and its neighbors, or between node $j$ and its neighbors, and $|a_l|$. Paths exhibiting such characteristics are defined as $b_n(l)$ (**Supplementary Fig. 45b**). The interplay between $|a_l|$ and $|b_n(l)|$ can be calculated as follows:

$$|b_n(l)| = \sum_{h=0}^{\frac{n-l}{2}} \kappa_i^h \kappa_j^{\frac{n-l}{2}-h} |a_l| \qquad (5)$$

where $|b_n(l)|$ is the quantity of $b_n(l)$ in $n$-hop $P_l$ paths. $\kappa_i$ and $\kappa_j$ represent the degrees of node $i$ and node $j$. Thus, $n$-hop paths between node $i$ and node $j$ in the network are represented as:

$$|n - \text{hop}| = |a_n| + |b_n| + |c_n|$$

$$= |a_n| + \sum_{P_l \in n-\text{hop}} |b_n(P_l)| + |c_n|$$

$$= \begin{cases} \sum_{t=1}^{\frac{n}{2}} \sum_{h=0}^{\frac{n}{2}-t} \kappa_i^h \kappa_j^{\frac{n}{2}-t-h} |a_{2t}| + |c_n| & \text{if } n \text{ is even} \\ \sum_{t=1}^{\frac{n-1}{2}} \sum_{h=0}^{\frac{n-1}{2}-t} \kappa_i^h \kappa_j^{\frac{n}{2}-t-h} |a_{2t+1}| + |c_n| & \text{if } n \text{ is odd} \end{cases} \qquad (6)$$

where $c_n$ denotes the paths distinct from $a_n$ and $b_n$. $|n - \text{hop}|$ refers to the counts of $n$-hop paths. Our analysis for 2-hop, 4-hop, and 6-hop paths reveals that $|c_n|$ does not exhibit a direct correlation with $\kappa_i$ and $\kappa_j$ (**Supplementary text 4, Supplementary Fig. 46**). Eq. 6 reveals that $a_n$ and $b_n$ for even-hop paths consist of polygons with an odd number of sides, while for odd-hop paths, they comprise polygons with an even number of sides.

Subsequently, we explore further the physical significance of $\boldsymbol{D}_{topo}$, which can also be expressed in the form of a set of paths as shown in Eq. 4. The emphasis on even-hop paths arises from their ability to capture the degree information and the global connectivity information of nodes $i$ and $j$. Furthermore, the TopoLa distance $d_{topo}$ is demonstrated to correlate with the topological structure of nodes. Specifically, $a_l$ can be classified into two types: those overlapping with 2-hop paths ($g_l$) and the remaining ones ($s_l$), with $|a_l| = |g_l| + |s_l|$ (**Supplementary Fig. 45c**). The $|s_l|$ relates to the topological structure of nodes $i$ and $j$. For instance, identical topological structure between $i$ and $j$ renders $|s_l|$ to zero (which indicates that it measures the topological similarity of nodes $i$ and $j$). Thus, we can express TLd between cell $i$ and cell $j$ as:

$$d_{topo}(i,j) = \lim_{n \to \infty} \left[ \left( \sum_{t=1}^{\frac{n}{2}} \sum_{h=0}^{\frac{n}{2}-t} \frac{\kappa_i^h \kappa_j^{(\frac{n}{2}-t-h)}}{-(-\lambda)^{(\frac{n}{2}-t+1)}} \right) \left( \sum_{t=1}^{\frac{n}{2}} \frac{|a_{2t}|}{(-\lambda)^{t-1}} \right) + \sum_{t=1}^{\frac{n}{2}} \frac{c_{(2t)}}{\lambda^t} \right]$$

$$= \lim_{n \to \infty} \left[ \left( \sum_{t=1}^{\frac{n}{2}} \sum_{h=0}^{\frac{n}{2}-t} \frac{\kappa_i^h \kappa_j^{(\frac{n}{2}-t-h)}}{-(-\lambda)^{(\frac{n}{2}-t+1)}} \right) \left[ \sum_{t=1}^{\frac{n}{2}} \frac{|g_{2t}|}{(-\lambda)^{t-1}} - \sum_{t=2}^{\frac{n}{2}} \frac{|s_{2t}|}{(-\lambda)^{t-1}} \right] + \sum_{t=1}^{\frac{n}{2}} \frac{c_{(2t)}}{\lambda^t} \right] \quad (7)$$

where $\sum_{t=1}^{\frac{n}{2}} \frac{|g_{2t}|}{(-\lambda)^{t-1}}$ and $\sum_{t=1}^{\frac{n}{2}} \frac{c_{(2t)}}{\lambda^t}$ are unaffected by node's topological structure, while $\sum_{t=2}^{\frac{n}{2}} \frac{|s_{2t}|}{(-\lambda)^{t-1}}$ acts as a topology-dependent penalty term.

**Spatial convolution via topology-encoded latent hyperbolic geometry**

We introduce a technique, Spatial convolution via topology-encoded latent hyperbolic geometry (TopoConv), which utilizes TLd to convolve neighboring cells to enhance cell representations. This process aims to reconfigure the cells' topological structure based on their latent space distances. TopoConv takes an original network (cell representations) as input to generate an enhanced network $\boldsymbol{A}^*$ without supervision or prior knowledge. The formula is as follows:

$$\boldsymbol{A}^*(j) = \sum_{k=1}^{N} \boldsymbol{D}_{topo}(k,j) \boldsymbol{A}(k) \quad (8)$$

where $\boldsymbol{A}^*(j)$ is the representation of the $j$-th cell. $N$ is the number of cells. In the formula, TopoConv essentially uses TLd as a weight to combine all cell representations into a new representation. We refer to this process as the convolution of neighboring cells. The reason for using "neighboring cells" is that, in latent hyperbolic space, most cells are not closely positioned, and TLd weights can differ by several orders of magnitude. We subsequently analyzed the changes in properties of the original network after adding geometric structural features. In summary, the enhanced network $\boldsymbol{A}^*$ has the following two properties:

1. The singular value gap of $\boldsymbol{A}^*$ is larger than that of $\boldsymbol{A}$ (**Supplementary text 2.2**, **Theorem 3**). while the left and right singular vectors of $\boldsymbol{A}^*$ remain the same as those of $\boldsymbol{A}$ (**Supplementary text 2.2**, **Theorem 3**). Specifically, we proved that when $|\sigma_k| \in [\sqrt{\lambda}, +\infty)$, the corresponding singular value gap $(\sigma_{k-1}^* - \sigma_k^*)$ of $\boldsymbol{A}^*$ is larger than the singular value gap $(\sigma_{k-1} - \sigma_k)$ of $\boldsymbol{A}$, where $\sigma_k$ is the $k$-th largest singular value of $\boldsymbol{A}$ and $\sigma_k^* = \sigma_k \frac{\sigma_k^2}{\sigma_k^2 + \lambda}$ is the $k$-th largest singular value of $\boldsymbol{A}^*$. The increase in the singular value gap and the invariance of the left and right singular matrices are prerequisites for our analysis. These conditions allow us to apply matrix perturbation theory to analyze how Gaussian noise affects the properties of both the original and enhanced networks.

2. The error upper bound on the distance of $\boldsymbol{A}^*$ is less than or equal to that of $\boldsymbol{A}$ (**Supplementary text 2.2**, **Theorem 4**), which suggests that $\boldsymbol{A}^*$ is more robust against Gaussian noises, errors, and data missing. By matrix perturbation theory, when $\sigma_{k-1} - \sigma_k > 2\|\boldsymbol{H}\|$, the error upper bound of the enhanced network is $\min \left\{ \frac{2\|\boldsymbol{H}\|}{\frac{\sigma_{k-1}^2}{\sigma_{k-1}^2 + \lambda}\sigma_{k-1} - \frac{\sigma_k^2}{\sigma_k^2 + \lambda}\sigma_k}, 1 \right\}$, which is less than or equal to the error upper bound of the original network $\min \left\{ \frac{2\|\boldsymbol{H}\|}{\sigma_{k-1} - \sigma_k}, 1 \right\}$ (**Supplementary Fig. 47**), where $\boldsymbol{H}$ is the Gaussian noise matrix, satisfying $\widetilde{\boldsymbol{A}} = \boldsymbol{A} + \boldsymbol{H} \in \mathbb{R}^{n \times m}$, with $\widetilde{\boldsymbol{A}}$ being the perturbed version of $\boldsymbol{A}$. It means that the perturbation to the left and right singular matrices is smaller when the Gaussian noise matrix is added to the enhanced network $A^*$, compared to when the same noise matrix is added to the original network $A$. In other words, the same Gaussian noise has a smaller impact on $A^*$ compared to $A$.

Cells enhance their representations by combining the geometric features of neighboring cells within a latent hyperbolic space, bringing them closer to geometrically similar cells. According to matrix perturbation theory, the enhanced cell representations are less affected by Gaussian noise, indicating that the distance between geometrically dissimilar cells increases. Consequently, we provide a multi-faceted explanation of the principles underlying TopoLa's enhancement of cell representations.

**Data description**

We use the TopoLa framework to investigate seven crucial cell analysis tasks across diverse datasets, encompassing various technologies, species, and tissues. Detailed information on the datasets and preprocessing steps is provided below.

**Clustering of single-cell RNA-seq data.** We utilize a diverse set of 30 scRNA-seq datasets generated from different sequencing platforms (**Supplementary Table 2**), encompassing a wide range of biological samples and experimental conditions. These datasets include human pancreatic cells (Baron_human1-4), mouse pancreatic cells (Baron_mouse), mouse embryonic stem cells (Biase, Buettner, Kolodz, Kumar, Leng), human pluripotent stem cells (Chu_celltime, Chu_celltype), human tumor and immune cells (Chung), human brain cells (Darmanis, Pollen), mouse cells from different stages (Deng, Goolam), mouse Natural killer T cells (Engel), human hematopoietic stem cells (Grover), human myxoid liposarcoma cells (Karlsson), non-malignant cells in head and neck cancer (Puram), mouse leukemia cell line and primary CD8+ T-cells (Robert), mouse circulating tumor cells (Ting), mouse lung epithelial cells (Treutlein), human embryonic stem cells (Yan), human induced pluripotent stem cells (Yeo), mouse cerebral cortex cells (Zelsel), mouse haematopoietic stem cells (Zhou), and Jurkat cells (Jurkat from 10X Genomics). These datasets, sourced from reputable databases such as GEO[44], E-MTAB[45], and 10X Genomics[46], provide a robust and comprehensive foundation for the clustering analysis of scRNA-seq data, enabling us to evaluate the performance of our model across various biological contexts and sequencing platforms.

**Single-cell multi-batch integration.** The evaluation employed three integration datasets: the COVID-19 dataset[47], the peripheral blood mononuclear cell (PBMC) 10k dataset[48], and the perirhinal cortex dataset[49]. The COVID-19 dataset, sourced from the NCBI GEO (GSE1459261) and derived from the work by Lotfollahi et al.[47], comprises 62,469 cells from 18 batches. This dataset includes merged data from lungs, PBMCs, and bone marrow samples, normalized using Scanpy with 2,000 highly variable genes selected for analysis. Initially containing 274,346 cells and 18,474 genes, it was subsampled to 20,000 cells for this study. For reference mapping evaluation, 12 batches (15,997 cells) were used as the reference, and six batches (4,003 cells) as the query. Cell type labels were taken from the original study.

The PBMC 10k dataset, processed and deposited on Figshare, includes data from two batches of human PBMCs obtained from a healthy donor. Reprocessed by Gayoso et al.[48], this dataset consists of 3,346 differentially expressed genes across two batches: the first with 7,982 cells and the second with 4,008 cells. Annotated using Seurat, the cell groups include B cells, CD4+ T cells, CD8+ T cells, CD14+ monocytes, dendritic cells, NK cells, FCGR3A+ monocytes, megakaryocytes, and others.

The perirhinal cortex dataset features adolescent mouse brain data from two batches, drawn from a larger study by Siletti et al.[49], originally containing 606 high-quality samples across ten diverse brain regions. The selected batches from the perirhinal cortex consist of 8,465 cells and 9,070 cells, respectively, incorporating a broad range of 59,357 genes. Annotations of ten unique cell types from the original study were used.

**Single-cell multi-omics integration.**   We benchmark scGPT for single-cell multi-omic integration using paired data, specifically comparing scGPT against scGPT+TopoLa on the BMMC dataset[50]. The BMMC dataset contains paired single-cell RNA and protein abundance measurements from bone marrow mononuclear cells (BMMCs), sequenced using the CITE-seq protocol. This dataset includes samples from 12 healthy human donors, distributed across 12 batches, comprising a total of 90,261 cells. Measurements were taken from 13,953 genes and 134 surface proteins, with annotations for 45 detailed immune cell subtypes.

**Single-cell rare cell identification.** We adopt a comprehensive set of 20 scRNA-seq datasets (**Supplementary Table 3**), each representing diverse biological samples and experimental conditions across multiple species and tissue types. In this study, we utilized a comprehensive set of 20 scRNA-seq datasets, each representing diverse biological samples and experimental conditions across multiple species and tissue types. The datasets include peripheral blood mononuclear cells (10X PBMC from 10X Genomics), airway epithelial cells (Airway), single cell RNA-seq of C. elegans (Cao), human tumor and immune in primary breast cells (Chung), human brain (Darmanis), early-stage mouse embryos (Deng), early mouse embryonic cells (Goolam), hippocampal neurons (hippocampus), murine lymph node cells (iLNs), human colorectal tumors (Li), liver cells (livers), pancreatic cells (pancrea), human developing cerebral cortex cells (Pollen), head and neck tumor cells (Puram), human tonsil immune cells (Tonsil from Broad Institute), kidney immune cells (UUOkidney), mouse hair follicles stem cells (Yang), mouse cortex and hippocampus (Zelsel), human embryonic stem cells (Koh), and human liver cells (MacParland). These datasets, sourced from reputable databases such as GEO[44], 10X Genomics[46], ArrayExpress[44], sci-RNA-seq[51], and the Broad Institute Single Cell Portal[52], provide a robust foundation for evaluating the performance of our model in identifying rare cell populations within complex and heterogeneous biological systems.

**Spatially informed clustering of ST**. For spatially informed clustering, we involve five spatial gene expression datasets. The first dataset is the DLPFC with 12 tissue slices acquired using 10x Visium (http://research.libd.org/spatialLIBD/)[31]. Each slice contains between 3,460 and 4,789 spots, capturing a total of 33,538 genes. Manual annotations identify five to seven regions per slice, including the DLPFC layers and white matter.

**Vertical integration of multiple tissue slices.** For the multiple sample integration task, we employ vertical integration of the dataset, sourced from graphST[11], mouse breast cancer tissue samples. The dataset comprises pairs of vertically adjacent slices from the same tissue. The number of spots per slice varies between 1,868 and 1,950.

**Spatially informed clustering of** scRNA-seq and ST data. For spatially informed clustering of multi-modal scRNA-seq and ST data, we utilize the DLPFC dataset[31] acquired using 10x Visium, which includes 12 tissue slices, each containing between 3,460 and 4,789 spots[31]. This dataset captures a total of 33,538 genes and is manually annotated to identify key spatial regions, including cortical layers and white matter. Additionally, scRNA-seq data from 10x Chromium on post-mortem brain tissue, encompassing 44 cell types, is used as a reference[53]. The snRNA-seq data are available through The Rush Alzheimer's Disease Center (RADC) Research Resource Sharing Hub and Synapse.

## Data availability

The datasets generated and analyzed during this study are publicly available. Public datasets used in this study can be accessed through various repositories. For clustering of scRNA-seq data, datasets are available from public repositories using the accession codes listed in **Supplementary Table 2**, including NCBI Gene Expression Omnibus (GEO) [https://www.ncbi.nlm.nih.gov/geo/], ArrayExpress [https://www.ebi.ac.uk/arrayexpress/], and the Sequence Read Archive (SRA) [https://www.ncbi.nlm.nih.gov/sra]. The mouse bladder cell dataset (Han) is accessible via the Mouse Cell Atlas project [https://bis.zju.edu.cn/MCA/index.html], the 10X PBMC dataset is available from 10X Genomics [https://www.10xgenomics.com/datasets], and the worm neuron dataset (Cao) was obtained from the sci-RNA-seq platform [http://atlas.gs.washington.edu/worm-rna/docs/]. For single-cell multi-batch integration, the processed COVID-19 dataset can be accessed at [https://github.com/theislab/scarches-reproducibility], the PBMC 10k dataset from scVI tools [https://scvi-tools.org/], and the perirhinal cortex dataset from the CELLxGENE Human Brain Cell Atlas version 1.0 [https://cellxgene.cziscience.com/collections/283d65eb-dd53-496d-adb7-7570c7caa443]. For single-cell multi-omic integration, the BMMC dataset is available from GEO with the accession number GSE194122. For rare cell identification, all datasets are accessible from the public repositories listed in **Supplementary Table 3**, including GEO, ArrayExpress, and SRA, with the Jurkat dataset available from 10X Genomics [https://www.10xgenomics.com/datasets/50-percent-

50-percent-jurkat-293-t-cell-mixture-1-standard-1-1-0], and the worm neuron dataset (Cao) from the sci-RNA-seq platform [http://atlas.gs.washington.edu/worm-rna/docs/]. The preprocessed human tonsil data (Tonsil) and Crohn's disease data are available from the Broad Institute Single Cell Portal [https://singlecell.broadinstitute.org/single_cell/study/SCP2169/slide-tags-snrna-seq-on-human-tonsil] and [https://singlecell.broadinstitute.org/single_cell/study/SCP359/ica-ileum-lamina-propria-immunocytes-sinai], respectively. For spatially informed clustering of ST, the human DLPFC data can be accessed via 10X Visium [http://spatial.libd.org/spatialLIBD/]. For vertical integration of multiple tissue sections of ST, the mouse breast cancer tissue samples are available at [https://zenodo.org/record/6925603#.YuM5WXZBwuU]. Finally, for spatially informed clustering of scRNA-seq and ST, the human DLPFC data is accessible at [http://spatial.libd.org/spatialLIBD/], and human post-mortem brain single-nucleus RNA-seq reference data can be accessed via Synapse [https://www.synapse.org/#!Synapse:syn18485175].

## Code availability

All code produced in this study is freely available in supplementary materials or open repositories, with the code for methods development and validation accessible on GitHub (https://github.com/kaizheng-academic/TopoLa)

## Funding

This work was supported in part by the National Key Research and Development Program of China (No.2021YFF1201200); the National Natural Science Foundation of China under Grants (Nos. 62350004, 62332020, U22A2041); the Project of Xiangjiang Laboratory (No. 23XJ01011). The

## Author contributions

## Competing interests

The authors declare no competing interests.

**Supplementary Information** for

**TopoLa: A Universal Framework for Enhancing Cellular Research through Topology-Encoded Latent Hyperbolic Geometry**

Zheng *et al.*

 **Supplementary Figures**

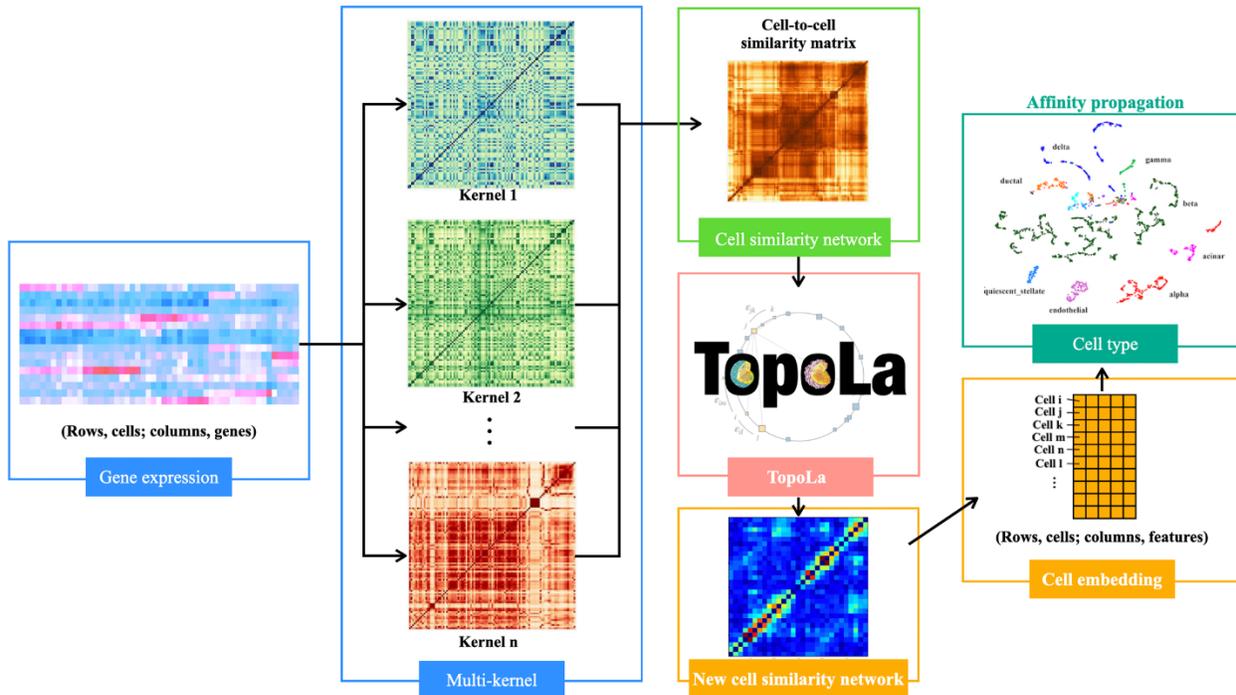

6
7 **Supplementary Fig. 1** Enhance cell representations of SIMLR using the TopoLa framework. The
8 pipeline begins with a gene expression matrix as input. SIMLR+TopoLa calculates multiple
9 kernels to learn intercellular similarities. This similarity matrix is then input into the TopoLa
10 framework, refining cellular similarities. These similarities are subsequently processed through
11 dimensionality reduction techniques to generate cellular representations. Finally, cell clustering is
12 performed using the affinity propagation algorithm.

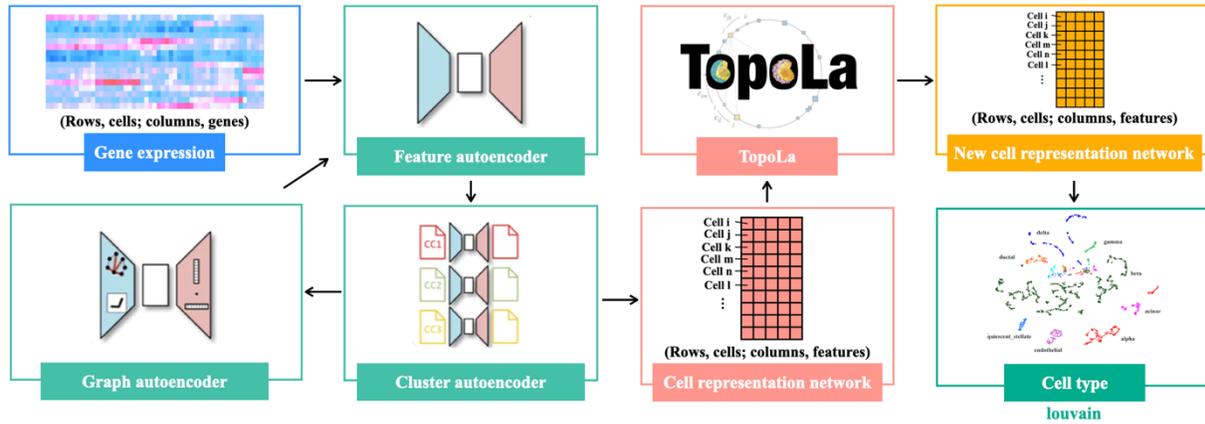

**Supplementary Fig. 2** Enhance cell representations of scGNN using the TopoLa framework. Firstly, the gene expression matrix is used as input. A feature autoencoder learns dimensional representations of the input as embeddings, which are subsequently used to construct and refine a cell graph. A graph autoencoder then learns the topological embeddings of the cell graph for clustering. Within each cell type, a distinct clustering autoencoder is used to reconstruct gene expression values. The framework iteratively uses the reconstructed expressions as new input until convergence is achieved. Subsequently, an imputation autoencoder regularizes the imputed gene expression values based on the cell-cell relationships learned from the cell graph, applying this regularization to the original preprocessed expression matrix. The final learned cell graph is used as the cell representation input for the TopoLa framework, where the enhanced cell graph is fed into the Louvain algorithm for cell clustering.

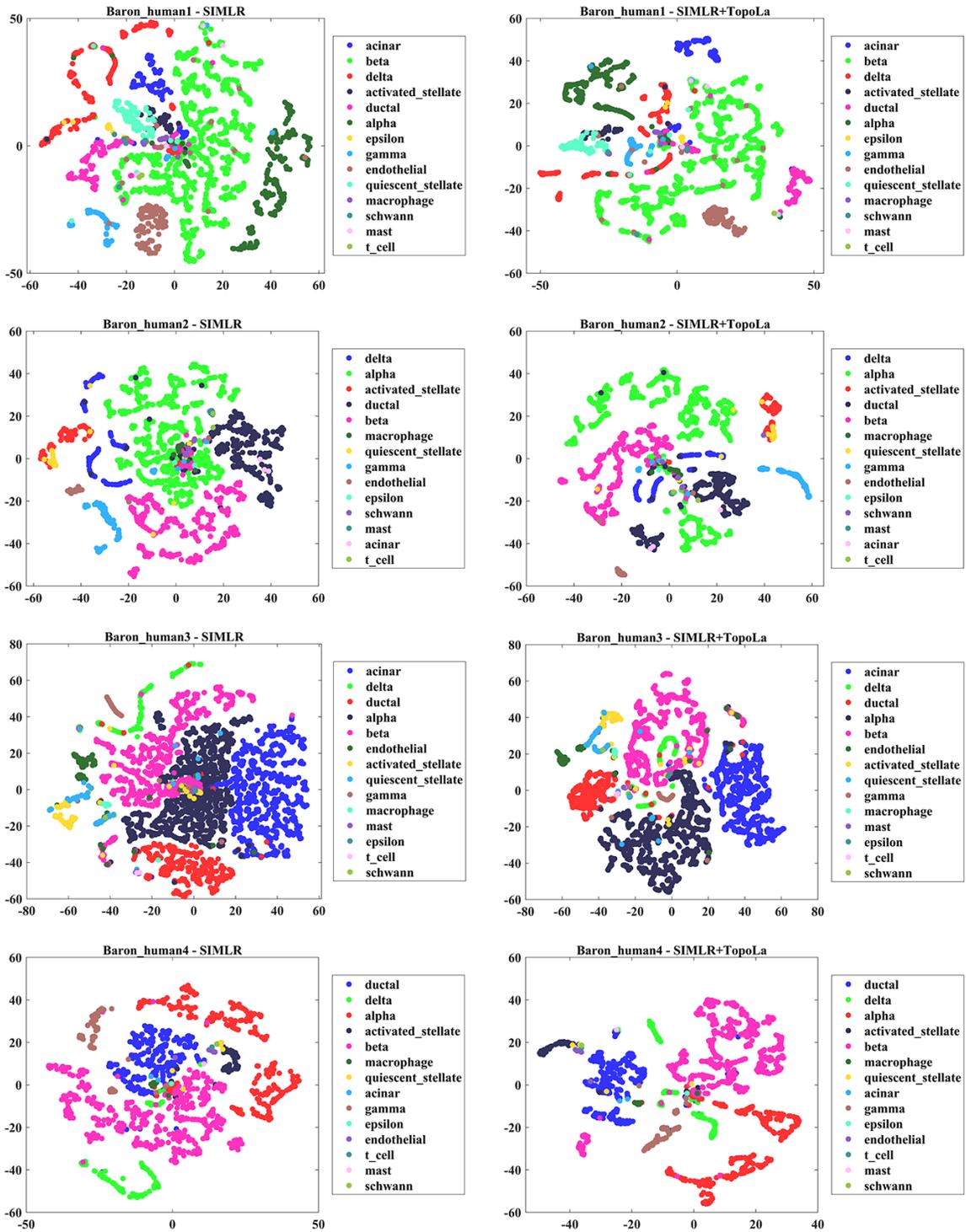

**Supplementary Fig. 3** t-SNE visualizations of cell type representations in the Baron_human1, Baron_human2, Baron_human3, and Baron_human4 datasets using SIMLR and SIMLR+TopoLa. The plots depict the quality of cell representations achieved by both models, offering visual confirmation of TopoLa's enhancement capabilities.

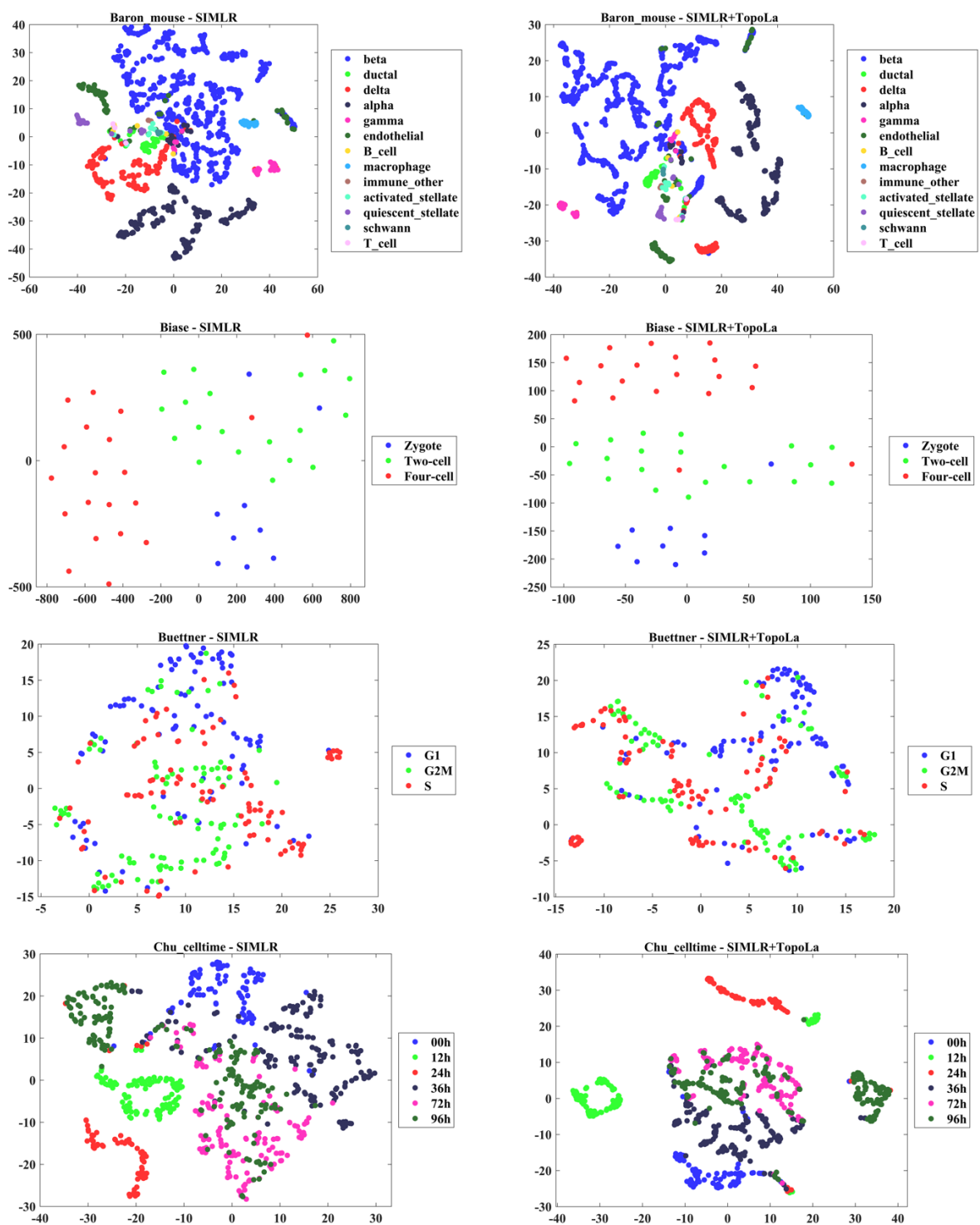

**Supplementary Fig. 4** t-SNE visualizations of cell type representations in the Baron_mouse, Biase, Buettner, and Chu_celltime datasets using SIMLR and SIMLR+TopoLa. The plots depict the quality of cell representations achieved by both models, offering visual confirmation of TopoLa's enhancement capabilities.

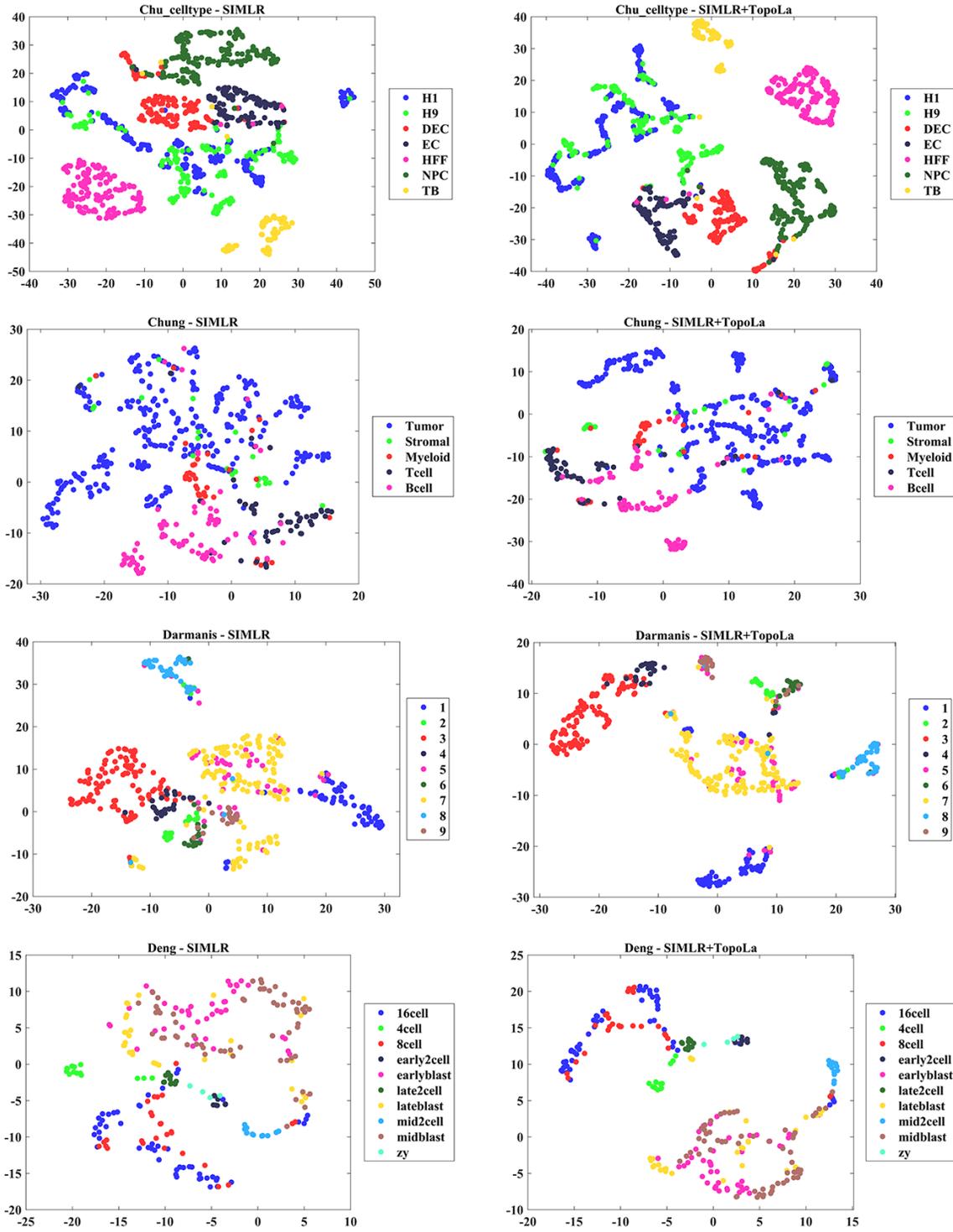

**Supplementary Fig. 5** t-SNE visualizations of cell type representations in the Chu_celltype, Chung, Darmains, and Deng datasets using SIMLR and SIMLR+TopoLa. The plots depict the quality of cell representations achieved by both models, offering visual confirmation of TopoLa's enhancement capabilities.

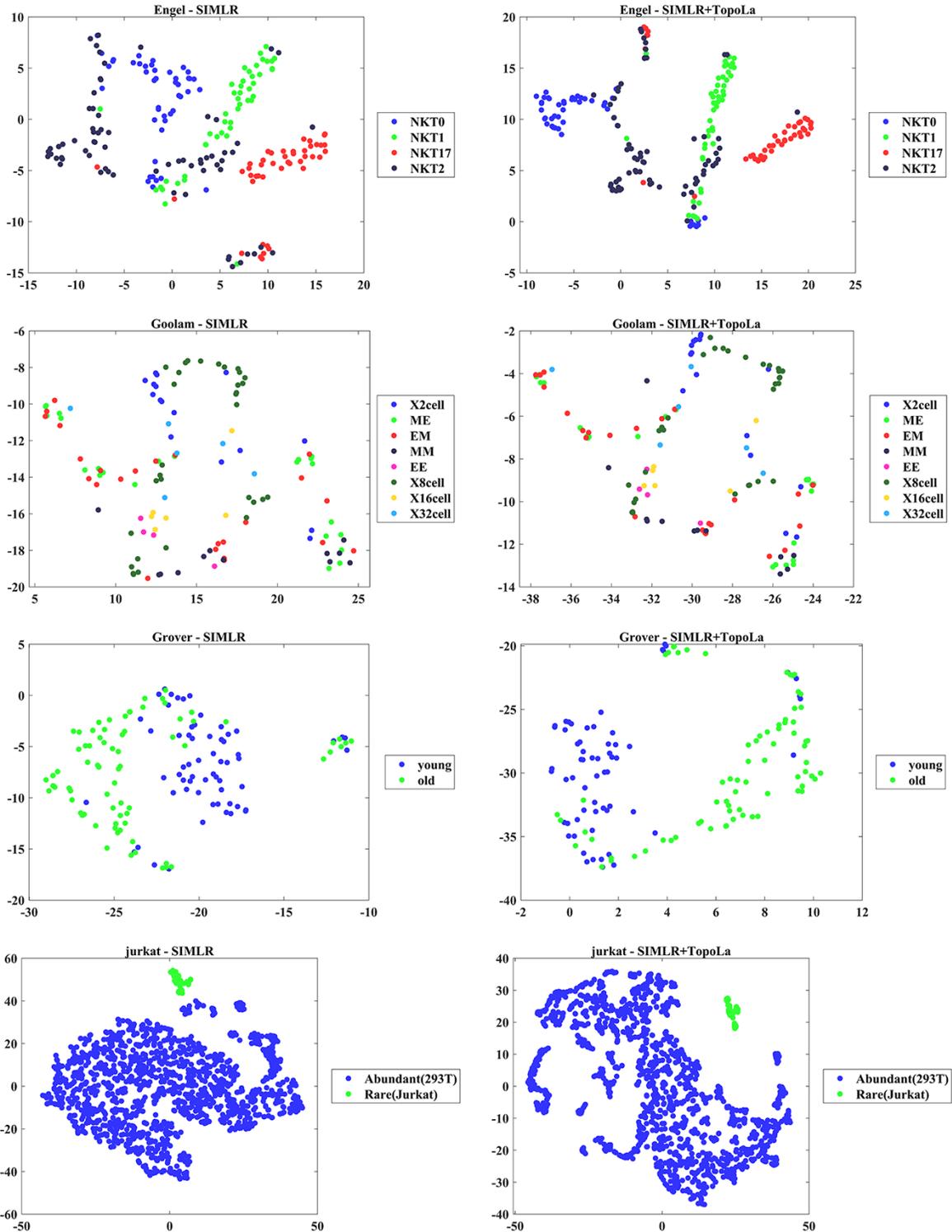

42 **Supplementary Fig. 6** t-SNE visualizations of cell type representations in the Engel, Goolam,
43 Grover, and Jurkat datasets using SIMLR and SIMLR+TopoLa. The plots depict the quality of cell
44 representations achieved by both models, offering visual confirmation of TopoLa's enhancement
45 capabilities.

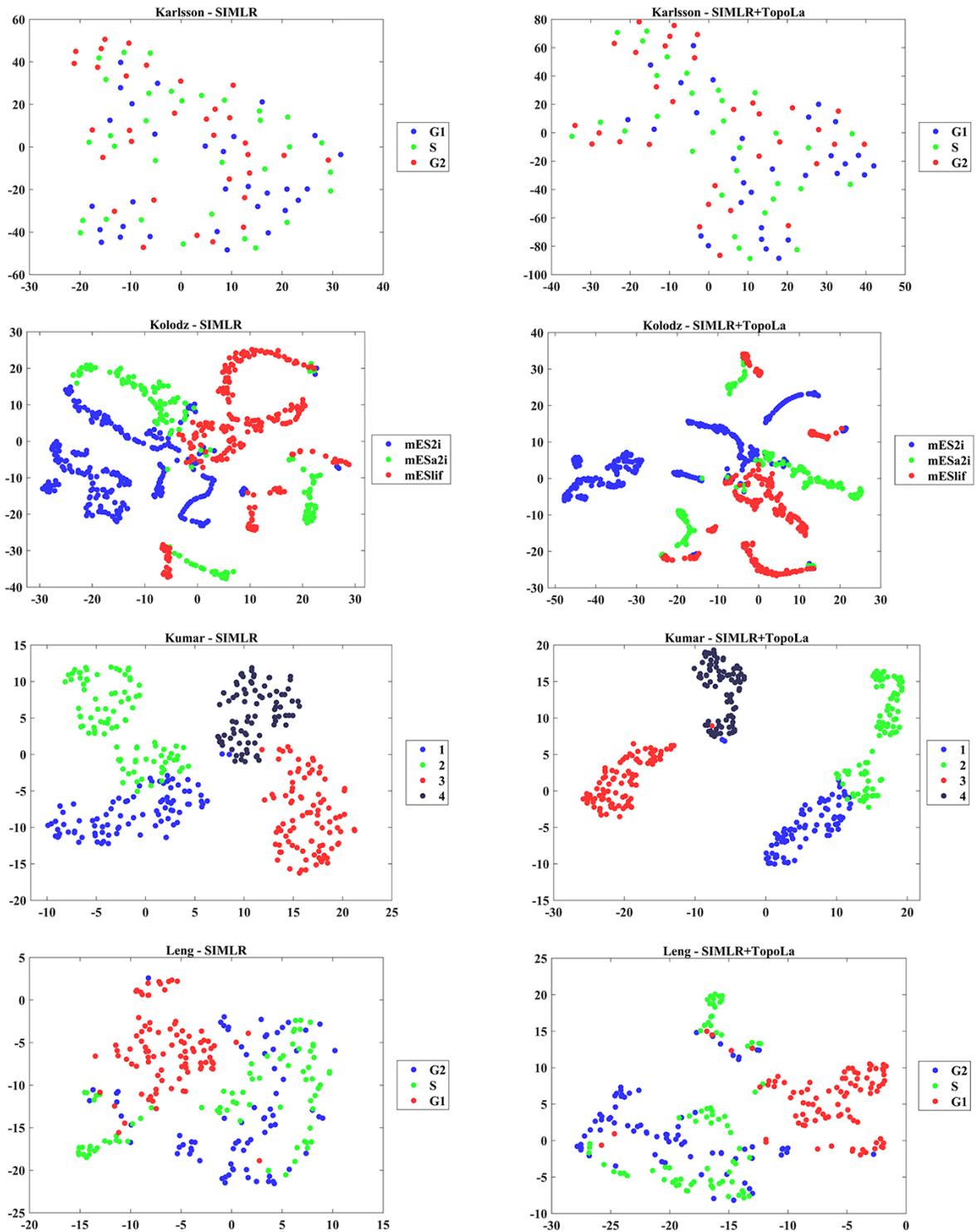

**Supplementary Fig. 7** t-SNE visualizations of cell type representations in the Karlsson, Kolodz, Kumar, and Leng datasets using SIMLR and SIMLR+TopoLa. The plots depict the quality of cell representations achieved by both models, offering visual confirmation of TopoLa's enhancement capabilities.

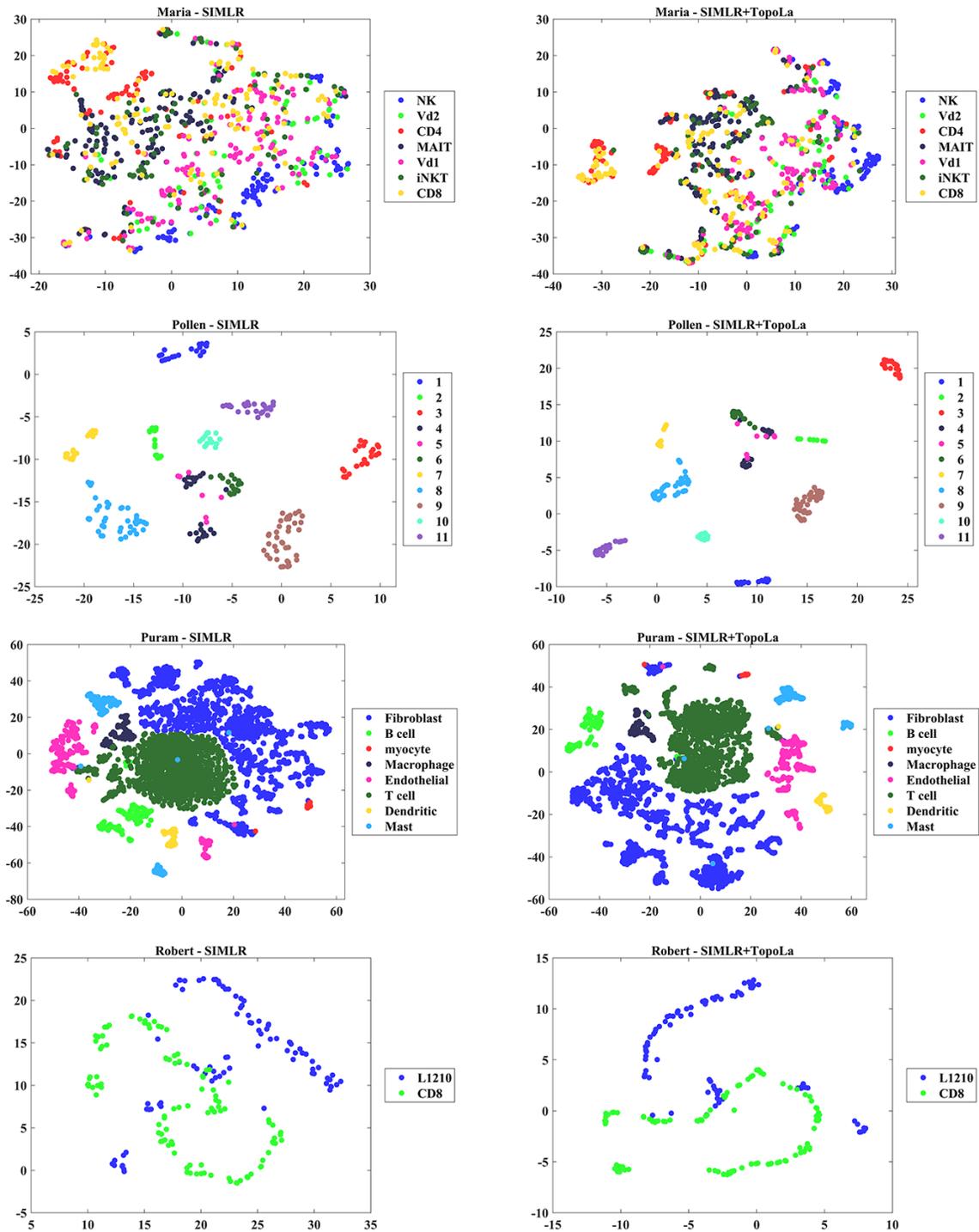

52
53  **Supplementary Fig. 8** t-SNE visualizations of cell type representations in the Maria, Pollen,
54  Puram, and Robert datasets using SIMLR and SIMLR+TopoLa. The plots depict the quality of
55  cell representations achieved by both models, offering visual confirmation of TopoLa's
56  enhancement capabilities.

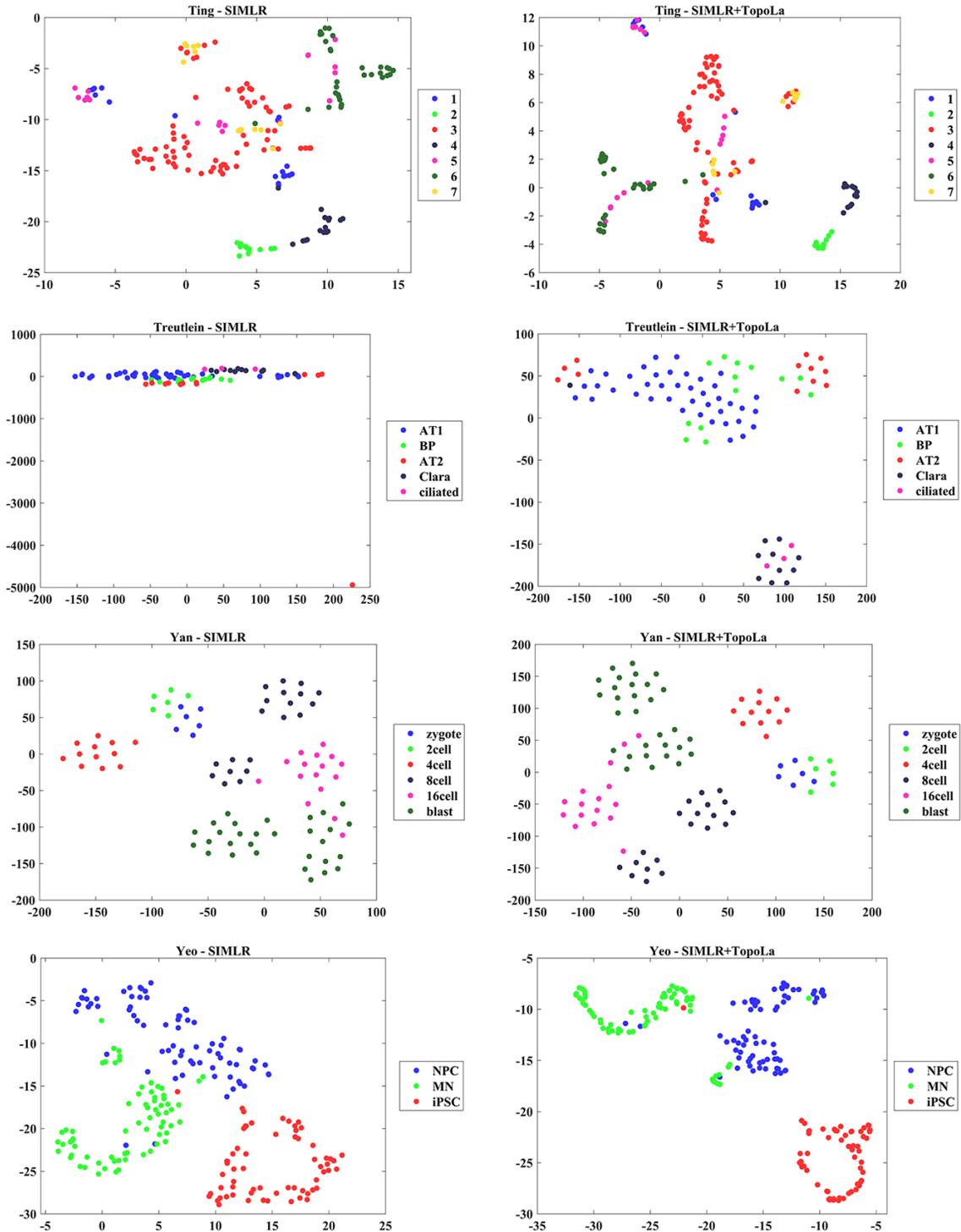

**Supplementary Fig. 9** t-SNE visualizations of cell type representations in the Ting, Treutlein, Yan, and Yeo datasets using SIMLR and SIMLR+TopoLa. The plots depict the quality of cell representations achieved by both models, offering visual confirmation of TopoLa's enhancement capabilities.

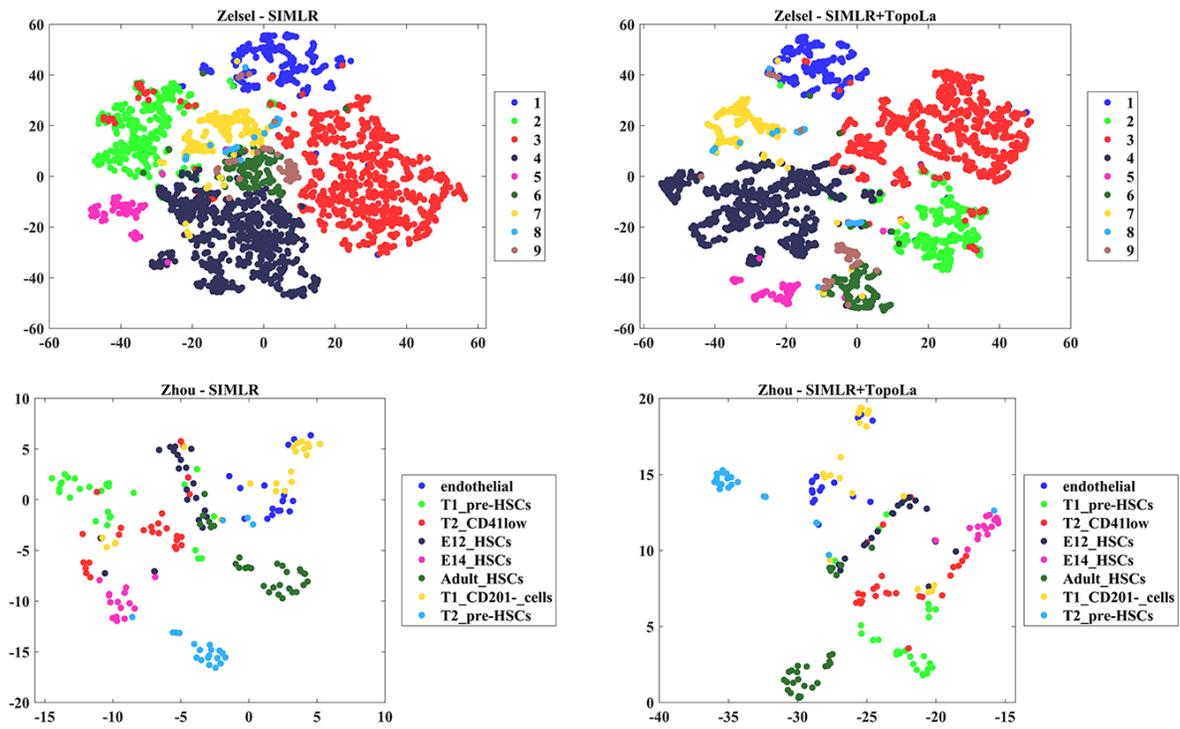

**Supplementary Fig. 10** t-SNE visualizations of cell type representations in the Zelsel and Zhou datasets using SIMLR and SIMLR+TopoLa. The plots depict the quality of cell representations achieved by both models, offering visual confirmation of TopoLa's enhancement capabilities.

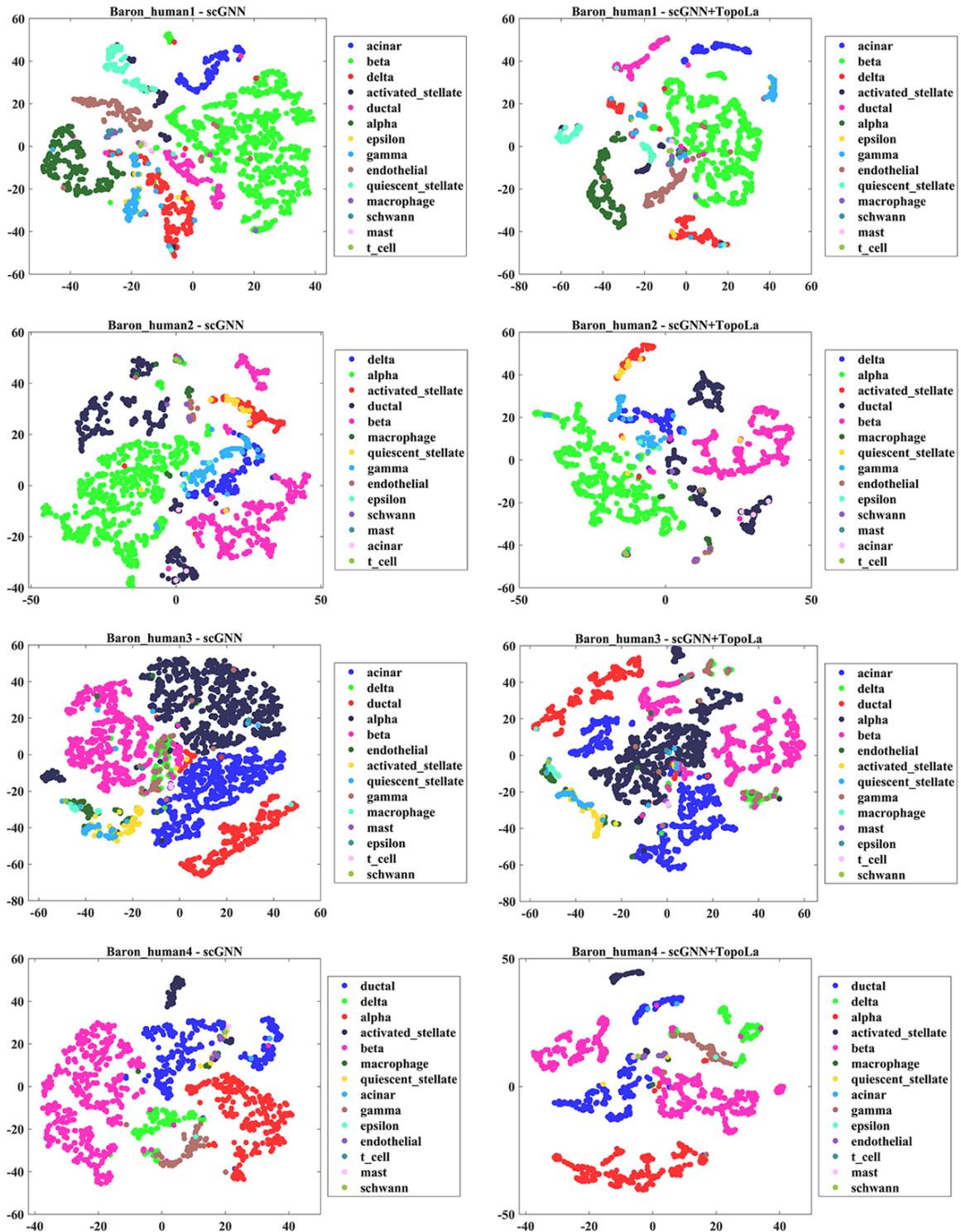

**Supplementary Fig. 11** t-SNE visualizations of cell type representations in the Baron_human1, Baron_human2, Baron_human3, and Baron_human4 datasets using scGNN and scGNN+TopoLa. The integration of TopoLa with scGNN enhances clustering performance and improves the dispersion of different cell types.

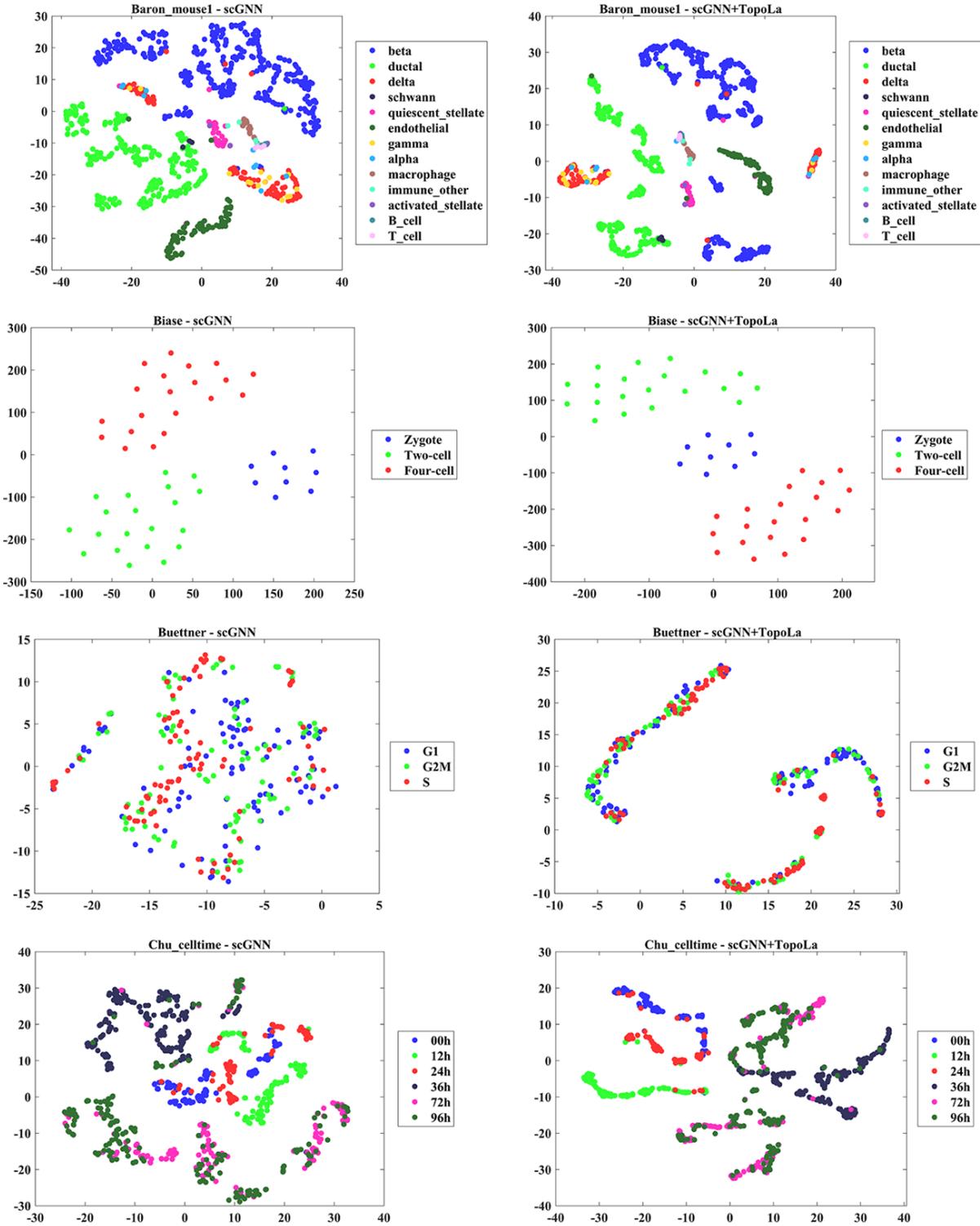

**Supplementary Fig. 12** t-SNE visualizations of cell type representations in the Baron_mouse, Biase, Buettner, and Chu_celltime datasets using scGNN and scGNN+TopoLa. The integration of TopoLa with scGNN enhances clustering performance and improves the dispersion of different cell types.

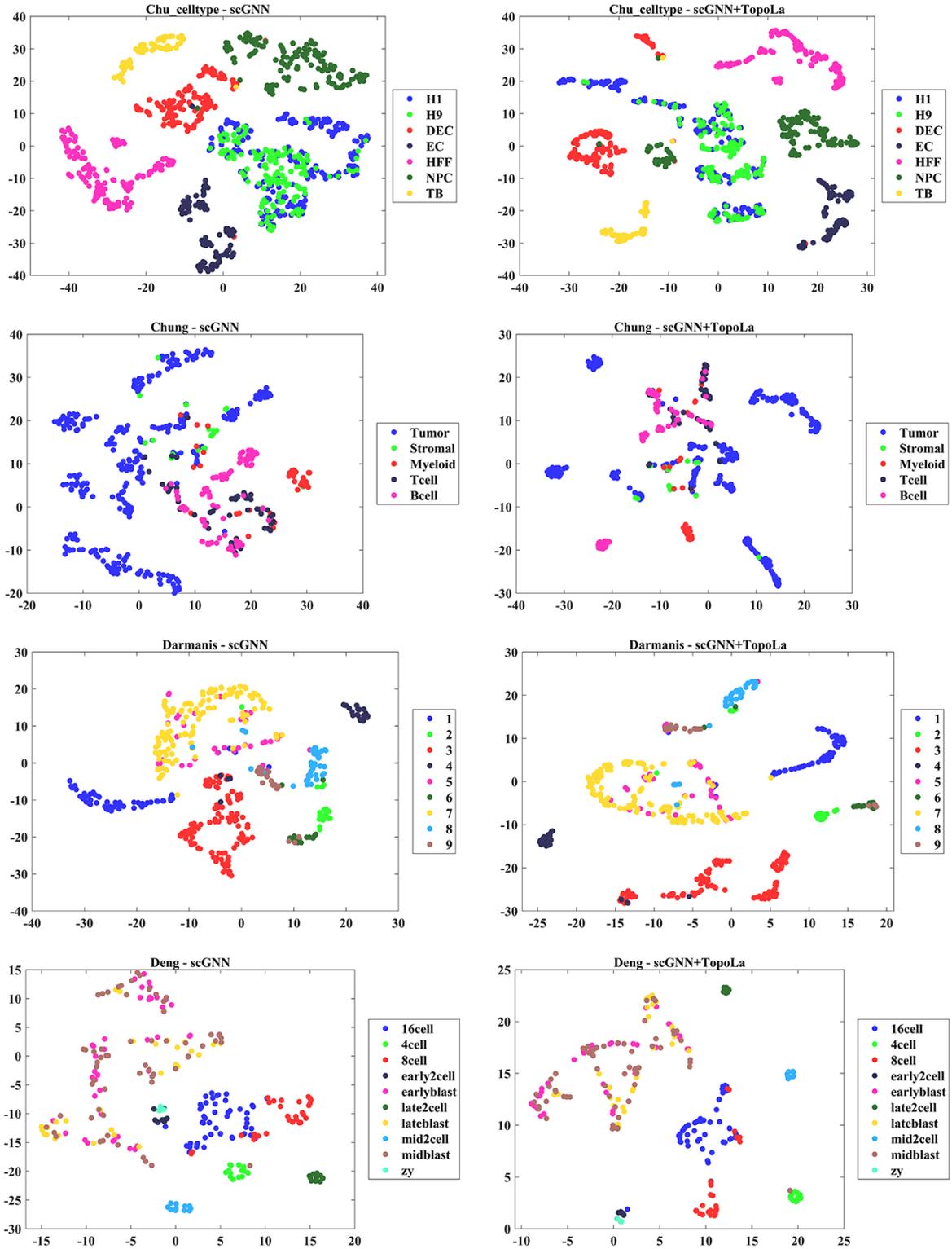

78
79 **Supplementary Fig. 13** t-SNE visualizations of cell type representations in the Chu_celltype,
80 Chung, Darmanis, and Deng datasets using scGNN and scGNN+TopoLa. The integration of
81 TopoLa with scGNN enhances clustering performance and improves the dispersion of different
82 cell types.

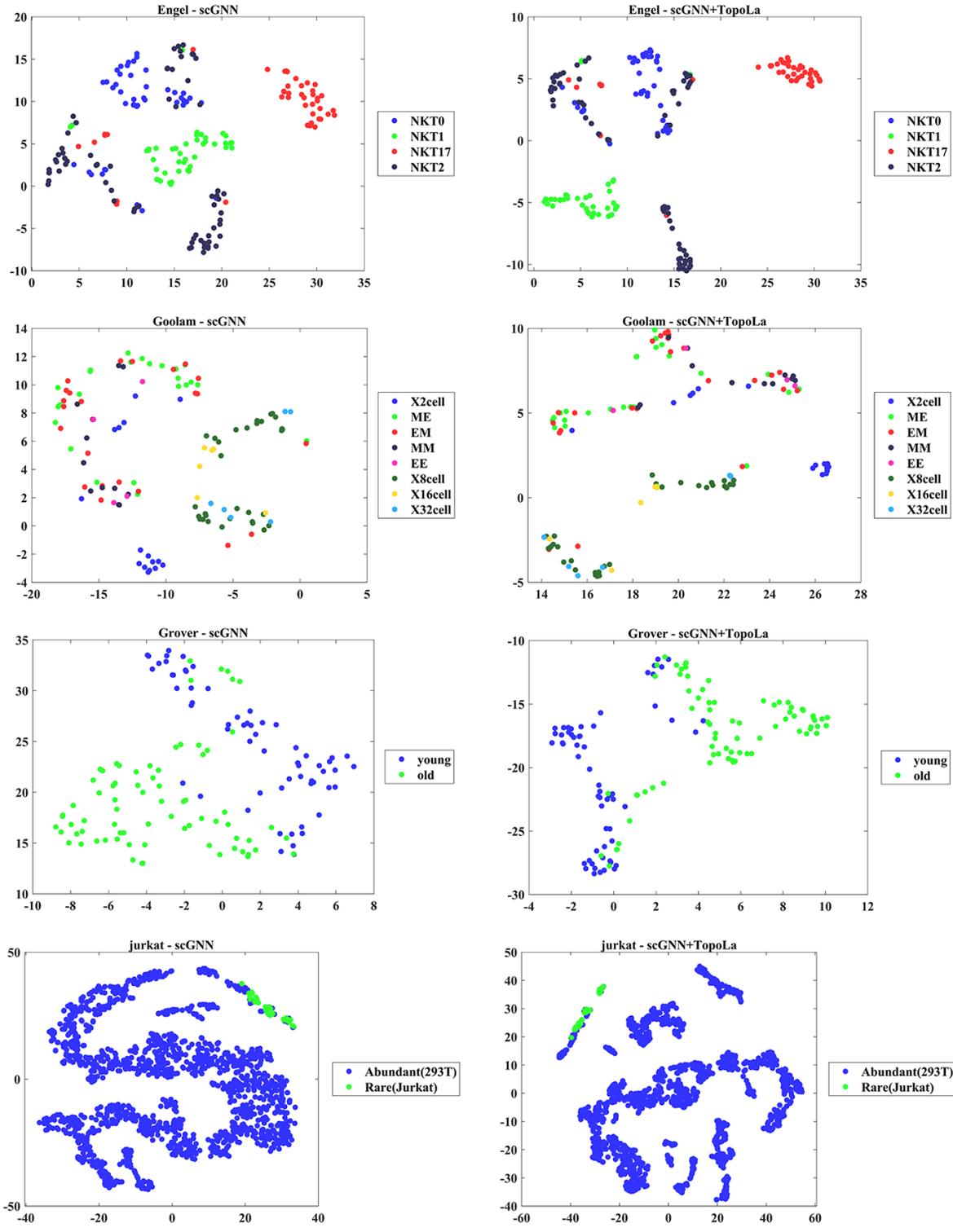

84 **Supplementary Fig. 14** t-SNE visualizations of cell type representations in the Engel, Goolam,
85 Grover, and Jurkat datasets using scGNN and scGNN+TopoLa. The integration of TopoLa with
86 scGNN enhances clustering performance and improves the dispersion of different cell types.

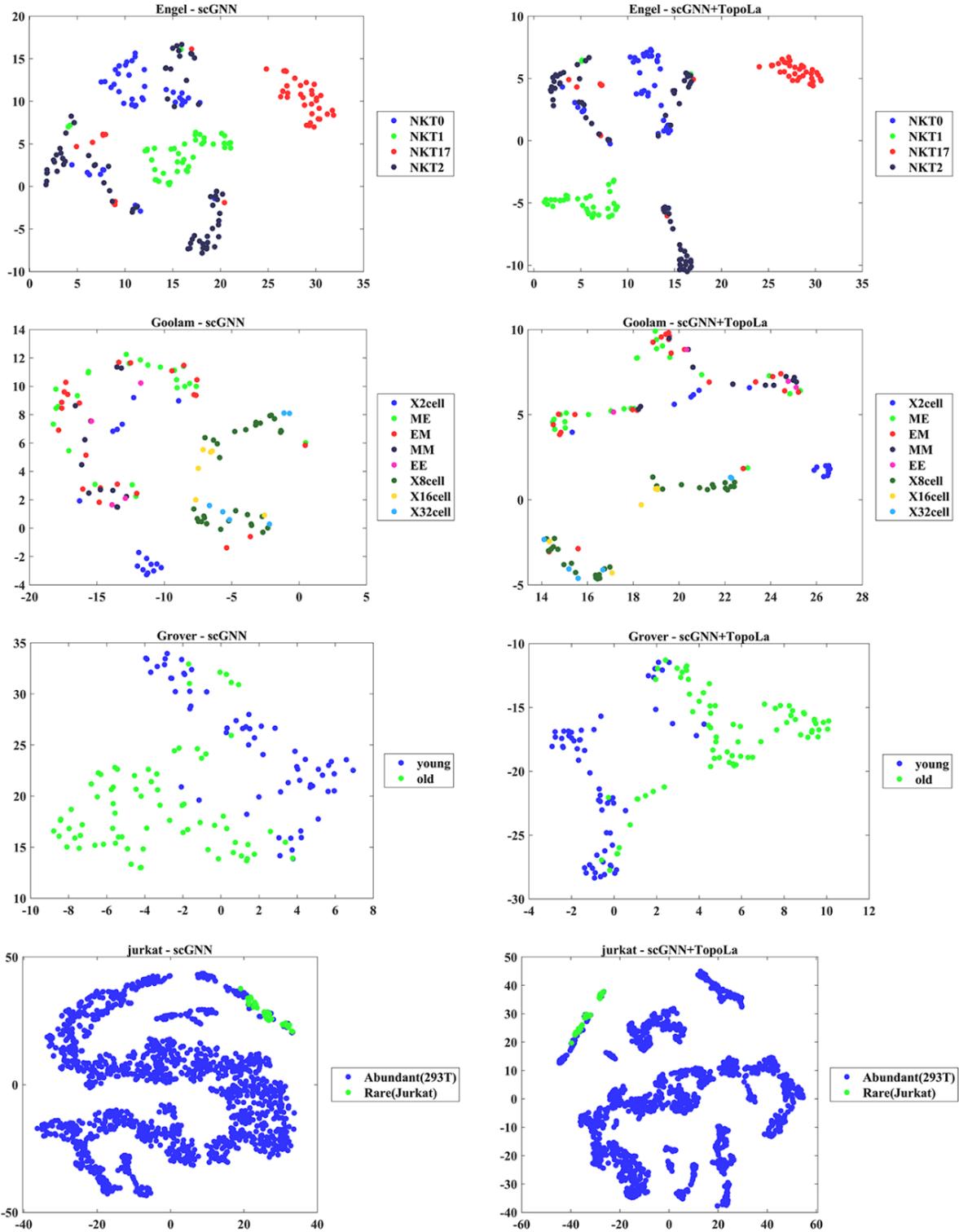

**Supplementary Fig. 15** t-SNE visualizations of cell type representations in the Karlsson, Kolodz, Kumar, and Leng datasets using scGNN and scGNN+TopoLa. The integration of scGNN with TopoLa enhances clustering performance and improves the dispersion of different cell types.

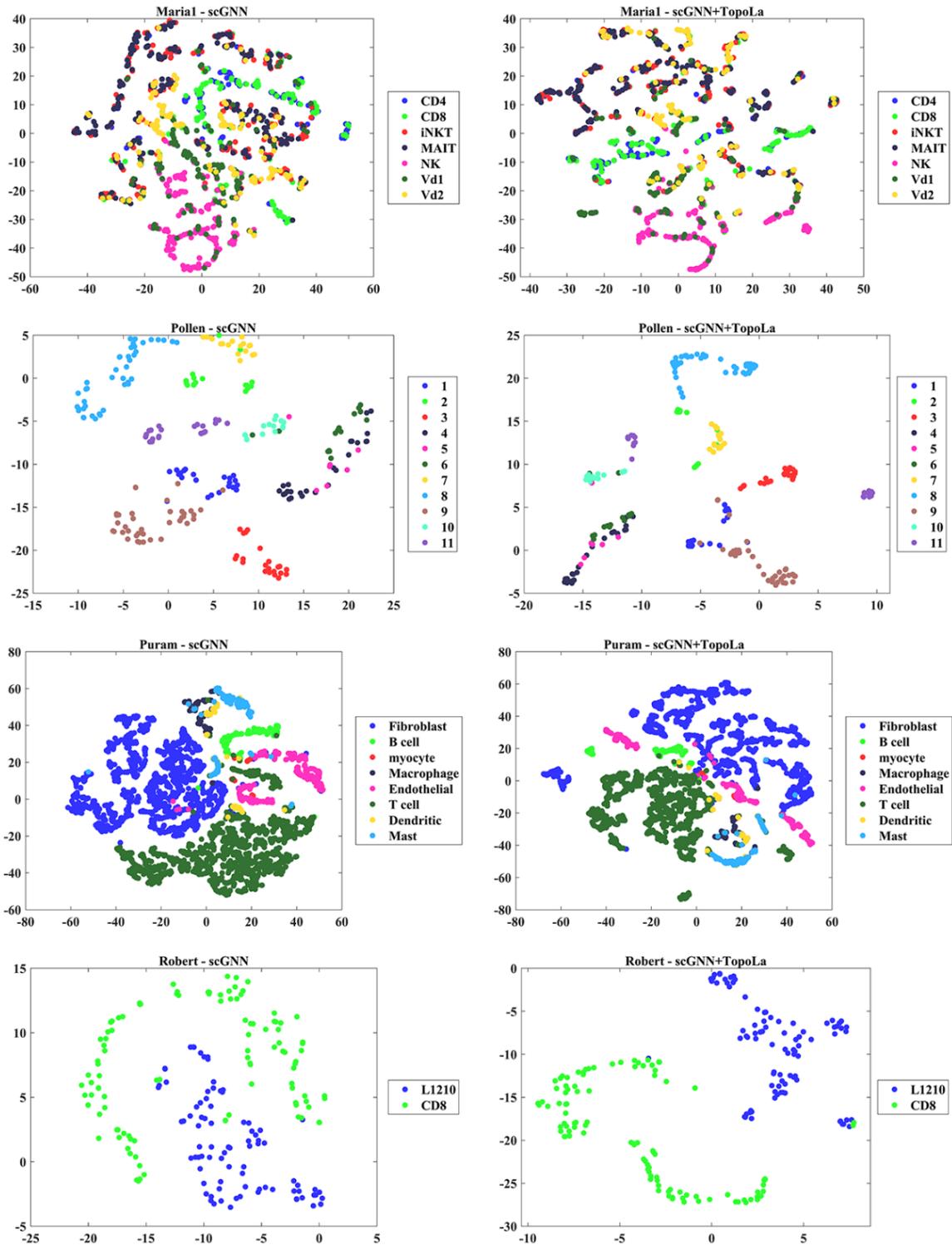

**Supplementary Fig. 16** t-SNE visualizations of cell type representations in the Maria, Pollen, Puram, and Robert datasets using scGNN and scGNN+TopoLa. The integration of TopoLa with scGNN enhances clustering performance and improves the dispersion of different cell types.

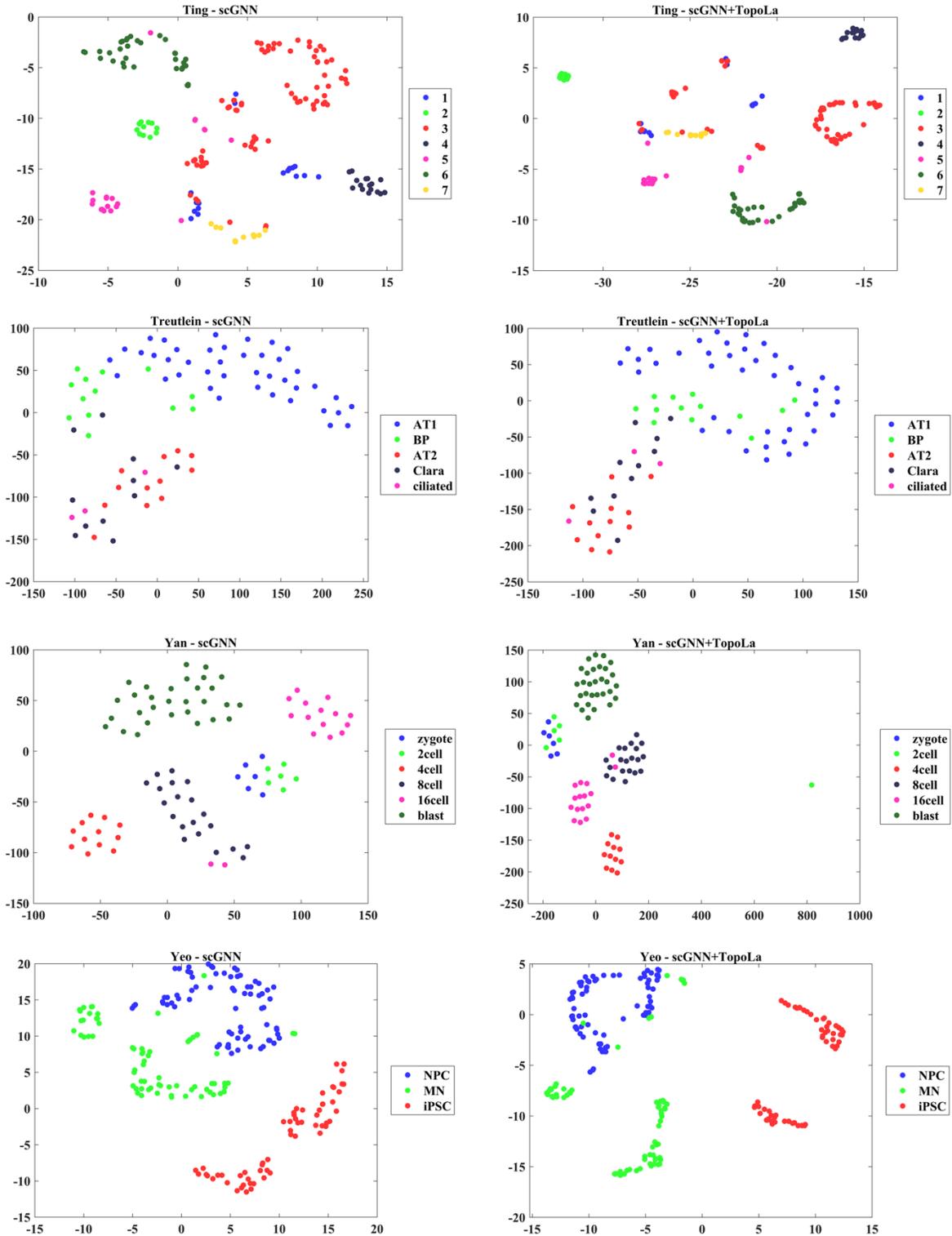

**Supplementary Fig. 17** t-SNE visualizations of cell type representations in the Ting, Treutlein, Yan, and Yeo datasets using scGNN and scGNN+TopoLa. The integration of scGNN with TopoLa enhances clustering performance and improves the dispersion of different cell types.

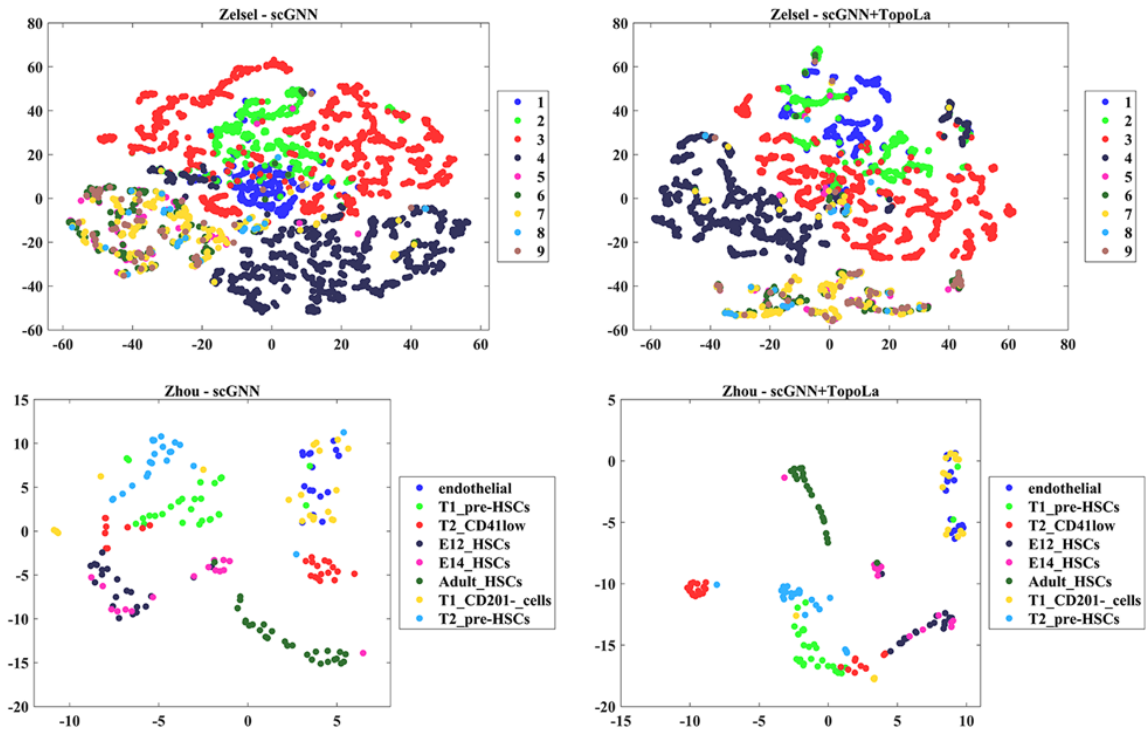

**Supplementary Fig. 18** t-SNE visualizations of cell type representations in the Zelsel and Zhou datasets using scGNN and scGNN+TopoLa. The integration of scGNN with TopoLa enhances clustering performance and improves the dispersion of different cell types.

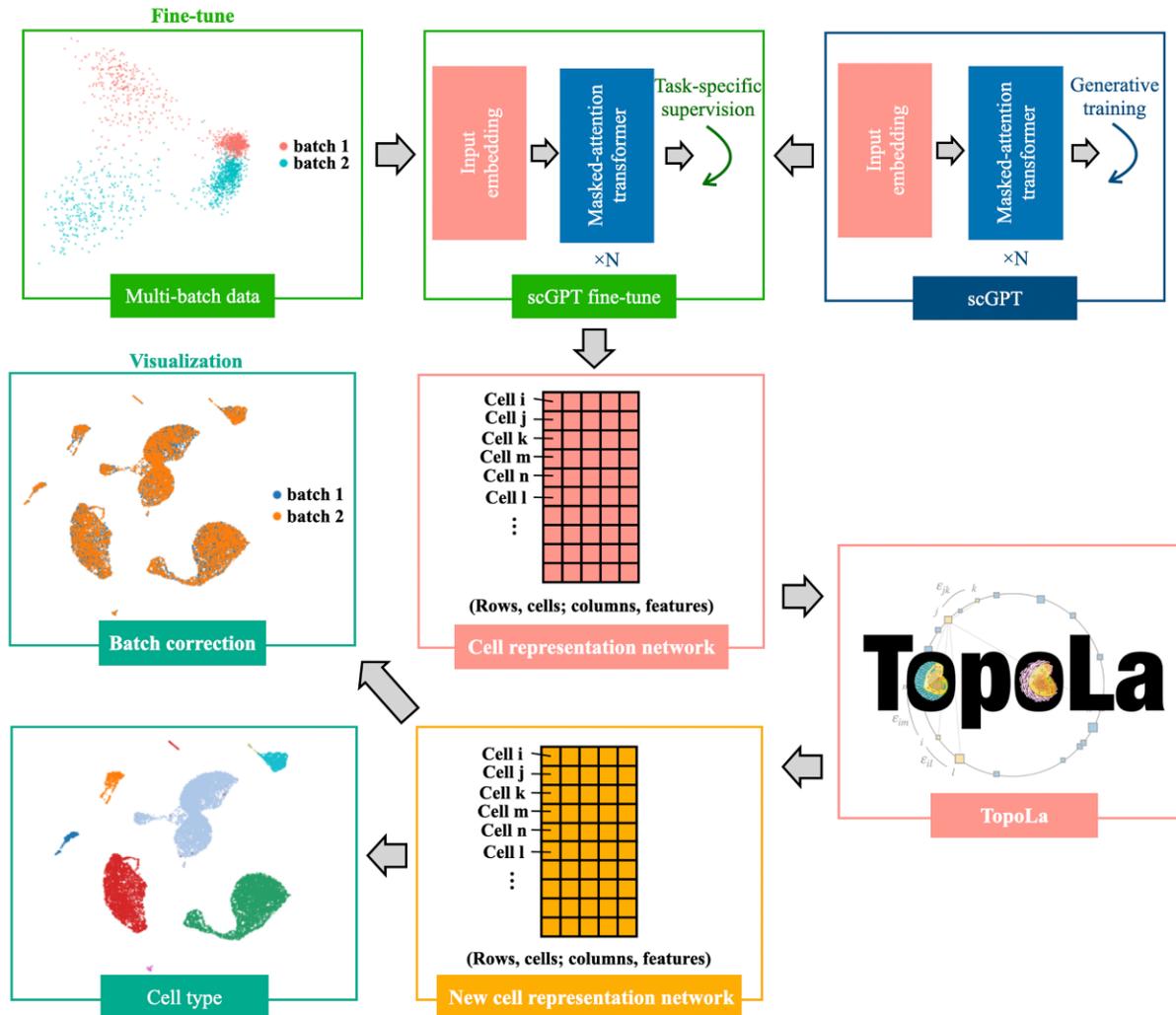

103
104    **Supplementary Fig. 19** Enhance cell representations of scGPT for multi-batch integration using
105    the TopoLa framework. The model begins with generative pretraining on large-scale scRNA-seq
106    data from the cell atlases. The core components of scGPT consist of stacked transformer blocks,
107    equipped with specialized attention masks for generative training. Next, the pretrained model
108    parameters are fine-tuned using multi-batch single-cell data. A cell graph, constructed from the
109    learned cell embeddings, is then input into the TopoLa framework to obtain an enhanced cell graph.
110    Finally, this enhanced cell graph is input into the Louvain algorithm for cell clustering.
111

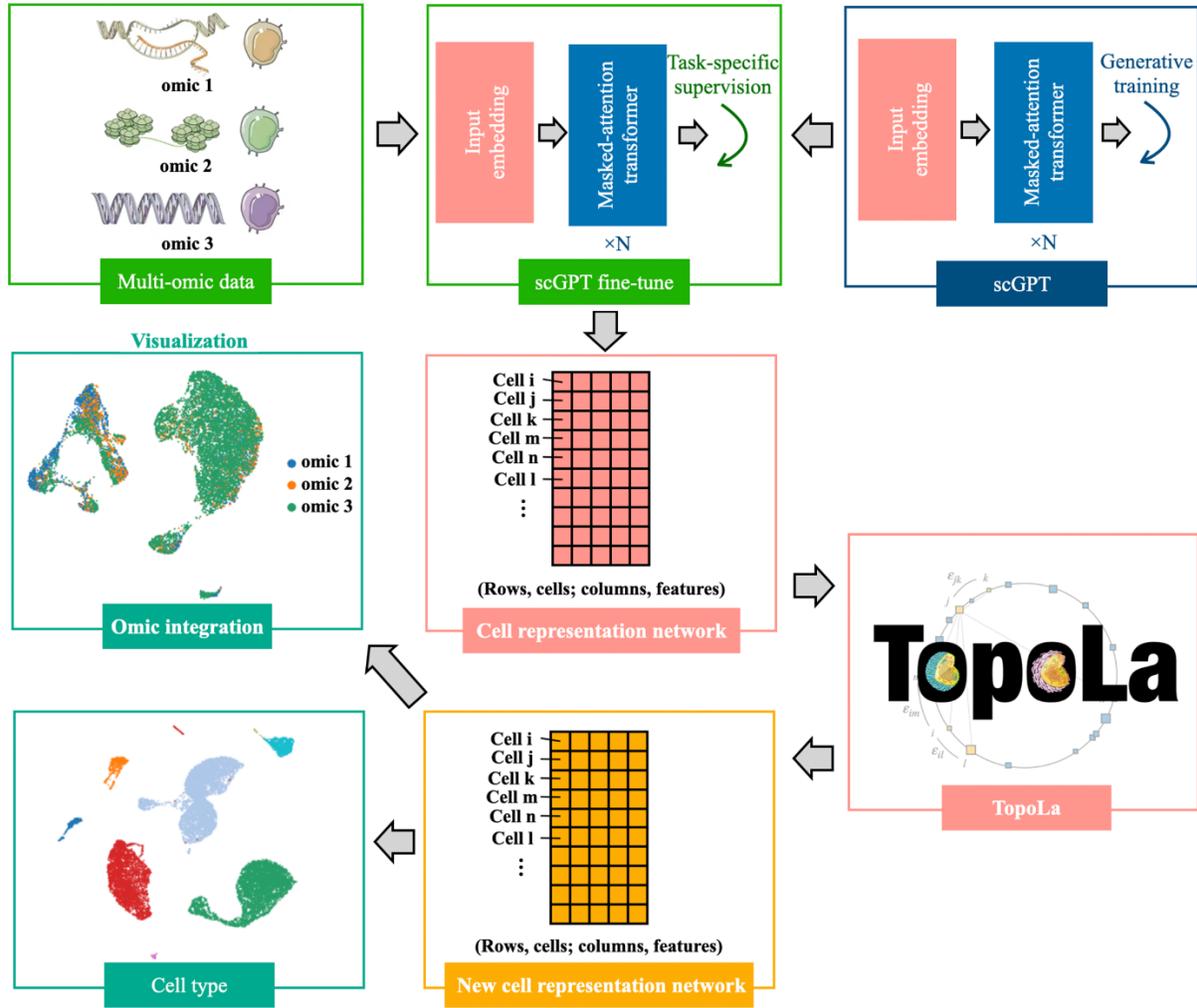

**Supplementary Fig. 20** Enhance cell representations of scGPT for multi-omic integration using
the TopoLa framework. The model begins with generative pretraining on large-scale scRNA-seq
data from the cell atlases. Next, the pretrained model parameters are fine-tuned using multi-omics
single-cell data. A cell graph, constructed from the learned cell embeddings, is then input into the
TopoLa framework to obtain an enhanced cell graph. Finally, this enhanced cell graph is input into
the Louvain algorithm for cell clustering.

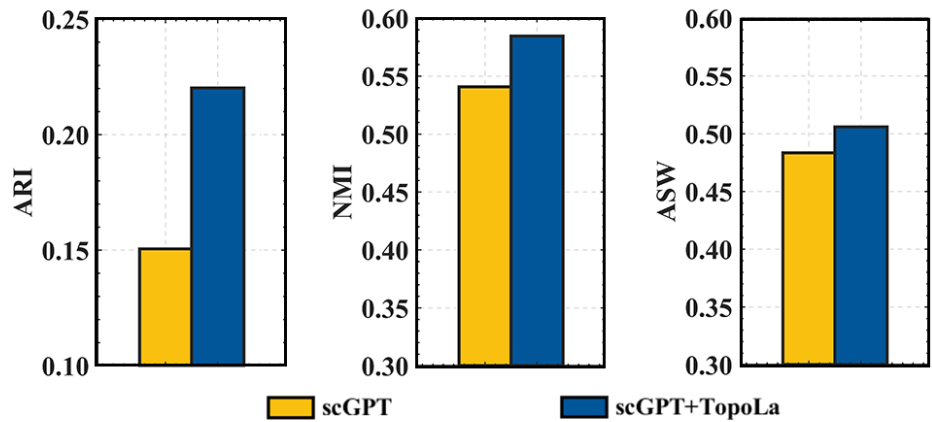

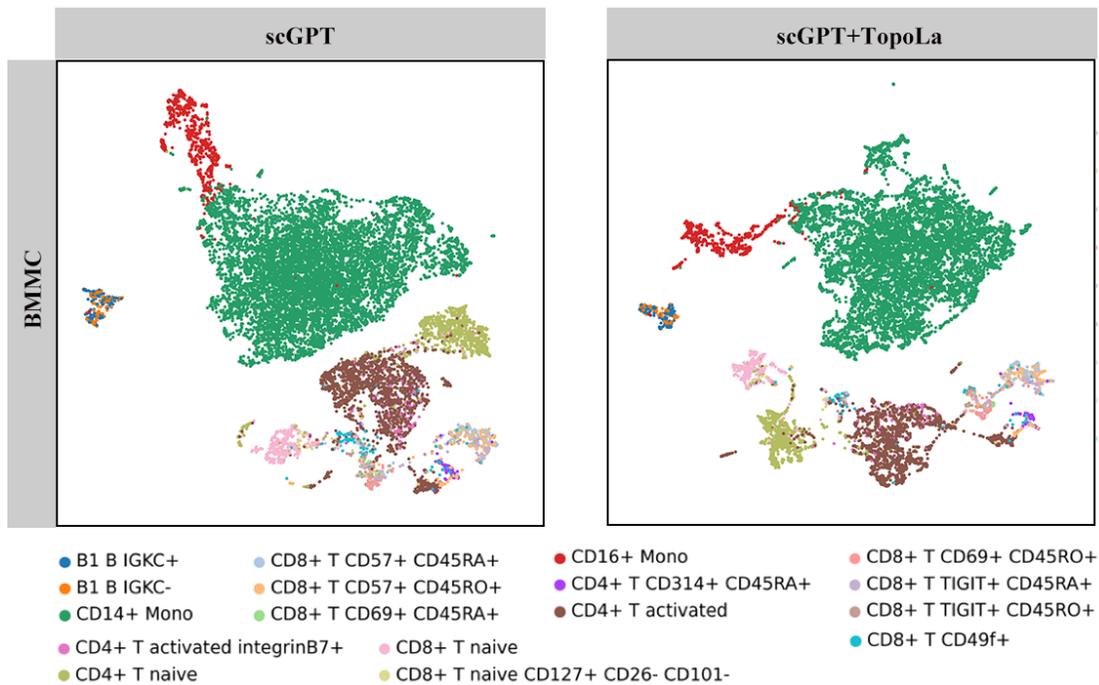

**Supplementary Fig. 21** Enhanced multi-omics integration using TopoLa on single-cell RNA-seq data. **a** Quantitative performance metrics (ARI, NMI, and ASW) on the BMMC dataset. **b** UMAP visualizations of the BMMC dataset comparing the performance of scGPT and scGPT+TopoLa, with cells labeled by their true cell types.

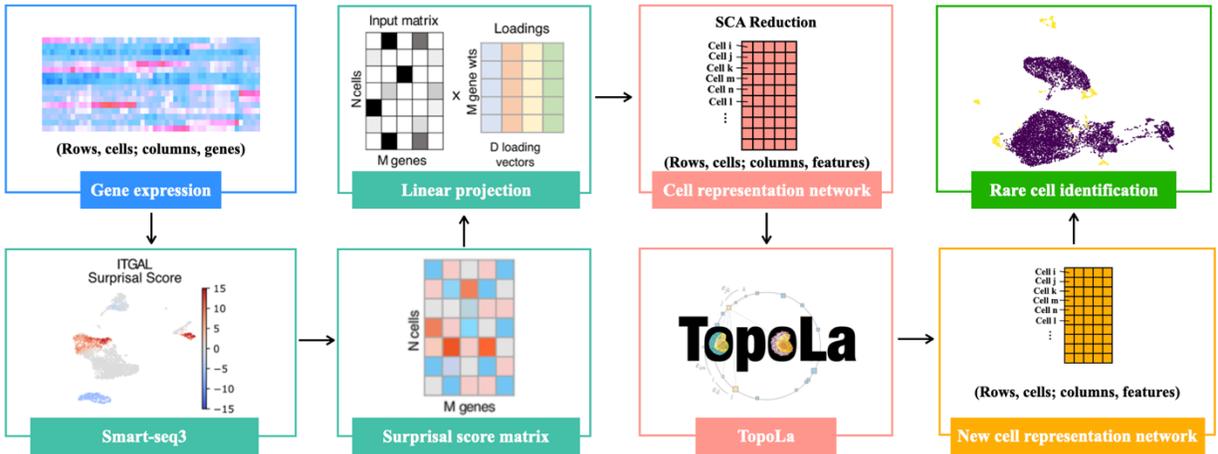

**Supplementary Fig. 22** Enhance cell representations of SCA using the TopoLa framework. The model begins with the gene expression matrix as input. A surprisal score matrix is then generated using Smart-seq3. Singular value decomposition (SVD) is performed on the surprisal scores across all genes, producing right-eigenvectors that capture key axes of variation within the data. The input transcript count matrix is linearly projected onto these axes to generate a representation for each cell, serving as the initial cell representation. These cell representations are then input into the TopoLa framework to obtain enhanced representations, which are subsequently used for rare cell identification.

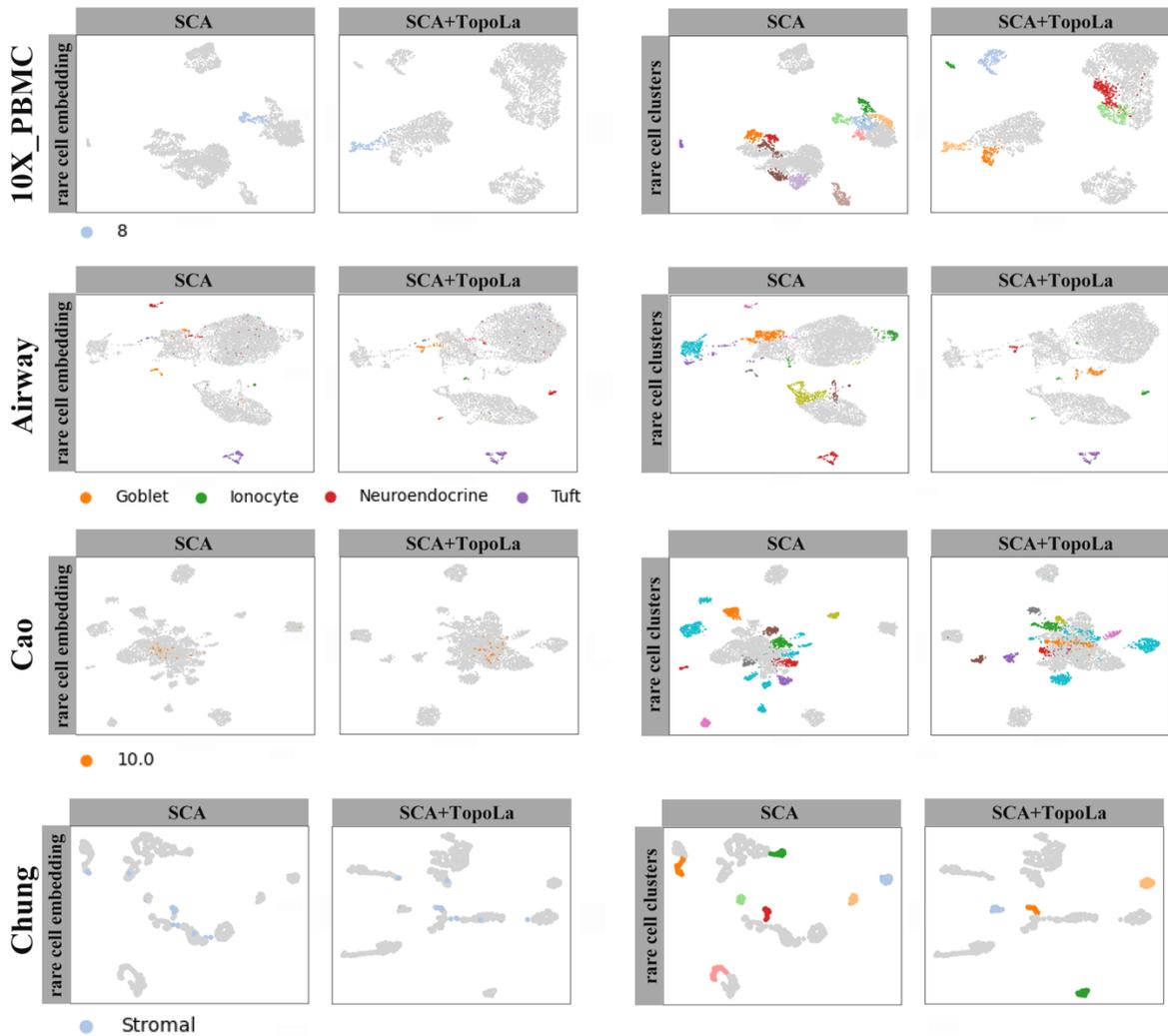

134
**Supplementary Fig. 23** UMAP visualizations analysis of results of SCA and SCA+TopoLa in
10X_PBMC, Airway, Cao, and Chung datasets. The first column displays visualizations of rare
cell embeddings obtained by SCA and SCA+TopoLa. The second column shows visualizations of
rare cell clustering results obtained by SCA and SCA+TopoLa.

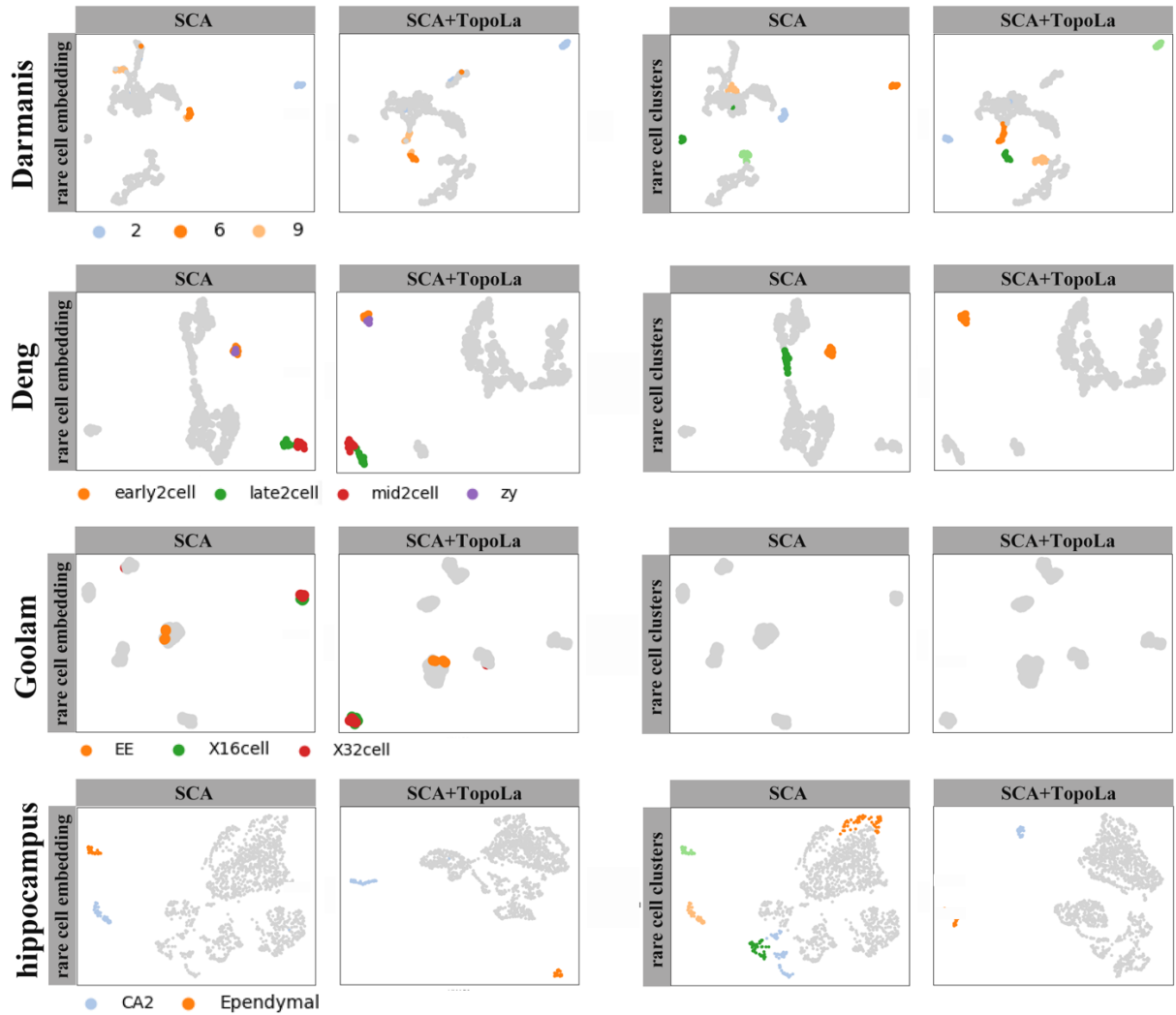

**Supplementary Fig. 24** UMAP visualizations analysis of results of SCA and SCA+TopoLa in Darmanis, Deng, Goolam, and Hippocampus datasets. The first column displays visualizations of rare cell embeddings obtained by SCA and SCA+TopoLa. The second column shows visualizations of rare cell clustering results obtained by SCA and SCA+TopoLa.

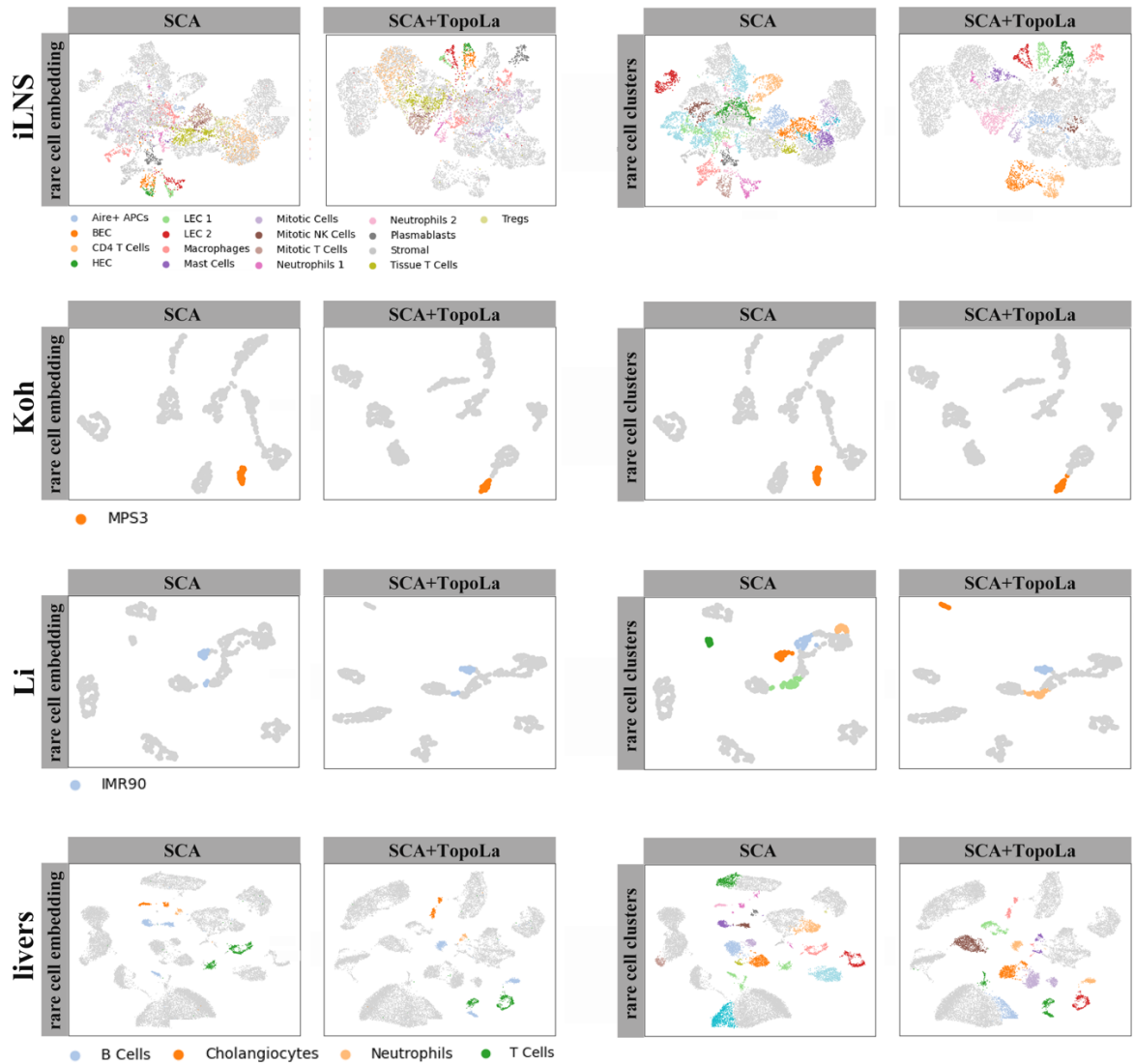

144
145 **Supplementary Fig. 25** UMAP visualizations analysis of results of SCA and SCA+TopoLa in
146 iLNs, Koh, Li, and livers datasets. The first column displays visualizations of rare cell embeddings
147 obtained by SCA and SCA+TopoLa. The second column shows visualizations of rare cell
148 clustering results obtained by SCA and SCA+TopoLa.

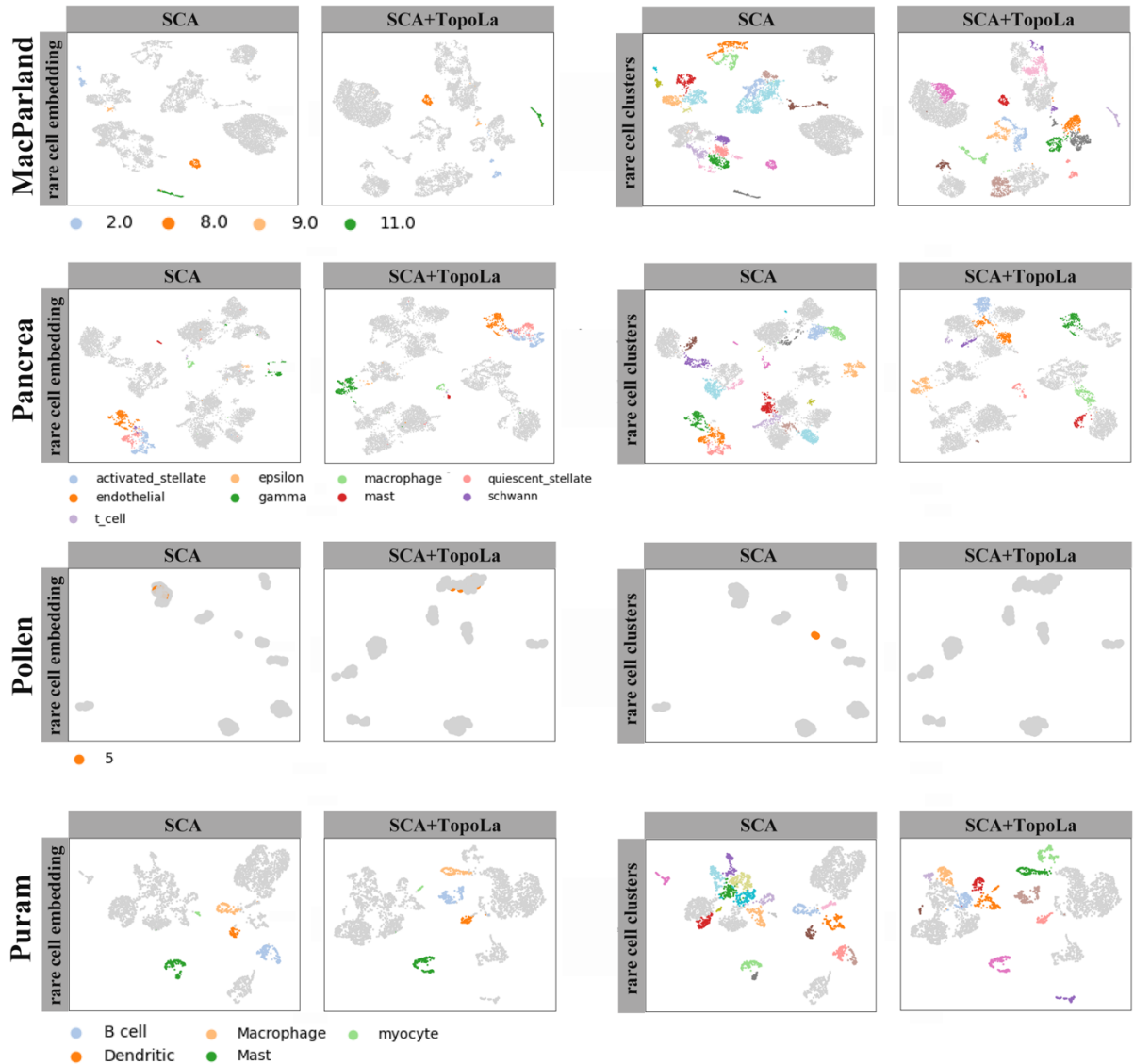

149
**Supplementary Fig. 26** UMAP visualizations analysis of results of SCA and SCA+TopoLa in MacParland, Pancrea, Pollen, and Puram datasets. The first column displays visualizations of rare cell embeddings obtained by SCA and SCA+TopoLa. The second column shows visualizations of rare cell clustering results obtained by SCA and SCA+TopoLa.

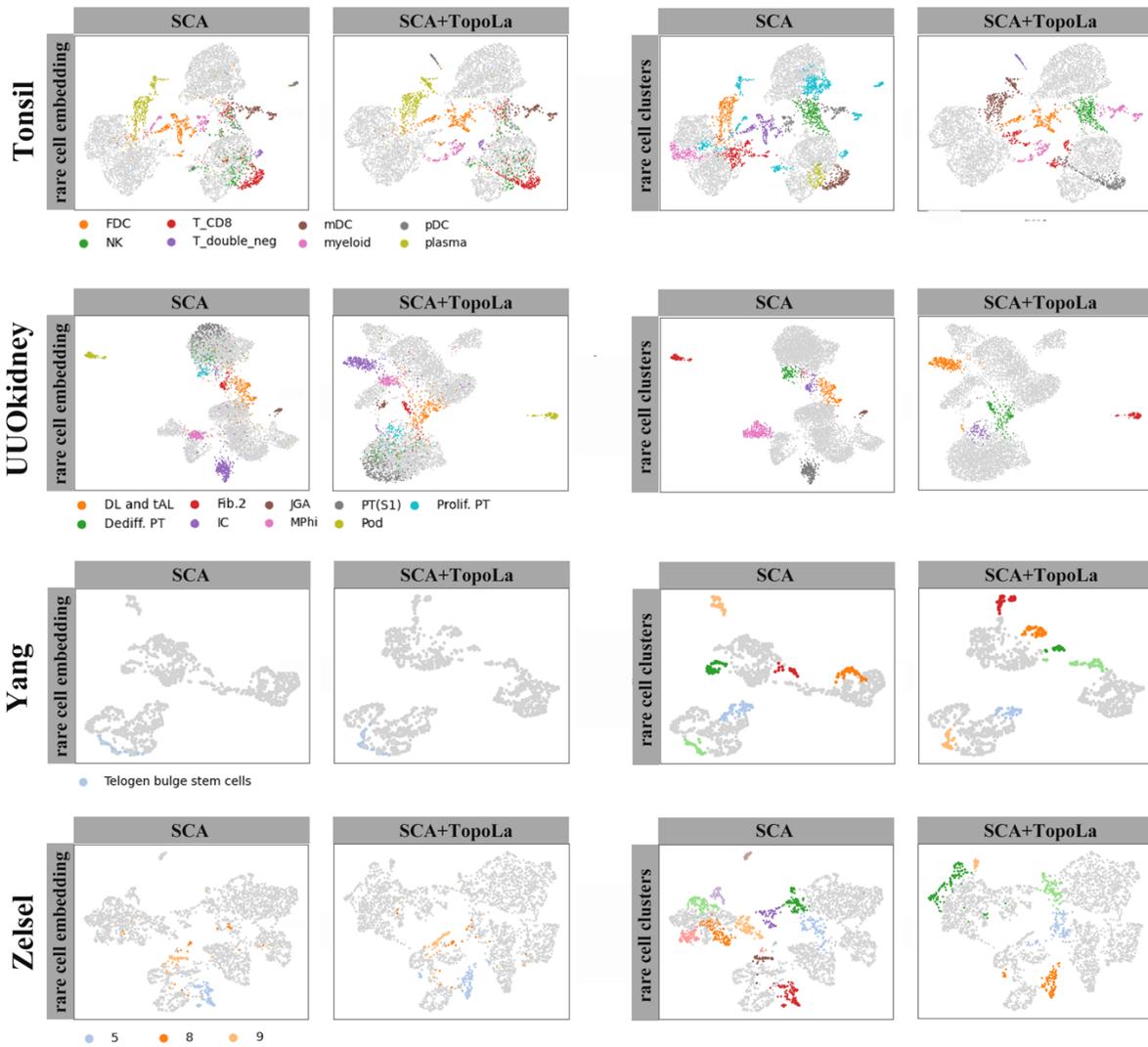

154
**Supplementary Fig. 27** UMAP visualizations analysis of results of SCA and SCA+TopoLa in
Tonsil, UUOkidney, Yang, and Zelsel datasets. The first column displays visualizations of rare
cell embeddings obtained by SCA and SCA+TopoLa. The second column shows visualizations of
rare cell clustering results obtained by SCA and SCA+TopoLa.

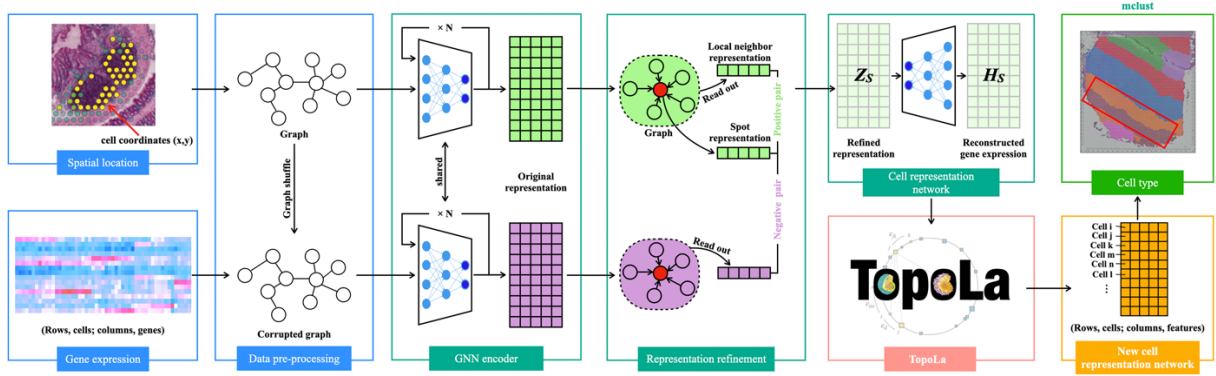

**Supplementary Fig. 28** Overview of GraphST+TopoLa for spatial informed clustering. The model begins with preprocessed spatial gene expression data and a neighborhood graph constructed using spatial coordinates as inputs. A graph-based self-supervised contrastive learning approach is then employed to learn latent representations that preserve key features from gene expression profiles, spatial location, and local context information. These latent representations are mapped back to the original feature space to reconstruct the cell representation matrix. The matrix is subsequently input into the TopoLa framework to obtain refined cell representations. Finally, the cells are clustered based on their refined representations using the mclust algorithm.

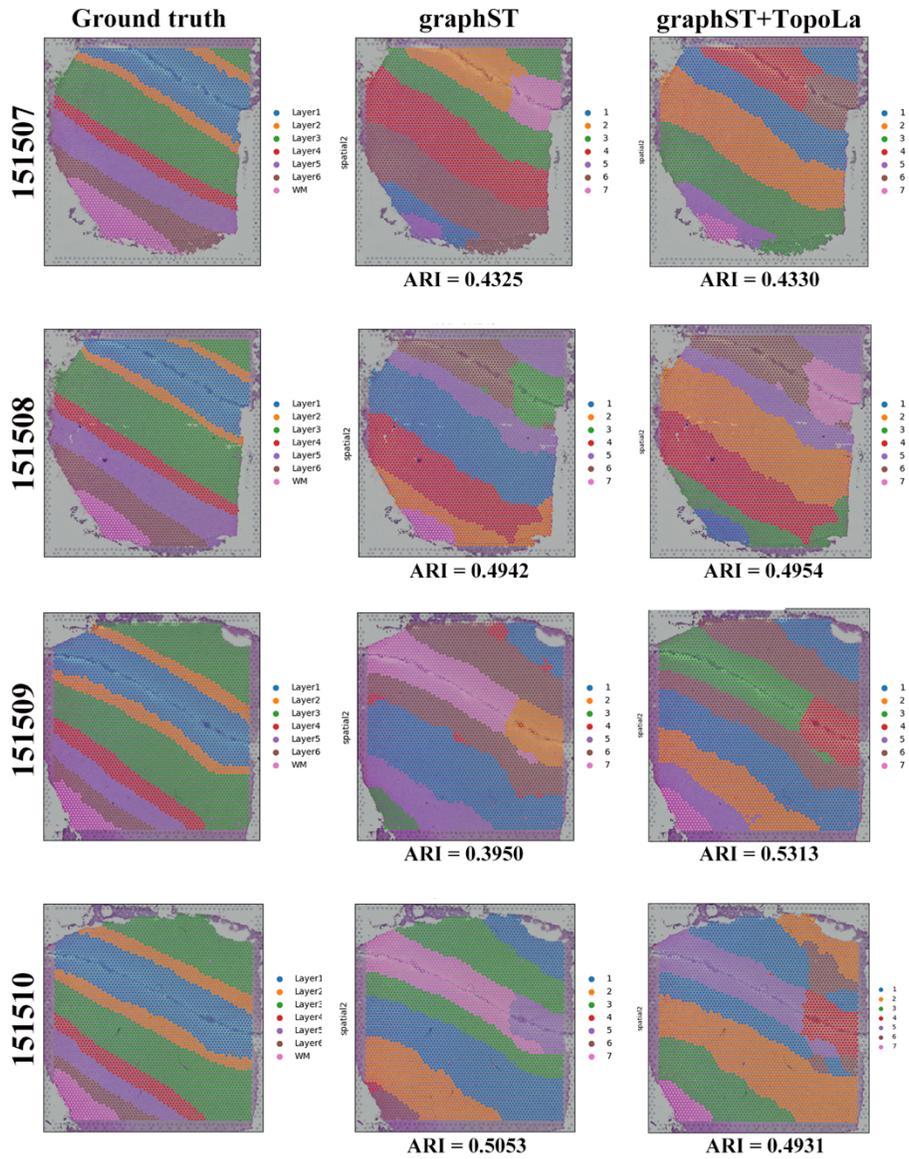

169
**Supplementary Fig. 29** Clustering results produced by GraphST and GraphST+TopoLa on slices
151507, 151508, 151509, and 151510 from the DLPFC dataset.

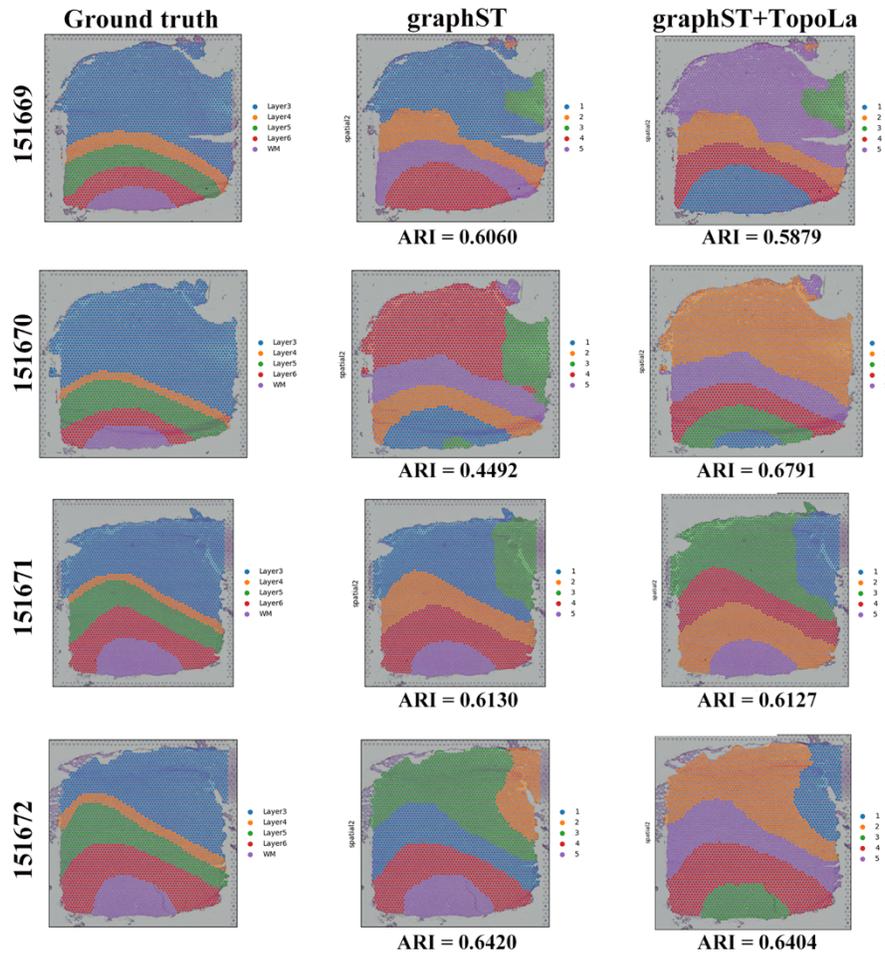

172
173 **Supplementary Fig. 30** Clustering results produced by GraphST and GraphST+TopoLa on slices
174 151669, 151670, 151671, and 151672 from the DLPFC dataset.

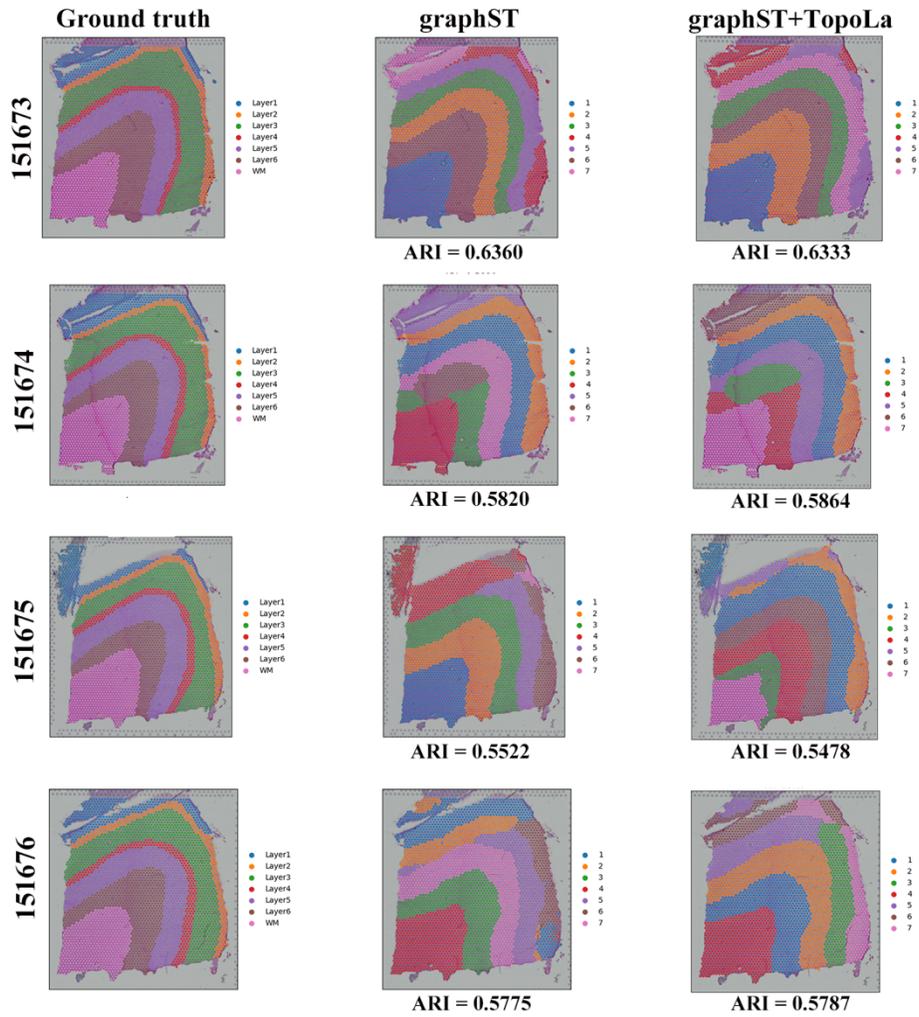

175
176 **Supplementary Fig. 31** Clustering results produced by GraphST and GraphST+TopoLa on slices
177 151673, 151674, 151675, and 151676 from the DLPFC dataset.

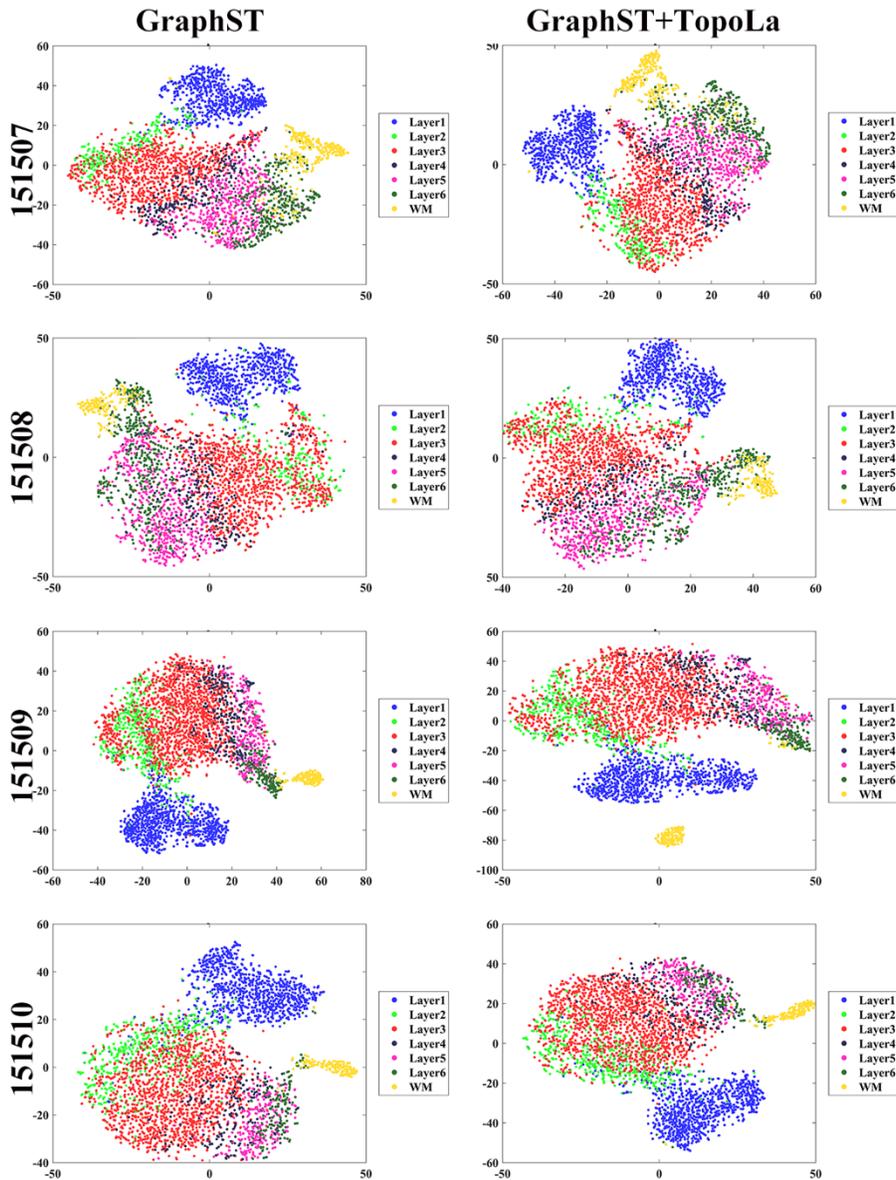

178
179 **Supplementary Fig. 32** t-SNE visualizations of cell embeddings from slices 151507, 151508,
180 151509, and 151510, comparing GraphST results before (left) and after (right) TopoLa
181 enhancement. The embeddings are visualized using t-SNE, with distinct colors representing
182 different cell types, showcasing the improved segregation of cell types.

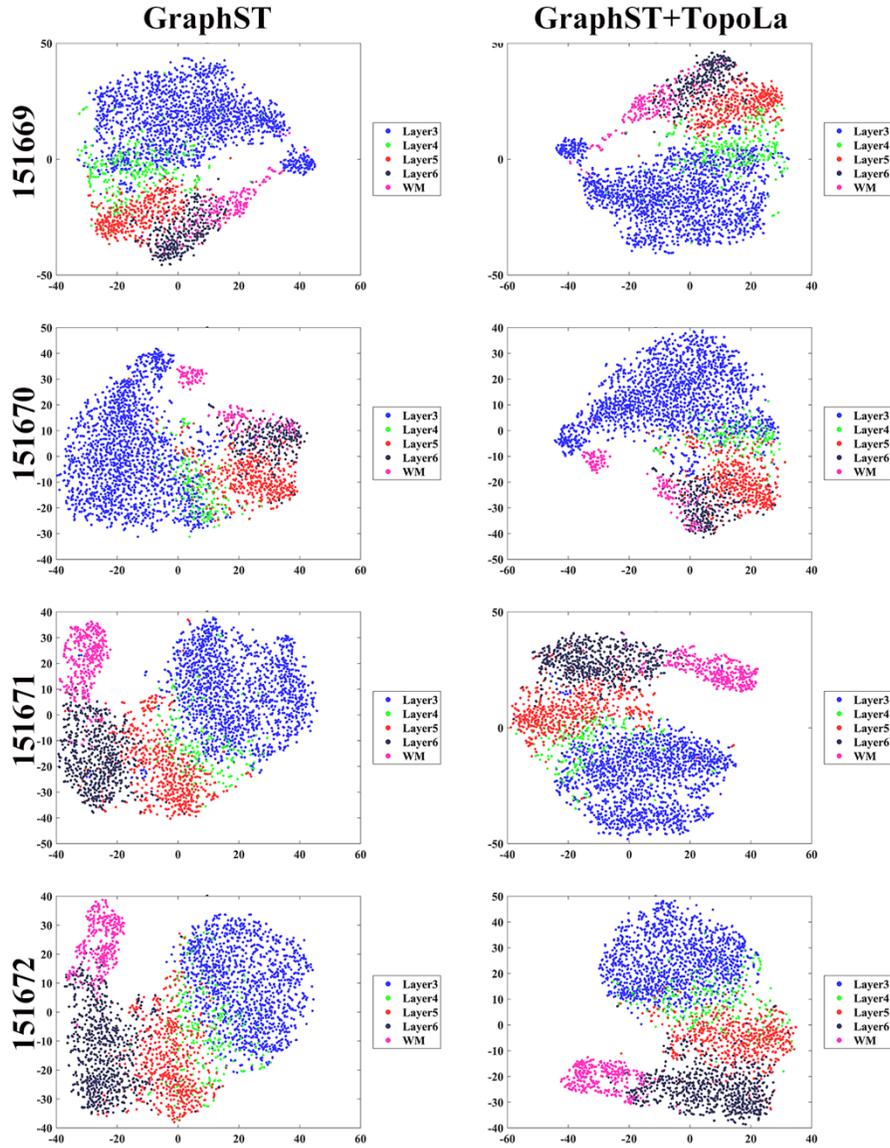

**Supplementary Fig. 33** t-SNE visualizations of cell embeddings from slices 151669, 151670, 151671, and 151672, comparing GraphST results before (left) and after (right) TopoLa enhancement. The embeddings are visualized using t-SNE, with distinct colors representing different cell types, showcasing the improved dispersion of cell types.

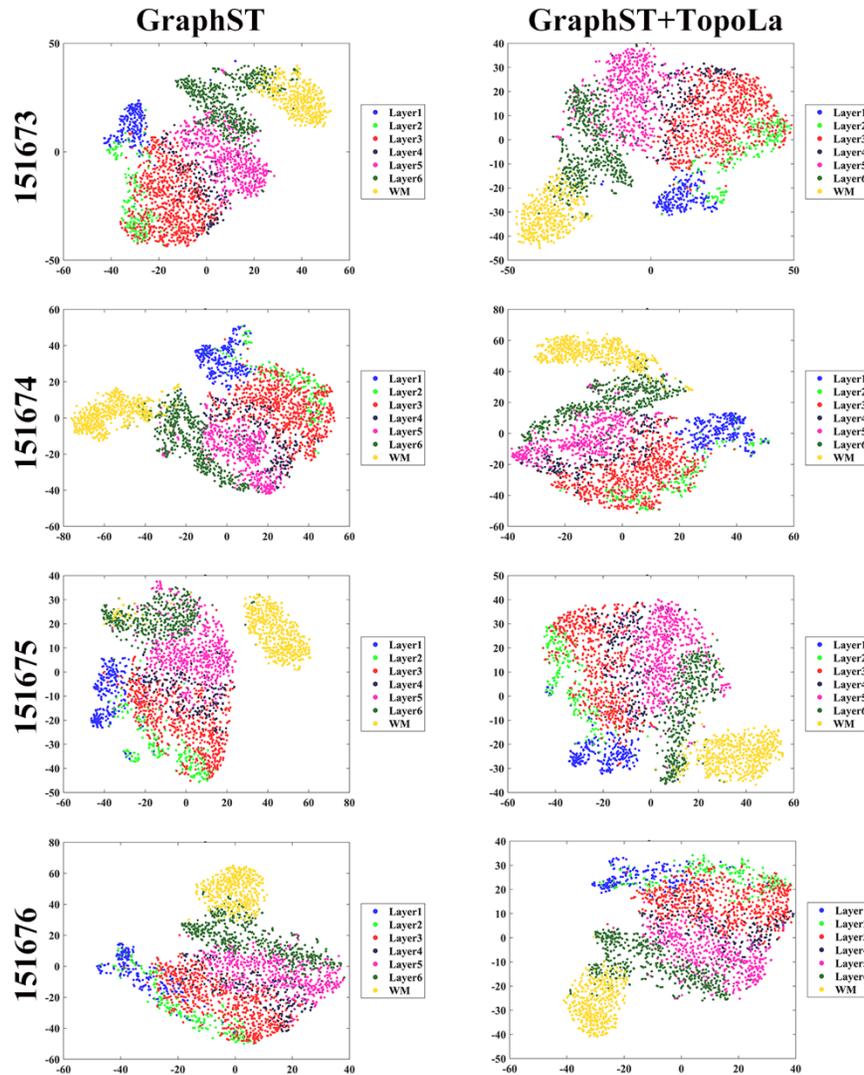

**Supplementary Fig. 34** t-SNE visualizations of cell embeddings from slices 151673, 151674, 151675, and 151676, comparing GraphST results before (left) and after (right) TopoLa enhancement. The embeddings are visualized using t-SNE, with distinct colors representing different cell types, showcasing the improved dispersion of cell types.

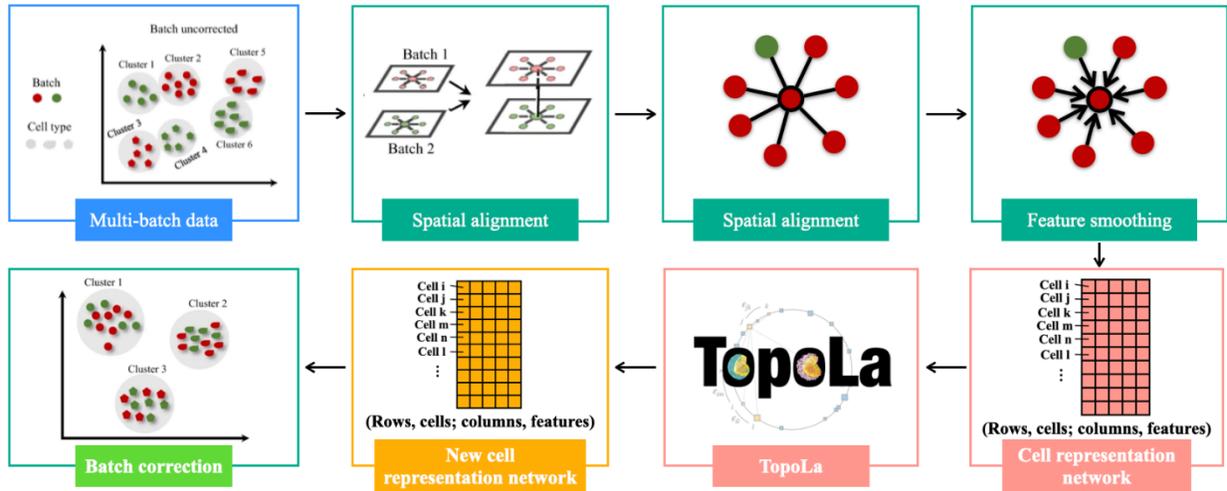

**Supplementary Fig. 35** Overview of GraphST+TopoLa for multi-batch integration. First, hematoxylin and eosin (H&E) stained images from two or more samples are aligned to establish spatial correspondence across samples. A shared neighborhood graph is constructed, accounting for both intra- and inter-sample neighbors, which enables feature smoothing across the entire dataset. Second, GraphST+TopoLa corrects batch effects by smoothing features across samples to generate cell representations. Third, these cell representations are input into the TopoLa framework to obtain enhanced representations for further batch effect correction.

**Supplementary Fig. 36** Enhanced batch effect correction and spatial clustering with GraphST+TopoLa. **a** Aligned ST images of mouse breast cancer samples generated using the PASTE algorithm. **b** UMAP projections of two consecutive slices before batch correction, showing pronounced batch effects. **c** Batch-corrected UMAP projections and clustering results for GraphST and GraphST+TopoLa. The first column displays the improved mixing of cells across batches with GraphST+TopoLa, while the second column is colored by true cell types. **d** Quantitative metrics (iLISI and ASW) show superior performance of GraphST+TopoLa compared to GraphST, indicating enhanced batch integration and cluster consistency.

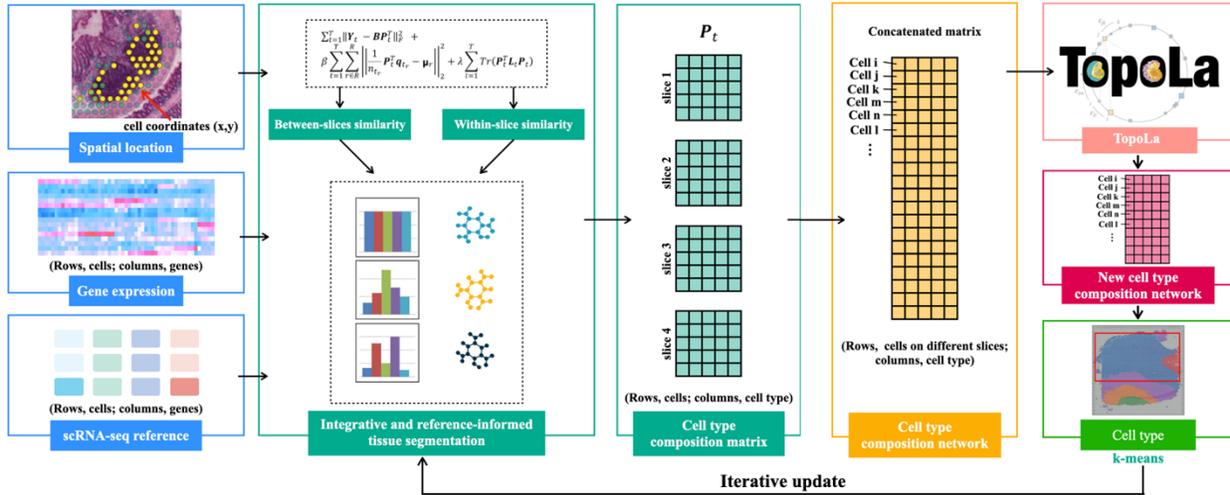

**Supplementary Fig. 37** Overview of the IRIS+TopoLa workflow. First, IRIS+TopoLa requires as input spatial transcriptomics (ST) data measured across multiple tissue slices with spatial localization information, along with scRNA-seq reference data from the same tissue containing cell-type-specific gene expression information. Second, IRIS+TopoLa integrates scRNA-seq data into the ST data and consolidates ST data across multiple tissue slices to perform comprehensive reference domain detection and obtain cell representations. Third, these cell representations are fed into the TopoLa framework to obtain enhanced cell representations. Finally, clustering is performed based on the enhanced cell representations to identify spatial domains within the tissue.

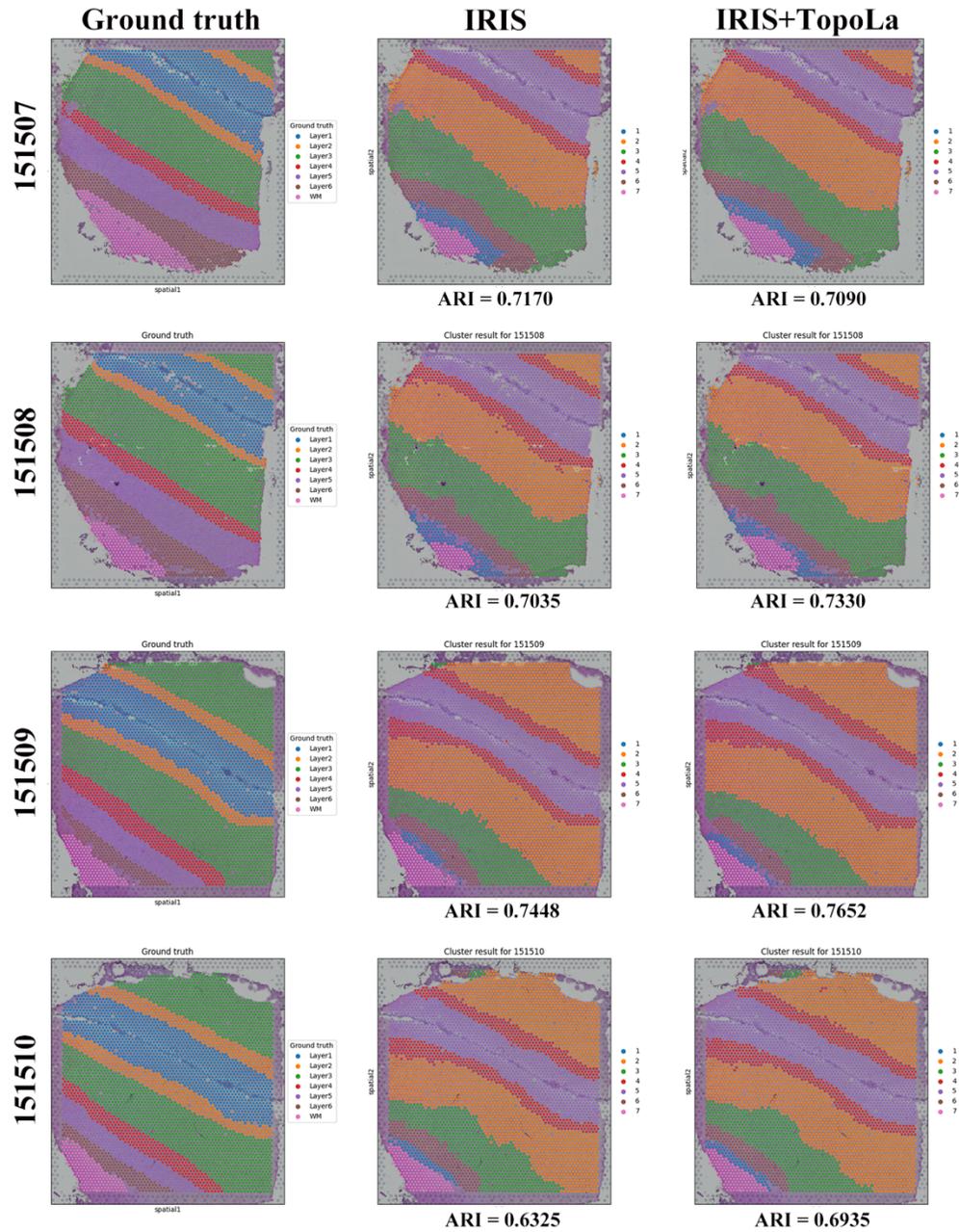

220
**Supplementary Fig. 38** Clustering results produced by IRIS and IRIS+TopoLa on slices 151507,
151508, 151509, and 151510 from the DLPFC dataset.

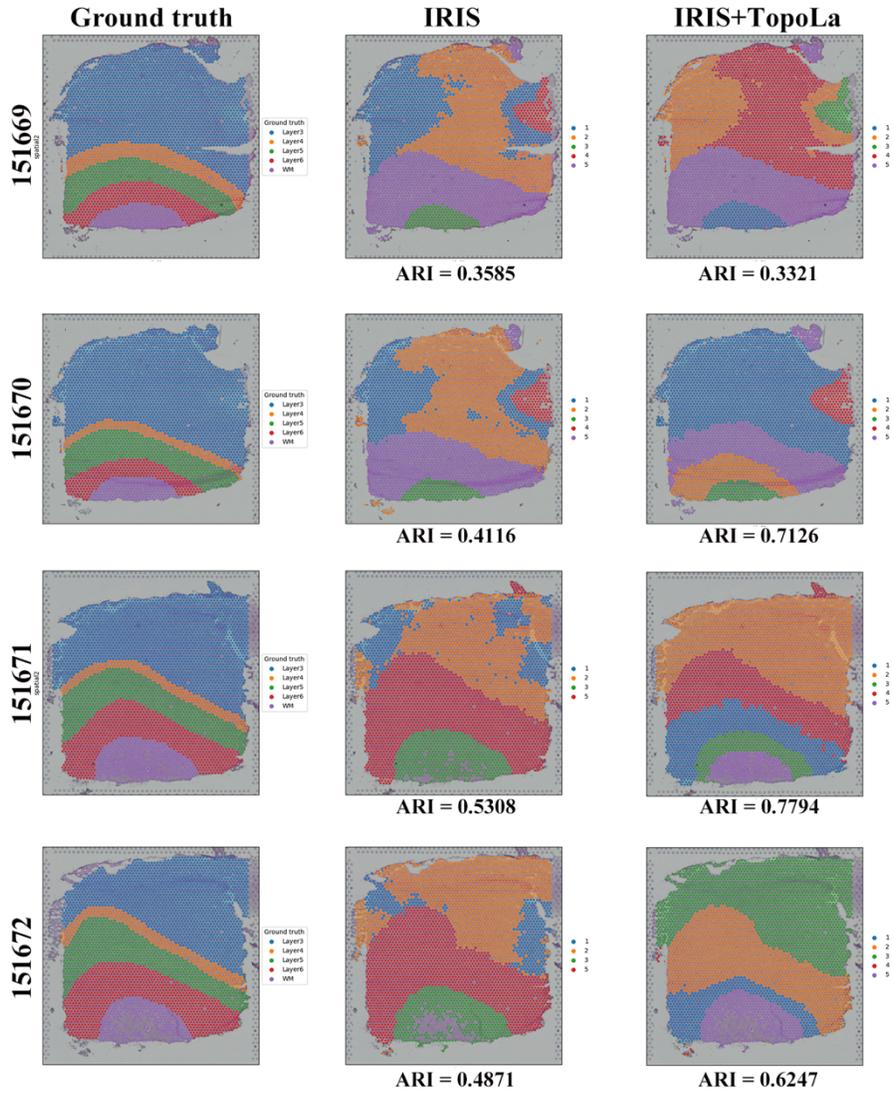

223
**Supplementary Fig. 39** Clustering results produced by IRIS and IRIS+TopoLa on slices 151669,
151670, 151671, and 151672 from the DLPFC dataset.

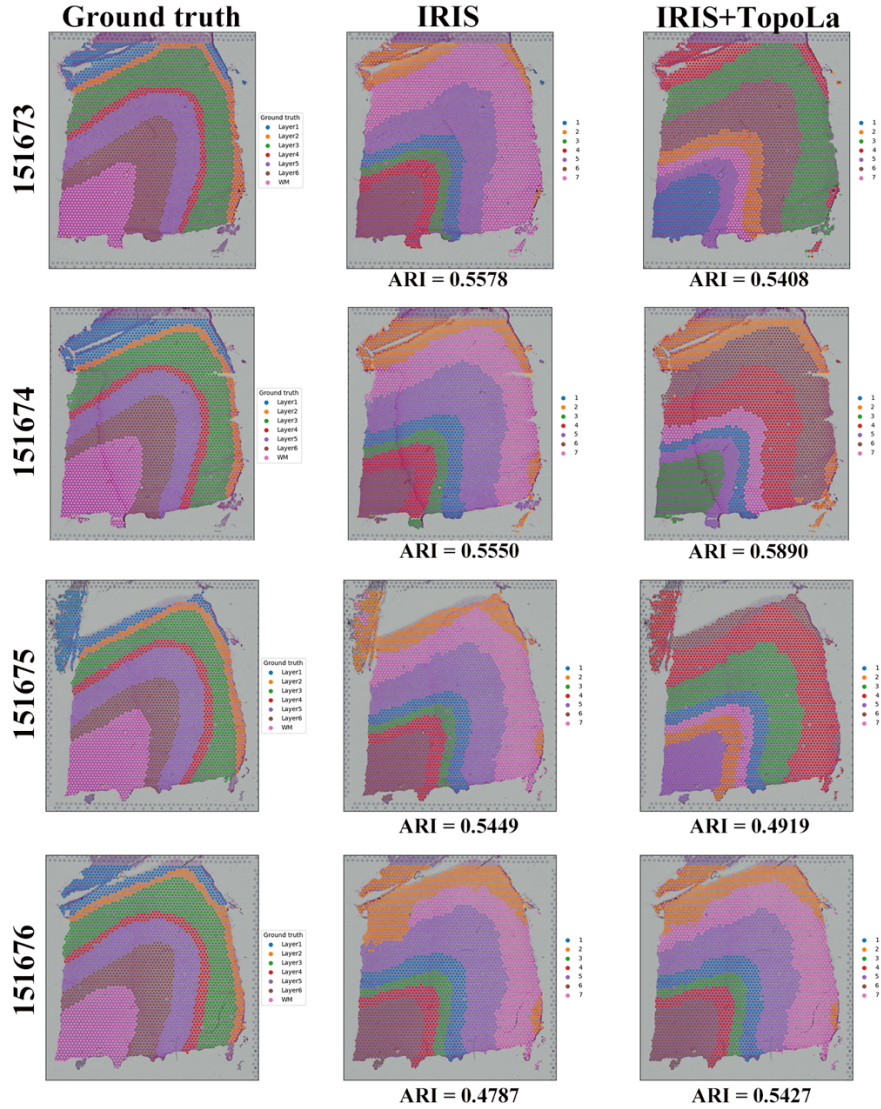

226
227 **Supplementary Fig. 40** Clustering results produced by IRIS and IRIS+TopoLa on slices 151673,
228 151674, 151675, and 151676 from the DLPFC dataset.

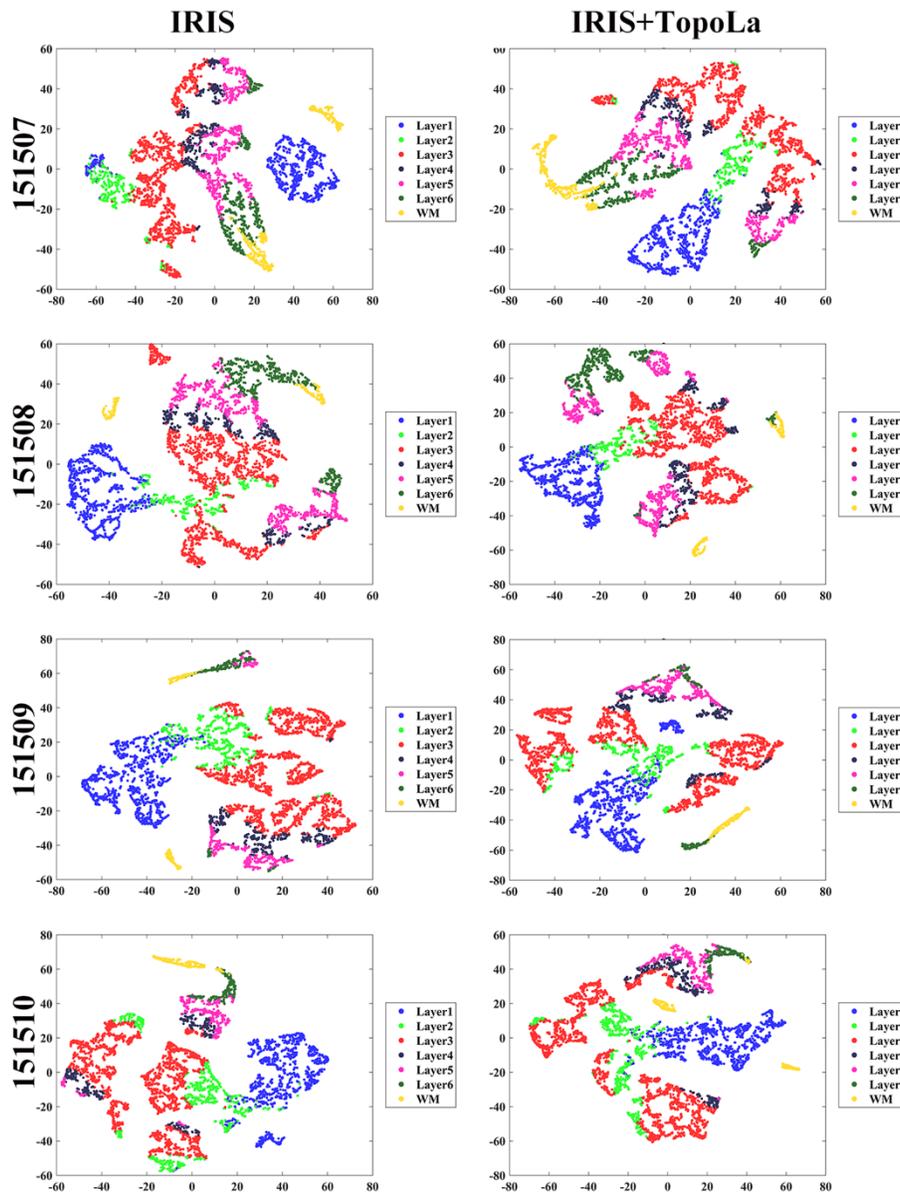

230 **Supplementary Fig. 41** t-SNE visualizations of cell embeddings from slices 151507, 151508,
231 151509, and 151510, comparing IRIS results (left) and IRIS+TopoLa (right). The embeddings are
232 visualized using t-SNE, with distinct colors representing different cell types.

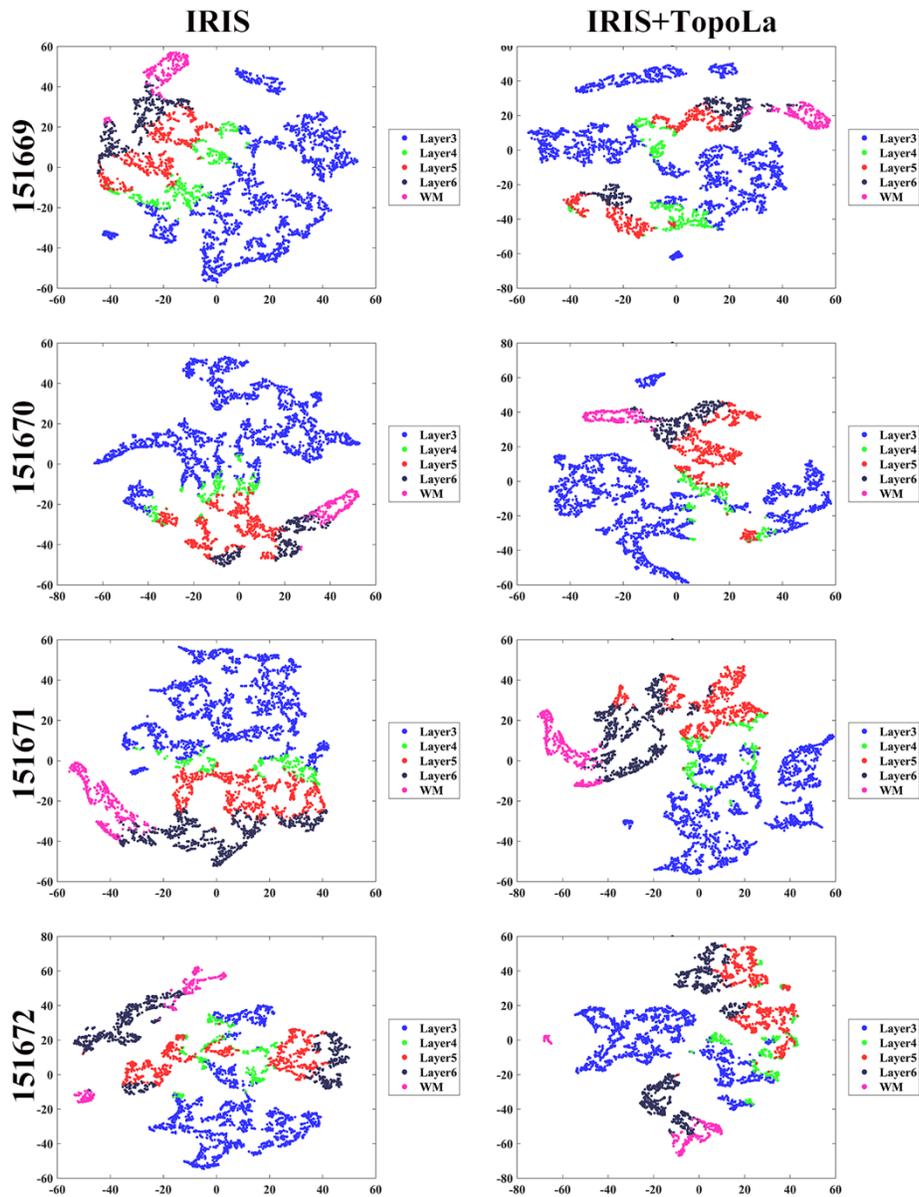

233
234 **Supplementary Fig. 42** t-SNE visualizations of cell embeddings from slices 151669, 151670,
235 151671, and 151672, comparing IRIS results (left) and IRIS+TopoLa (right). The embeddings are
236 visualized using t-SNE, with distinct colors representing different cell types.

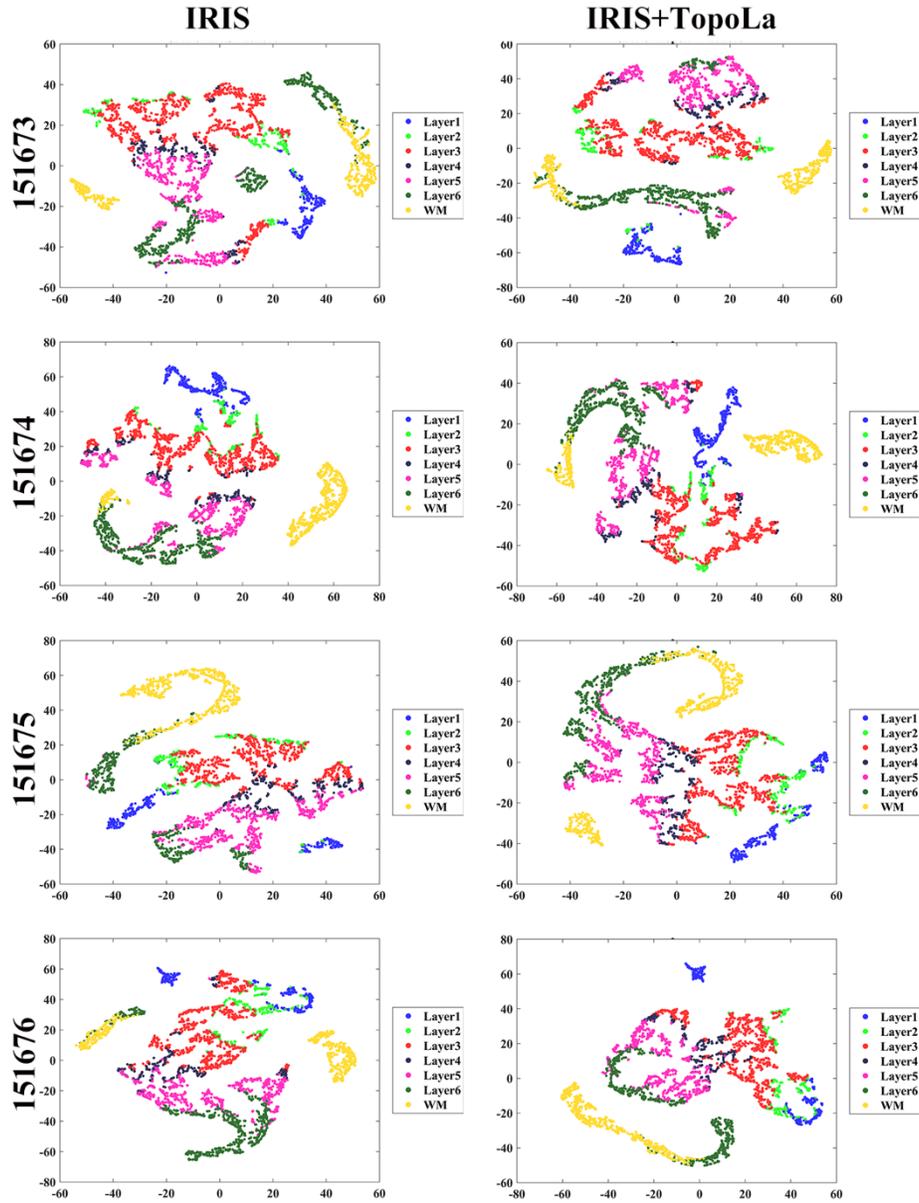

237
238 **Supplementary Fig. 43** t-SNE visualizations of cell embeddings from slices 151673, 151674,
239 151675, and 151676, comparing IRIS results (left) and IRIS+TopoLa (right). The embeddings are
240 visualized using t-SNE, with distinct colors representing different cell types.

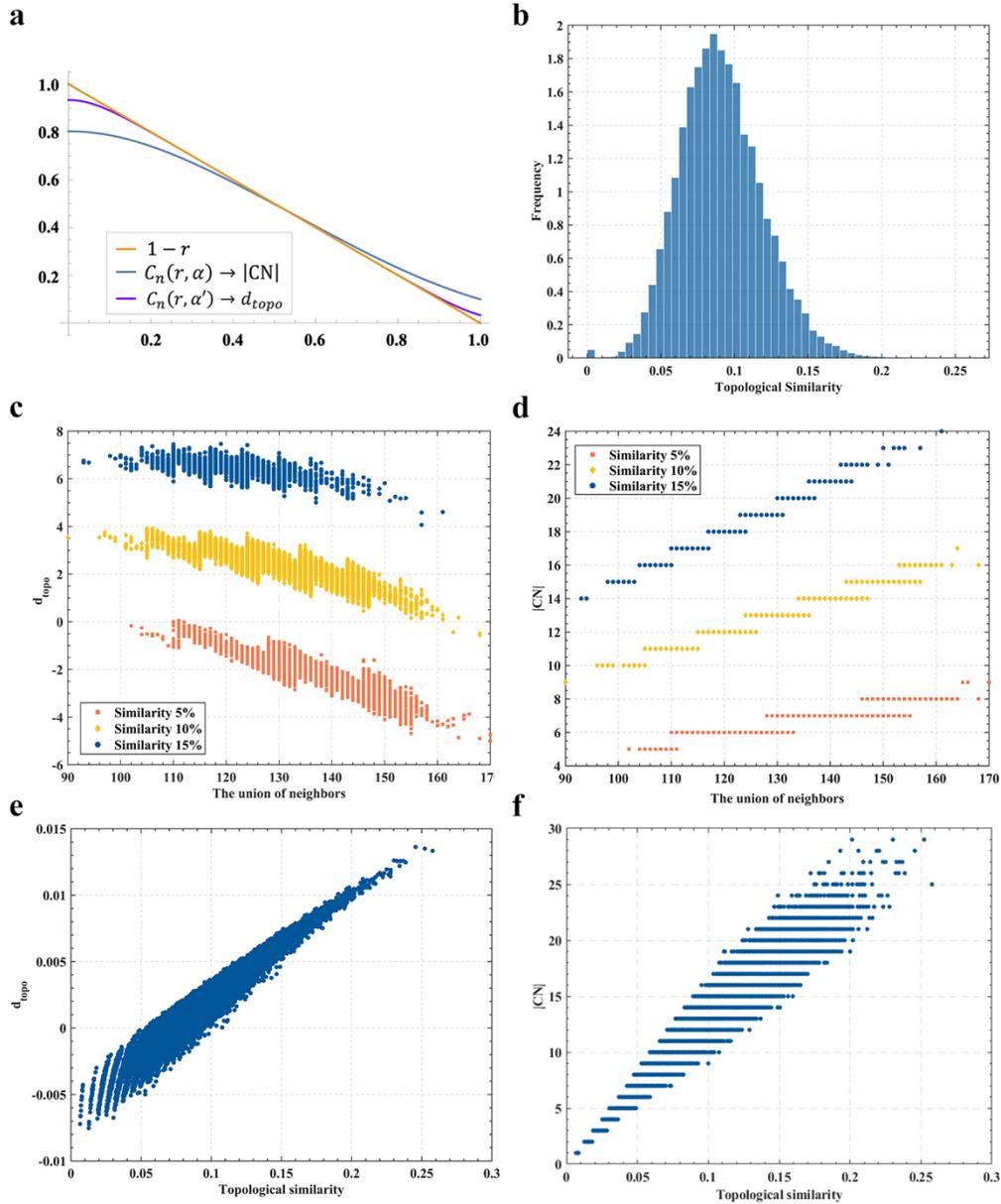

**Supplementary Fig. 44** The properties of energy distance measures. **a** The relationships between energy distance and its measures. Utilizing the maximum entropy principle, $|CN|$ and $d_{topo}$ can be expressed as functions of energy distance $r$, where $C_n(\cdot)$ denotes the integration of all 2-hop paths between the two involving nodes. The distinction between $|CN|$ and $d_{topo}$ lies in weighting, with $d_{topo}$ exhibiting a higher logarithm of thermodynamic activity and thus closer to the energy distance (approaching the line of $1-r$). **b** The topological similarity distribution between nodes in a randomly generated undirected graph ($\mathbb{R}^{500\times500}$), calculated using the Jaccard index. **c** The relationship between the number of neighbors and $d_{topo}$ for the sampled node pairs, with corresponding topological similarities around 5%, 10%, and 15%. **d** The relationship between the number of neighbors and $|CN|$ for the sampled node pairs, with corresponding topological similarities around 5%, 10%, and 15%. **e** The relationship between topological similarity and $d_{topo}$. **f** The relationship between the topological similarity and $|CN|$.

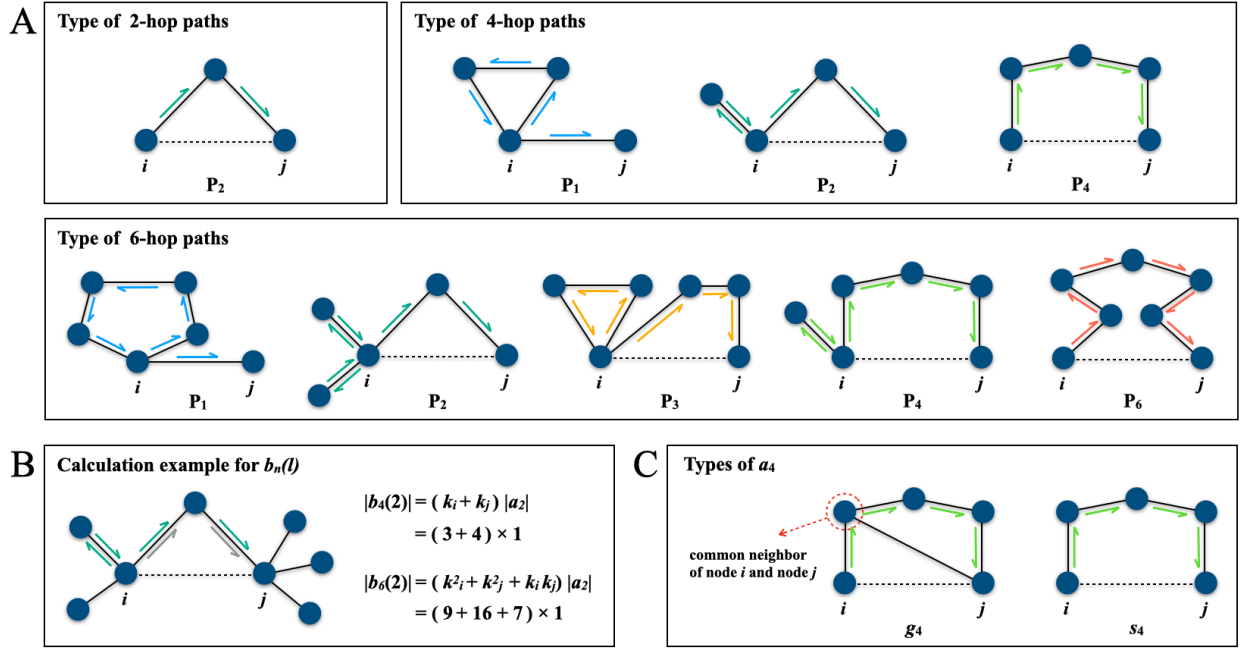

**Supplementary Fig. 45** Types of even-hop paths. **a** Paths can be abstracted into polygons. For instance, a 2-hop path can be visualized as a triangle $P_2$. When considering loops in paths, multi-hop path can be depicted as a collection of various polygons. 4-hop paths consist of straight lines, triangles, and pentagons. 6-hop paths consist of straight lines, triangles, quadrangles, pentagons, and heptagons **b** The calculation example for $|b_n(l)|$. $b_n(l)$ denotes $n$-hop $P_l$-type paths that solely loop between node $i$ and its neighbors, or between node $j$ and its neighbors. $\kappa_i$ and $\kappa_j$ are the degree of node $i$ and $j$. $|a_l|$ is the quantity of $l$-hop paths without loop. When $l$ is fixed, $|b_n(l)|$ can be computed using $\kappa_i$ and $\kappa_j$, along with $|a_l|$. This implies that $|b_n(l)|$ encapsulates information regarding $\kappa_i$ and $\kappa_j$. **c** Polygons $a_l$ can be classified into two types: $a_l$ overlapping with $P_2$ ($g_l$) and the remainder ($s_l$). The quantity of $s_l$ exhibits a direct correlation with the topological similarity of the two nodes.

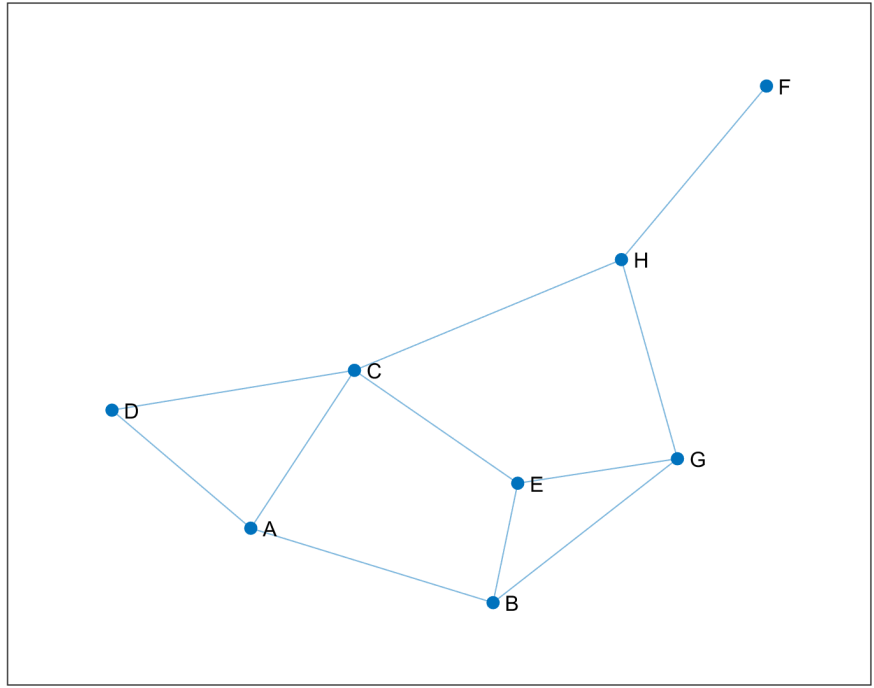

**Supplementary Fig. 46** Network example illustrating the analysis of $n$-hop path types and their statistics. Between nodes D and E, the counts for 2-hop, 4-hop, and 6-hop paths are 1, 10, and 89, while the respective quantities of $|P_3|$, $|P_5|$, and $|P_7|$ are 1, 3, and 2. The degrees of nodes B and D are 3 and 2, with the value of $\kappa_{ij}$ being 4.

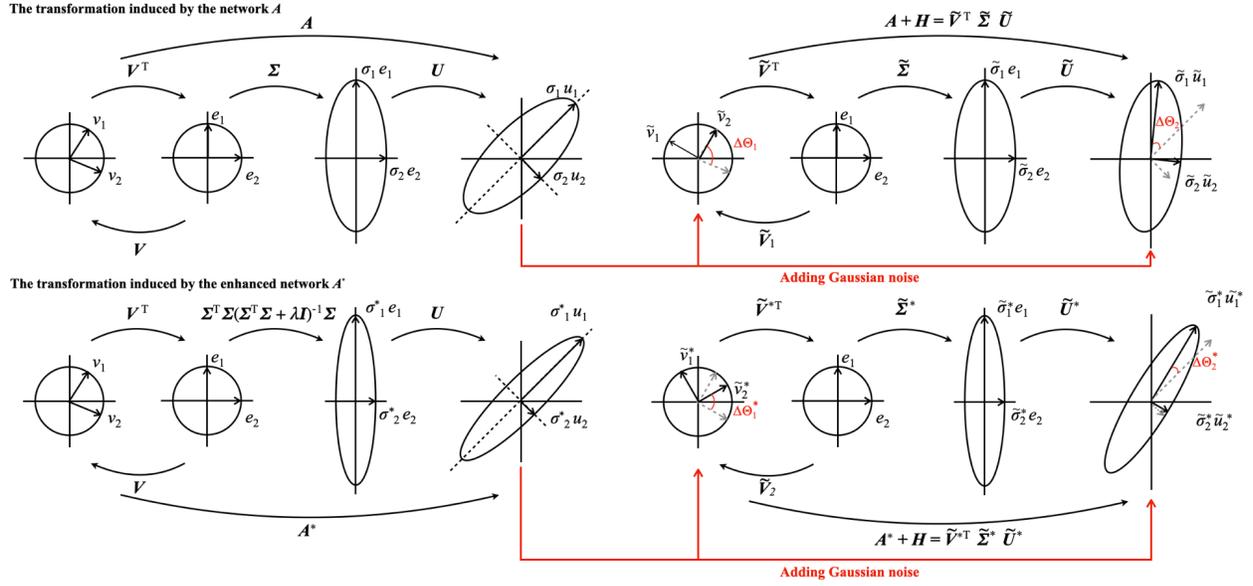

271
**Supplementary Fig. 47** TopoConv enhances complex networks by adding singular value gaps.
The principle is illustrated using a graphical representation of linear transformation. In this case,
increasing the singular value gap reduces the weight of non-principal component data, as depicted
in the figure where $\sigma_2^*$ is smaller than $\sigma_2$, leading to a narrower $\boldsymbol{A}^*$. Importantly, TopoLa preserves
the left and right singular vectors of the matrix, ensuring that the underlying structure of the matrix
is not altered. Through matrix perturbation analysis, when identical Gaussian noise $\boldsymbol{H}$ is added to
both $\boldsymbol{A}$ and $\boldsymbol{A}^*$, the error upper bound of $\boldsymbol{A}^*$ is less than or equal to that of $\boldsymbol{A}$, that is
$\max\{\sin(\Delta\Theta_1^*), \sin(\Delta\Theta_2^*)\} \leq \max\{\sin(\Delta\Theta_1), \sin(\Delta\Theta_2)\}$ (**supplementary text 2.2**, **Theorem 4**).

# Supplementary Tables

**Supplementary Table 1** Incorporating TopoLa improves the performance of many tasks in scRNA-seq and spatial transcriptomics

| Method | Journal | Published time | Biological research | Dataset | Average improvement (metrics) |
|--------|---------|----------------|---------------------|---------|-------------------------------|
| SIMLR | Nature Methods | 2017.5 | Clustering of scRNA-seq data | Baron_human1, Baron_human2, Baron_human3, Baron_human4, Baron_mouse1, Biase, Buettner, Chu_celltime, Chu_celltype, Chung, Darmanis, Deng, Engel, Goolam, Grover, Jurkat, Karlsson, Kolodz, Kumar, Leng, Maria, Pollen, Puram, Robert, Ting, Treutlein, Yan, Yeo, Zelsel, Zhou | 5.8% (ARI) 2.4% (NMI) |
| scGNN | Nature Communications | 2021.5 | Clustering of scRNA-seq data | Baron_human1, Baron_human2, Baron_human3, Baron_human4, Baron_mouse1, Biase, Buettner, Chu_celltime, Chu_celltype, Chung, Darmanis, Deng, Engel, Goolam, Grover, Jurkat, Karlsson, Kolodz, Kumar, Leng, Maria, Pollen, Puram, Robert, Ting, Treutlein, Yan, Yeo, Zelsel, Zhou | 2.6% (ARI) 6.1% (NMI) |
| scGPT | Nature Methods | 2024.1 | Single-cell multi-batch integration | PBMC, Cortex, Covid | 15.4% (ARI) 3.2% (NMI) |
| scGPT | Nature Methods | 2024.1 | Single-cell multi-omic integration | BMMC | 45.7% (ARI) 8.1% (NMI) |
| SCA | Genome Biology | 2023.8 | Rare cell identification | 10X PBMC, Airway, Cao, Chung, Darmanis, Deng, Goolam, hippocampus, iLNs, Koh, Li, livers, MacParland, pancrea, Pollen, Puram, Tonsil, UUOkidney, Yang, Zelsel | 28.1% (F1 score) |
| graphST | Nature Communications | 2023.5 | spatially informed clustering of ST | DLPFC: 151507, 151508, 151509, 151510, 151669,151670, 151671, 151672, 151673, 151674, 151675,151676 | 20.9% (ARI) 4.3% (NMI) |
| graphST | Nature Communications | 2023.5 | Vertical integration of multiple tissue slices | Mouse breast cancer tissue | 59.8% (ASW) |
| IRIS | Nature Methods | 2024.6 | spatially informed clustering with the integration of scRNA-seq and ST | DLPFC: 151507, 151508, 151509, 151510, 151669,151670, 151671, 151672, 151673, 151674, 151675,151676 | 11.7% (ARI) 3.5% (NMI) |

**Supplementary Table 2** Detailed information on the thirty datasets used in single cell clustering analysis

| Dataset | Cells | Features | Types | Accession | Description |
|---|---|---|---|---|---|
| Baron_human1 | 1937 | 20124 | 14 | GSE84133 | human pancreatic cells |
| Baron_human2 | 1724 | 20124 | 14 | GSE84133 | human pancreatic cells |
| Baron_human3 | 3605 | 20124 | 14 | GSE84133 | human pancreatic cells |
| Baron_human4 | 1303 | 20124 | 14 | GSE84133 | human pancreatic cells |
| Baron_mouse | 822 | 14877 | 13 | GSE84133 | mouse pancreatic cells |
| Biase | 49 | 25736 | 3 | GSE57249 | mouse embryonic stem cells |
| Buettner | 288 | 38293 | 3 | E-MTAB-2805 | mouse embryonic stem cells |
| Chu_celltime | 758 | 19176 | 6 | GSE75748 | human pluripotent stem cells |
| Chu_celltype | 1018 | 19095 | 7 | GSE75748 | human pluripotent stem cells |
| Chung | 515 | 20345 | 5 | GSE75688 | human tumor and immune cells |
| Darmanis | 466 | 22085 | 9 | GSE67835 | human brain cells |
| Deng | 259 | 22958 | 10 | GSE45719 | mouse cells from different stages |
| Engel | 203 | 23337 | 4 | GSE74596 | mouse Natural killer T cells |
| Goolam | 124 | 41388 | 8 | E-MTAB-3321 | mouse cells from different stages |
| Grover | 135 | 15180 | 2 | GSE70657 | human hematopoietic stem cells |
| Karlsson | 94 | 59745 | 3 | E-MTAB-6142 | human myxoid liposarcoma cells |
| Kolodz | 704 | 38653 | 3 | E-MTAB-2600 | mouse embryonic stem cells |
| Kumar | 361 | 22394 | 4 | GSE60749 | mouse embryonic stem cells |
| Leng | 247 | 19084 | 3 | GSE64016 | human embryonic stem cells |
| Maria | 759 | 33694 | 7 | GSE124731 | human innate T cells |
| Pollen | 249 | 6981 | 11 | GSM1832359 | human brain cells |
| Puram | 3363 | 23686 | 8 | GSE103322 | non-malignant cells in Head and Neck Cancer |
| Robert | 194 | 23418 | 2 | GSE74923 | mouse leukemia cell line andprimary CD8+ T-cells |
| Ting | 187 | 21583 | 7 | GSE51372 | mouse circulating tumor cells |
| Treutlein | 80 | 23271 | 5 | GSE52583 | mouse lung epithelial cells |
| Yan | 90 | 20214 | 6 | GSE36552 | human embryonic stem cells |
| Yeo | 206 | 20345 | 3 | GSE85908 | human induced pluripotentstem cells |
| Zelsel | 3005 | 19486 | 9 | GSE60361 | mouse cerebral cortex cells |
| Zhou | 181 | 23937 | 8 | GSE67120 | mouse haematopoietic stem cells |
| Jurkat | 1580 | 32738 | 2 | 10X PBMC | Jurkat cells |

285

286

287

**Supplementary Table 3** Detailed information on the twenty datasets used in rare cell identification analysis

| Dataset | Cells | Genes | Types | Accession | Description |
|---|---|---|---|---|---|
| 10X PBMC | 4271 | 16653 | 8 | 10X genomics | peripheral blood mononuclear cells |
| Airway | 7193 | 27716 | 7 | GSE103354 | mouse tracheal epithelium cells |
| Cao | 4186 | 13488 | 10 | sci-RNA-seq platform | worm neuron cells |
| Chung | 515 | 20345 | 5 | GSE75688 | human tumor and immune cells |
| Darmanis | 466 | 22085 | 9 | GSE67835 | human brain |
| Deng | 259 | 22431 | 10 | GSE45719 | mouse cells from different stages |
| Goolam | 124 | 41388 | 8 | E-MTAB-3321 | mouse cells from different stages |
| Hippocampus | 1402 | 25392 | 8 | GSE84371 | mouse hippocampus |
| iLNs | 9662 | 25929 | 24 | GSE139600 | murine iLNs |
| Koh | 498 | 60483 | 9 | GSM2257302 | human embryonic stem cells |
| Li | 561 | 57241 | 7 | GSE81861 | human colorectal tumors |
| Livers | 16015 | 19907 | 11 | GSE148339 | mouse frozen liver |
| MacParland | 8444 | 5000 | 11 | GSE115469 | human liver cells |
| Pancreas | 8569 | 20125 | 14 | GSE84133 | human pancreas |
| Pollen | 249 | 6982 | 11 | SRP041736 | human developing cortex cells |
| Puram | 3363 | 23686 | 8 | GSE103322 | non-malignant cells in Head and Neck Cancer |
| Tonsil | 5778 | 36601 | 13 | Broad Institute Single Cell Portal | human tonsil |
| UUOkidney | 6147 | 21516 | 17 | GSE119531 | mouse kidney from unilateral ureteral obstruction (UUO) |
| Yang | 1119 | 46609 | 6 | GSE90848 | mouse hair follicles stem cells |
| Zelsel | 3005 | 19486 | 9 | GSE60361s | mouse cortex and hippocampus |

**Supplementary Text**

# 1 Definition of Evaluation Metrics
## 1.1 The Adjusted Rand Index

In this article, we used the Adjusted Rand Index (ARI) to assess the degree of correspondence between the clustering results and the true class labels. The ARI is calculated using the following formula:

$$ARI = \frac{\sum_i \sum_j \binom{n_{ij}}{2} - \left[\sum_i \binom{a_i}{2} \sum_j \binom{b_j}{2}\right] / \binom{n}{2}}{\frac{1}{2}\left[\sum_i \binom{a_i}{2} + \sum_j \binom{b_j}{2}\right] - \left[\sum_i \binom{a_i}{2} \sum_j \binom{b_j}{2}\right] / \binom{n}{2}}. \tag{1}$$

where $X = \{X_1, X_2, X_3, \dots, X_r\}$ represents the set of $r$ clusters obtained by clustering, and $Y = \{Y_1, Y_2, Y_3, \dots, Y_s\}$ represents the set of the correct labels, and $n_{ij}$ refers to the number of samples at the intersection of $X_i$ and $Y_j$ that is, $n_{ij} = |X_i \cap Y_j|$. $a_i$ represents $\sum_k n_{ik}$ and $b_j$ represents $\sum_l n_{lj}$. The value range of ARI is $[-1, 1]$, the larger the value, the higher the consistency between the two sets.

## 1.2 Normalized Mutual Information

We introduced Normalized Mutual Information (NMI)[1] to evaluate the quality of clustering from another perspective. The NMI is defined as $I(U,V) / \max\{H(U), H(V)\}$ given two clustering results $U$ and $V$ on a set of data points. Here, $I(U,V)$ represents the mutual information between $U$ and $V$, while $H(U)$ represents the entropy of the clustering $U$. To compute the mutual information, we assume that $U$ has $p$ clusters and $V$ has $q$ clusters, and the calculation is performed as follows:

$$I(U,V) = \sum_{i=1}^{p} \sum_{j=1}^{q} \frac{|U_i \cap V_j|}{N} log \frac{|U_i \cap V_j|}{|U_i|/N \times |V_j|/N}, \tag{2}$$

The cardinality of the $p$-th cluster in $U$ is denoted by $|U_i|$, and $N$ represents the number of points. The entropy of each cluster assignment is calculated using the following formula:

$$H(U) = -\sum_{i=1}^{p} \frac{|U_i|}{N} log \frac{|U_j|}{N}, \tag{3}$$

$$H(V) = -\sum_{j=1}^{q} \frac{|V_j|}{N} log \frac{|V_j|}{N}, \tag{4}$$

The value range of NMI is $[0, 1]$, the larger the value, the higher the consistency between the two sets.

## 1.3 Average Silhouette Width

The silhouette width evaluates the relationship between a cell's within-cluster distances and its distances to the nearest cluster boundaries[2]. The average silhouette width (ASW) score is obtained by calculating the average of all the silhouette widths. This score can range from -1 to 1. A score of 1 indicates well-separated clusters, while scores closer to -1 or 0 suggest that the clusters are either overlapping or there is potential misclassification. When evaluating cell type clustering, the ASW score is computed based on the cell type labels. The following formula is used to calculate this score:

$$ASW = \frac{ASW_C + 1}{2} \tag{5}$$

where C represents the cell types. Higher scores indicate better cell-type clustering or batch-mixing performance.

### 1.4 F1 score

We used the standard classification metric, the F1 score, to evaluate the performance of rare cell identification. The F1 scores are calculated using true positives ($tp$), false positives ($fp$), and false negatives ($fn$). Cell types that comprise less than 5% of the total dataset are labeled as rare cells. The F1 score is calculated as follows:

$$F1 = \frac{2 \times Precision \times Recall}{Precision + Recall} \tag{6}$$

where the Precision and Recall scores are calculated as follows:

$$Precision = \frac{tp}{tp + fp} \tag{7}$$

$$Recall = \frac{tp}{tp + fn} \tag{8}$$

The metrics mentioned above are calculated using scikit-learn's implementations[3].

### 1.5 Integration Local Inverse Simpson's Index

We are using the integration Local Inverse Simpson's Index (iLISI) adapted from Korsunsky *et al*[4]. to evaluate batch-mixing performance. This metric operates on all full feature, embedding, and kNN integration outputs, using shortest path-based distance computation on single-cell kNN graphs. By default, the function returns a value scaled between 0 and 1. The metrics mentioned above are calculated using scib's implementations[5].

## 2 Theoretical Analysis

### 2.1 The Topology-encoded Latent Hyperbolic Geometry

This section mainly presents the proofs of the relevant theorems for TopoLa.

**Theorem 1.** Given a network $X \in \mathbb{R}^{d \times n}$, let $\langle t(x) \rangle$ be the expected number of triangles, and $\langle t'(x) \rangle$ be the expected number of weighted triangles. We have $\alpha' > \alpha$, where $\alpha'$ is the logarithm of thermodynamic activity corresponding to $\langle t'(x) \rangle$ and $\alpha$ is the logarithm of thermodynamic corresponding to $\langle t(x) \rangle$.

**Proof.** In network geometry, the expected numbers of triangles can be obtained through the integration of the graphon[6]:

$$\langle t(x) \rangle = \frac{1}{2} \iint_{\mathbb{R}^2} p(x, y) \, p(y, z) p(z, x) \, dy dz, \tag{9}$$

Under the assumption of redundancy in the current statistics, the non-redundant expectation $\langle t'(x) \rangle$, is anticipated to be less than its redundant counterpart, $\langle t(x) \rangle$. In this context, $t(x)$ denotes the quantity of triangles associated with node $x$. Utilizing the maximum-entropy principle[6], we derive the following formula:

$$\langle t'(x) \rangle = \frac{1}{2} \iint_{\mathbb{R}^2} p'(x,y)\, p'(y,z) p'(z,x)\, dy dz = \bar{t}' < \bar{t}. \tag{10}$$

Therefore,

$$\frac{1}{2} \iint_{\mathbb{R}^2} p'(x,y)\, p'(y,z) p'(z,x)\, dy dz < \frac{1}{2} \iint_{\mathbb{R}^2} p(x,y)\, p(y,z) p(z,x)\, dy dz, \tag{11}$$

When the network size is sufficiently large, the approximate solution for the graphon that maximizes entropy is the Fermi-Dirac graphon:

$$p^*(x,y) = \begin{cases} \dfrac{1}{1 + e^{2\alpha\left(r - \frac{1}{2}\right)}} & \text{if } 0 \le r \le 1, \\[4mm] \dfrac{1}{1 + e^{\alpha}} & \text{if } r > 1, \end{cases} \tag{12}$$

where $\alpha$ and $r$ are the rescaled inverse temperature and energy distance, respectively. Inserting the terms from Formula (12) into Formula (11), we obtain:

$$\iint_{\mathbb{R}^2} \left[ \frac{1}{1 + e^{2\alpha'\left(r - \frac{1}{2}\right)}} \right]^3 dy dz < \iint_{\mathbb{R}^2} \left[ \frac{1}{1 + e^{2\alpha\left(r - \frac{1}{2}\right)}} \right]^3 dy dz \tag{13}$$

Consequently, we deduce that $\alpha' > \alpha$. Based on prior research, the common neighbor integral corresponding to $\alpha'$ provides a more accurate representation of the latent space distance than that associated with $\alpha$[6]. From this, we demonstrate that the current triangle statistics exhibit redundancy. Given that the common-neighbor integral is defined as $\int_{\mathbb{R}} \left[ \frac{1}{1 + e^{2\alpha\left(r - \frac{1}{2}\right)}} \right]^2 dz$, the 2-hop paths (common neighbors) are also found to be imprecise when representing distances within the latent space and **Theorem 1** holds.

**Theorem 2.** Given a topological structure of node $y \in \mathbb{R}^d$, network $\boldsymbol{X} \in \mathbb{R}^{n \times d}$ and a parameter $\lambda$. Let $\boldsymbol{c}^*$ be the optimal solution to the least squares optimization with regularization (in vector form):

$$\min \frac{1}{\lambda} \|\boldsymbol{y} - \boldsymbol{cX}\|_F^2 + \|\boldsymbol{c}\|_F^2 \tag{14}$$

Then, we have

$$\frac{\left\| \boldsymbol{c}_i^* - \boldsymbol{c}_j^* \right\|_F^2}{\|\boldsymbol{y}\|_F^2} \le \frac{1}{\lambda} \left\| \boldsymbol{x}_i^T - \boldsymbol{x}_j^T \right\|_F^2 \tag{15}$$

**Proof.** Inspired by prior work, we conducted the following inference[7]. Let $L(c) = \frac{1}{\lambda} \|\boldsymbol{y} - \boldsymbol{cX}\|_F^2 + \|\boldsymbol{c}\|_F^2$. Since $\boldsymbol{c}^*$ is the optimal solution to problem, it satisfies

$$\frac{\partial L(c)}{\partial c_k}\bigg|_{c=c^*} = 0 \tag{16}$$

Thus, we have

$$-\frac{2}{\lambda}\boldsymbol{x}_i^T(\boldsymbol{y}-\boldsymbol{c}^*X) + 2\boldsymbol{c}_i^* = 0, \tag{17}$$

$$-\frac{2}{\lambda}\boldsymbol{x}_j^T(\boldsymbol{y}-\boldsymbol{c}^*X) + 2\boldsymbol{c}_j^* = 0. \tag{18}$$

Formulas (17) and (18) give us

$$\boldsymbol{c}_i^* - \boldsymbol{c}_j^* = \frac{1}{\lambda}\left(\boldsymbol{x}_i^T - \boldsymbol{x}_j^T\right)(\boldsymbol{y}-\boldsymbol{c}^*\boldsymbol{X}). \tag{19}$$

Since $\boldsymbol{c}^*$ is optimal to Formula (15), we get

$$\frac{1}{\lambda}\|\boldsymbol{y}-\boldsymbol{c}^*\boldsymbol{X}\|_F^2 + \|\boldsymbol{c}^*\|_F^2 = L(\boldsymbol{c}^*) \leq L(0) = \|\boldsymbol{y}\|_F^2 \tag{20}$$

Thus, we have $\frac{1}{\lambda}\|\boldsymbol{y}-\boldsymbol{c}^*\boldsymbol{X}\|_F^2 < \|\boldsymbol{y}\|_F^2$. Then Formula (20) implies

$$\frac{\left\|\boldsymbol{c}_i^* - \boldsymbol{c}_j^*\right\|_F^2}{\|\boldsymbol{y}\|_F^2} \leq \frac{1}{\lambda}\left\|\boldsymbol{x}_i^T - \boldsymbol{x}_j^T\right\|_F^2 \tag{21}$$

From Formula (21), we discern that the difference in distances between two nodes in the latent space relative to other nodes correlates with the divergence in their topological structures. That is, nodes with highly similar topological structures occupy proximate positions within the latent space and **Theorem 2** holds.

### 2.2 Spatial convolution via topology-encoded latent hyperbolic geometry

This section mainly presents the proofs of the relevant theorems for TopoConv.

**Theorem 3.** If $|\sigma_j| \in [\sqrt{\lambda}, +\infty)$, the singular value gap $(\sigma_{j-1} - \sigma_j)$ of $\boldsymbol{A}$ is smaller than the singular value gap $(\sigma_j^* - \sigma_{j+1}^*)$ of $\boldsymbol{A}^*$.

***Proof.*** Let $\boldsymbol{A}$ represent the original network and $\boldsymbol{A}^*$ represent the enhanced network. Since the matrix $\boldsymbol{A} \in \mathbb{R}^{n \times m}$, where $n$ and $m$ are not necessarily equal, we can express it as $\boldsymbol{A} = \boldsymbol{U}\boldsymbol{\Sigma}\boldsymbol{V}^T$, where $\boldsymbol{U}$ is the left singular vector and the eigenvector of $\boldsymbol{A}\boldsymbol{A}^T$, $\boldsymbol{V}$ is the right singular vector and the eigenvector of $\boldsymbol{A}^T\boldsymbol{A}$, and $\boldsymbol{\Sigma}$ is the singular value matrix with $\boldsymbol{\Sigma}_{i,i} = \sigma_i$. Consequently, we can obtain the change in singular values as follows:

$$\boldsymbol{A}^* = \boldsymbol{A}\boldsymbol{A}^T(\boldsymbol{A}\boldsymbol{A}^T + \lambda\boldsymbol{I})^{-1}\boldsymbol{A}$$

$$= [\boldsymbol{U}\boldsymbol{\Sigma}^T\boldsymbol{\Sigma}\boldsymbol{U}^T][(\boldsymbol{U}^T)^{-1}(\boldsymbol{\Sigma}^T\boldsymbol{\Sigma} + \lambda\boldsymbol{I})^{-1}\boldsymbol{U}^{-1}]\boldsymbol{U}\boldsymbol{\Sigma}\boldsymbol{V}^T$$

$$= \boldsymbol{U}\boldsymbol{\Sigma}^T\boldsymbol{\Sigma}(\boldsymbol{\Sigma}^T\boldsymbol{\Sigma} + \lambda\boldsymbol{I})^{-1}\boldsymbol{\Sigma}\boldsymbol{V}^T \tag{22}$$

As shown in the formula above, the left and right singular vectors remain unchanged, while the singular values become $\boldsymbol{\Sigma}^*_{i,i} = \sigma_i \frac{\sigma_i^2}{\sigma_i^2+\lambda}$. We need to prove that $|\sigma_{j-1} - \sigma_j| \leq |\sigma_{j-1}^* - \sigma_j^*|$, where $j \geq 2$ and $\sigma_j^*$ is a singular value of $\boldsymbol{A}^*$. Since $\sigma_j^* = \sigma_j \frac{\sigma_j^2}{\sigma_j^2+\lambda}$, the inequality can be turned into:

$$\sigma_{j-1} - \sigma_{j-1} \frac{\sigma_{j-1}^2}{\sigma_{j-1}^2 + \lambda} \leq \sigma_j - \sigma_j \frac{\sigma_j^2}{\sigma_j^2 + \lambda}. \tag{23}$$

Since $\sigma_{j-1} \geq \sigma_j$, the remaining task is to prove that $g(x) = x - \frac{x^3}{x^2+\lambda}$ is a decreasing function. We obtain the derivative of $g(x)$ as follows:

$$\frac{\partial g(x)}{\partial x} = 1 - \frac{3x^2(x^2+\lambda) - 2x^4}{(x^2+\lambda)^2} \leq 0$$

$$= 1 - \frac{x^4 + 3\lambda x^2}{(x^2+\lambda)^2} \leq 0. \tag{24}$$

When $\lambda \geq 0$, we can get $|x| \geq \sqrt{\lambda}$.

$$1 - \frac{x^4 + 3\lambda x^2}{(x^2+\lambda)^2} \leq 0, \tag{25}$$

$$1 - \frac{3x^2(x^2+\lambda) - 2x^4}{(x^2+\lambda)^2} \leq 0. \tag{26}$$

We set $g(x) = x - \frac{x^3}{x^2+\lambda}$, and its derivative is $g'(x) = 1 - \frac{3x^2(x^2+\lambda)-2x^4}{(x^2+\lambda)^2}$. Since $1 - \frac{3x^2(x^2+\lambda)-2x^4}{(x^2+\lambda)^2} \leq 0$, $g(x)$ is a decreasing function. We can get $\sigma_{j-1} - \sigma_j \leq \sigma_{j-1}\frac{\sigma_{j-1}^2}{\sigma_{j-1}^2+\lambda} - \sigma_j\frac{\sigma_j^2}{\sigma_j^2+\lambda}$. Therefore, we derive $|\sigma_{j-1} - \sigma_j| \leq |\sigma_{j-1}^* - \sigma_j^*|$ from $|x| \geq \sqrt{\lambda}$ and **Theorem 3** holds.

**Theorem 4.** Suppose that $\widetilde{\boldsymbol{A}} = \boldsymbol{A} + \boldsymbol{H} \in \mathbb{R}^{n \times m}$, $\widetilde{\boldsymbol{A}}^* = \boldsymbol{A}^* + \boldsymbol{H} \in \mathbb{R}^{n \times m}$, $\boldsymbol{H}$ has *i.i.d.* $N(0,\sigma^2)$ entries, $\boldsymbol{A}^* = \boldsymbol{A}\boldsymbol{A}^T(\boldsymbol{A}\boldsymbol{A}^T + \lambda\boldsymbol{I})^{-1}\boldsymbol{A}$. Then

$$sup\{dist(\boldsymbol{A}, \widetilde{\boldsymbol{A}})\} \geq sup\{dist(\boldsymbol{A}^*, \widetilde{\boldsymbol{A}}^*)\} \tag{27}$$

where $\sigma_k \geq \sqrt{\lambda}$, $\sigma_k$ is the singular value of $\boldsymbol{A}$, and $dist(\boldsymbol{A}, \widetilde{\boldsymbol{A}})$ refers to the sin $\Theta$ distance between singular spaces of $\boldsymbol{A}$ and $\widetilde{\boldsymbol{A}}$[8]. $\widetilde{\boldsymbol{A}} := \boldsymbol{A} + \boldsymbol{H}$, where $\boldsymbol{H}$ is the noise. Similarly, $\widetilde{\boldsymbol{A}}^* := \boldsymbol{A}^* + \boldsymbol{H}$.

*Proof.* We prove **Theorem 4** in terms of **Lemma 1**:

For rectangular matrices $\boldsymbol{A}, \boldsymbol{H} \in \mathbb{R}^{n \times m}$, we can write the conformal SVD of the original matrix $\boldsymbol{A}$ and its perturbed version $\widetilde{\boldsymbol{A}} = \boldsymbol{A} + \boldsymbol{H}$ as

$$\boldsymbol{A} = \boldsymbol{U}\boldsymbol{\Sigma}\boldsymbol{V}^T = (\boldsymbol{U}_k \ \boldsymbol{U}_\perp) \begin{pmatrix} \boldsymbol{\Sigma}_k & \\ & \boldsymbol{\Sigma}_\perp \end{pmatrix} \begin{pmatrix} \boldsymbol{V}_k^T \\ \boldsymbol{V}_\perp^T \end{pmatrix}, \tag{28}$$

$$\widetilde{A} = \widetilde{U}\widetilde{\Sigma}\widetilde{V}^T = (\widetilde{U}_k \ \ \widetilde{U}_\perp)\begin{pmatrix}\widetilde{\Sigma}_k & \\ & \widetilde{\Sigma}_\perp\end{pmatrix}\begin{pmatrix}\widetilde{V}_k^T \\ \widetilde{V}_\perp^T\end{pmatrix}. \tag{29}$$

where $U_k \in \mathbb{R}^{n\times(k-1)}$, $U_\perp \in \mathbb{R}^{n\times(n-k+1)}$, $V_k \in \mathbb{R}^{m\times(k-1)}$, $V_\perp \in \mathbb{R}^{m\times(m-k+1)}$, $[U_k, U_\perp] \in \mathbb{R}^{n\times n}$, $[V_k, V_\perp] \in \mathbb{R}^{m\times m}$ are orthogonal matrices. $\Sigma_k = \text{diag}\{\sigma_1, \sigma_2, ..., \sigma_{k-1}\} \in \mathbb{R}^{(k-1)\times(k-1)}$, $\Sigma_\perp = \text{diag}\{\sigma_k, \sigma_{k+1}, ..., \sigma_{\min(m,n)}\} \in \mathbb{R}^{(n-k+1)\times(m-k+1)}$. When $n \neq m$, $\Sigma_\perp$ is rectangular, and the extra columns/rows are padded with $0s$. The decomposition of $\widetilde{A}$ has a similar structure $\widetilde{\Sigma}_k = \text{diag}\{\tilde{\sigma}_1, \tilde{\sigma}_2, ..., \tilde{\sigma}_{k-1}\} \in \mathbb{R}^{(k-1)\times(k-1)}$, $\widetilde{\Sigma}_\perp = \text{diag}\{\tilde{\sigma}_k, \tilde{\sigma}_{k+1}, ..., \tilde{\sigma}_{\min(m,n)}\} \in \mathbb{R}^{(n-k+1)\times(m-k+1)}$. Then,

$$sup\{dist(A, \widetilde{A})\} = \max\{\|\sin\Theta(U_k, \widetilde{U}_k)\|, \|\sin\Theta(V_k, \widetilde{V}_k)\|\} \leq min\left\{\frac{2\|\widetilde{A} - A\|}{\sigma_{k-1} - \sigma_k}, 1\right\}. \tag{30}$$

where $\max\{\|\sin\Theta(U_k, \widetilde{U}_k)\|, \|\sin\Theta(V_k, \widetilde{V}_k)\|\}$ is the maximum value of the distance between the $\widetilde{V}_k$ and $V_k$, $\widetilde{U}_k$ and $U_k$, $\sin\Theta(U_k, \widetilde{U}_k) = \sin(\Delta\Theta_2)$, $\sin\Theta(V_k, \widetilde{V}_k) = \sin(\Delta\Theta_1)$. Suppose $\widetilde{A}^* = A^* + H$. Then the error upper bound of $A^*$ can be expressed as:

$$\max\{\|\sin\Theta(U_k^*, \widetilde{U}_k^*)\|, \|\sin\Theta(V_k^*, \widetilde{V}_k^*)\|\} \leq \min\left\{\frac{\|H\|}{\frac{\sigma_{k-1}^2}{\sigma_{k-1}^2 + \lambda}\sigma_{k-1} - \frac{\sigma_k^2}{\sigma_k^2 + \lambda}\sigma_k}, 1\right\}. \tag{31}$$

Therefore, when $\sigma_{k-1} - \sigma_k > 2\|H\|$, we can infer that the error upper bound of the enhanced network is smaller than the error upper bound of the original network. That is, the error upper bound on the distance between $\widetilde{A}^*$ and $A^*$ is less than or equal to the error upper bound on the distance between $\widetilde{A}$ and $A$. **Theorem 4** holds.

**Lemma 1**. We have the following uniform error bound on $\sin\Theta$ distance

$$\max\{\|\sin\Theta(U_k, \widetilde{U}_k)\|, \|\sin\Theta(V_k, \widetilde{V}_k)\|\} \leq \min\left\{\frac{2\|H\|}{\sigma_{k-1} - \sigma_k}, 1\right\} \tag{32}$$

***Proof.*** To calculate the error bound, we use a modified version of the $sin\Theta$ distance, as presented in **Lemma 4.7** in [8]. If $\sigma_{k-1} = \sigma_k$, then $max\{\|\sin\Theta(U_k, \widetilde{U}_k)\|, \|\sin\Theta(V_k, \widetilde{V}_k)\|\} \leq 1$, and the formula holds. When $\sigma_{k-1} > \sigma_k$, there are two cases: 1. $\sigma_{k-1} - \sigma_k > 2\|H\|$; 2. $\sigma_{k-1} - \sigma_k \leq 2\|H\|$. When $\sigma_{k-1} - \sigma_k > 2\|H\|$, combining with the Weyl inequality ($|\tilde{\sigma}_k - \sigma_k| \leq \|H\|$), we can obtain:

$$\tilde{\sigma}_{k-1} - \sigma_k > \sigma_{k-1} - \sigma_k - \|H\| > \frac{1}{2}(\sigma_{k-1} - \sigma_k) > 0. \tag{33}$$

This satisfies the assumptions of **Lemma 2** and thus we can conclude that:

$$\|\sin\Theta(U_k, \widetilde{U}_k)\| = \|U_\perp^T \widetilde{U}_k\| \leq \frac{\|H\|}{\sigma_{k-1} - \sigma_k - \|H\|} \leq \frac{2\|H\|}{\sigma_{k-1} - \sigma_k}. \tag{34}$$

467 When $\sigma_{k-1} - \sigma_k \le 2\|\boldsymbol{H}\|$, we can obtain:

468
$$\left\|\boldsymbol{U}_\perp^T \widetilde{\boldsymbol{U}}_k\right\| \le 1 \le \frac{2\|\boldsymbol{H}\|}{\sigma_{k-1} - \sigma_k}, \tag{35}$$

469 Combining the two cases, we can get the following formula:

470
$$\left\|\sin\Theta\left(\boldsymbol{U}_k, \widetilde{\boldsymbol{U}}_k\right)\right\|\} \le \min\left\{\frac{2\|\boldsymbol{H}\|}{\sigma_{k-1} - \sigma_k}, 1\right\}. \tag{36}$$

471 Similarly, we can show that

472
$$\left\|\sin\Theta\left(\boldsymbol{V}_k, \widetilde{\boldsymbol{V}}_k\right)\right\|\} \le \min\left\{\frac{2\|\boldsymbol{H}\|}{\sigma_{k-1} - \sigma_k}, 1\right\}. \tag{37}$$

473 Therefore, **Lemma 1** holds.

474 **Lemma 2**. If $\widetilde{\sigma}_{k-1} - \sigma_k > 0$ and $\sigma_{k-1} - \widetilde{\sigma}_k > 0$, then:

475
$$max\{\left\|\sin\Theta\left(\boldsymbol{U}_k, \widetilde{\boldsymbol{U}}_k\right)\right\|, \left\|\sin\Theta\left(\boldsymbol{V}_k, \widetilde{\boldsymbol{V}}_k\right)\right\|\} \le \min\left\{\frac{1}{\sigma_{k-1} - \widetilde{\sigma}_k}, \frac{1}{\widetilde{\sigma}_{k-1} - \sigma_k}\right\}\|\boldsymbol{H}\| \tag{38}$$

476 *Proof.* According to **Lemma 3**, we can obtain the following formula:

477
$$\boldsymbol{U}_\perp^T \widetilde{\boldsymbol{U}}_k = \boldsymbol{F}_U^{21} \circ \left(\boldsymbol{U}_\perp^T \boldsymbol{H} \widetilde{\boldsymbol{V}}_k \widetilde{\boldsymbol{\Sigma}}_k^T + \boldsymbol{\Sigma}_\perp \boldsymbol{V}_\perp^T \boldsymbol{H}^T \widetilde{\boldsymbol{U}}_k\right), \tag{39}$$

478
$$\boldsymbol{U}_k^T \widetilde{\boldsymbol{U}}_\perp = \boldsymbol{F}_U^{12} \circ \left(\boldsymbol{U}_k^T \boldsymbol{H} \widetilde{\boldsymbol{V}}_\perp \boldsymbol{\Sigma}_\perp^T + \boldsymbol{\Sigma}_k \boldsymbol{V}_k^T \boldsymbol{H}^T \widetilde{\boldsymbol{U}}_\perp\right). \tag{40}$$

479 According to **Lemma 4**, we have

480
$$\left\|\boldsymbol{U}_\perp^T \widetilde{\boldsymbol{U}}_k\right\| \le \frac{\widetilde{\sigma}_{k-1}}{\widetilde{\sigma}_{k-1}^2 - \sigma_k^2}\left\|\boldsymbol{U}_\perp^T \boldsymbol{H} \widetilde{\boldsymbol{V}}_k\right\| + \frac{\sigma_k}{\widetilde{\sigma}_{k-1}^2 - \sigma_k^2}\left\|\boldsymbol{V}_\perp^T \boldsymbol{H}^T \widetilde{\boldsymbol{U}}_k\right\|, \tag{41}$$

481
$$\left\|\boldsymbol{U}_\perp^T \widetilde{\boldsymbol{U}}_k\right\| \le \frac{\widetilde{\sigma}_{k-1}}{\widetilde{\sigma}_{k-1}^2 - \widetilde{\sigma}_k^2}\left\|\boldsymbol{H} \widetilde{\boldsymbol{V}}_k\right\| + \frac{\sigma_k}{\widetilde{\sigma}_{k-1}^2 - \sigma_k^2}\left\|\boldsymbol{H}^T \widetilde{\boldsymbol{U}}_k\right\|, \tag{42}$$

482
$$\left\|\boldsymbol{U}_\perp^T \widetilde{\boldsymbol{U}}_k\right\| \le \frac{\|\boldsymbol{H}\|}{\widetilde{\sigma}_{k-1} - \sigma_k}. \tag{43}$$

483 Similarly, we can get

484
$$\left\|\boldsymbol{U}_k^T \widetilde{\boldsymbol{U}}_\perp\right\| \le \frac{\widetilde{\sigma}_k}{\sigma_{k-1}^2 - \widetilde{\sigma}_k^2}\left\|\boldsymbol{U}_k^T \boldsymbol{H}\right\| + \frac{\sigma_{k-1}}{\widetilde{\sigma}_{k-1}^2 - \widetilde{\sigma}_k^2}\left\|\boldsymbol{H} \boldsymbol{V}_k\right\|, \tag{44}$$

485
$$\left\|\boldsymbol{U}_k^T \widetilde{\boldsymbol{U}}_\perp\right\| \le \frac{\|\boldsymbol{H}\|}{\sigma_{k-1} - \widetilde{\sigma}_k}. \tag{45}$$

486 Substituting the above formula into $\left\|\sin\Theta\left(\boldsymbol{U}_k, \widetilde{\boldsymbol{U}}_k\right)\right\| = \min\{\left\|\boldsymbol{U}_\perp^T \widetilde{\boldsymbol{U}}_k\right\|, \left\|\boldsymbol{U}_k^T \widetilde{\boldsymbol{U}}_\perp\right\|\}$[9], and **Lemma 2**
487 holds.

488

**Lemma 3**. Let $\widetilde{A} = A + H$, where $A, H \in \mathbb{R}^{n \times m}$. The conformal SVD of matrices $A$ and $\widetilde{A}$ are defined as:

$$A = U\Sigma V^T = (U_k \; U_\perp) \begin{pmatrix} \Sigma_k & \\ & \Sigma_\perp \end{pmatrix} \begin{pmatrix} V_k^T \\ V_\perp^T \end{pmatrix}. \tag{46}$$

$$\widetilde{A} = \widetilde{U}\widetilde{\Sigma}\widetilde{V}^T = (\widetilde{U}_k \widetilde{U}_\perp) \begin{pmatrix} \widetilde{\Sigma}_k & \\ & \widetilde{\Sigma}_\perp \end{pmatrix} \begin{pmatrix} \widetilde{V}_k^T \\ \widetilde{V}_\perp^T \end{pmatrix}. \tag{47}$$

If the singular values of $A$ and $\widetilde{A}$ satisfy $\sigma_{k-1} - \tilde{\sigma}_k > 0$ and $\tilde{\sigma}_{k-1} - \sigma_k > 0$, then the following expression holds:

$$U_\perp^T \widetilde{U}_k = F_U^{21} \circ \left( U_\perp^T H \widetilde{V}_k \widetilde{\Sigma}_k^T + \Sigma_\perp V_\perp^T H^T \widetilde{U}_k \right), \tag{48}$$

$$U_k^T \widetilde{U}_\perp = F_U^{12} \circ \left( U_k^T H \widetilde{V}_\perp \widetilde{\Sigma}_\perp^T + \Sigma_k V_k^T H^T \widetilde{U}_\perp \right), \tag{49}$$

$$V_\perp^T \widetilde{V}_k = F_V^{21} \circ \left( \widetilde{\Sigma}_\perp^T \widetilde{U}_\perp H \widetilde{V}_k + V_\perp^T H^T \widetilde{U}_k \widetilde{\Sigma}_k \right), \tag{50}$$

$$V_k^T \widetilde{V}_\perp = F_V^{12} \circ \left( \Sigma_k^T U_k^T H \widetilde{V}_\perp + V_k^T H^T \widetilde{U}_k \widetilde{\Sigma}_\perp \right). \tag{51}$$

where $\circ$ is Hadamard product. $F_U^{21} \in \mathbb{R}^{n-k+1,k-1}$ has entries $(F_U^{21})_{i,j} = \frac{1}{\tilde{\sigma}_j^2 - \sigma_{i+k-1}^2}, 1 \leq i \leq n - k + 1, 1 \leq j \leq k - 1$. $F_U^{12} \in \mathbb{R}^{k-1,n-k+1}$ has entries $(F_V^{12})_{i,j} = \frac{1}{\tilde{\sigma}_j^2 - \sigma_{i+k-1}^2}, 1 \leq i \leq m - k + 1, 1 \leq j \leq k - 1$. $\mathbf{F}_V^{12} \in \mathbb{R}^{k-1,m-k+1}$ has entries $(F_V^{12})_{i,j} = \frac{1}{\tilde{\sigma}_{j+k-1}^2 - \sigma_i^2}$, $1 \leq i \leq k - 1, 1 \leq j \leq m - k + 1$. Since $i > \min\{n, m\}$, we set $\sigma_i$ and $\tilde{\sigma}_i$ to be 0.

***Proof.*** Firstly, we will decompose the perturbation $H$ in the following two ways:

$$H = \widetilde{A} - A = \widetilde{U}\widetilde{\Sigma}\widetilde{V}^T - U\Sigma V^T$$

$$= (U + \Delta U)\widetilde{\Sigma}\widetilde{V}^T - U\Sigma(\widetilde{V} - \Delta V)^T$$

$$= U\widetilde{\Sigma}\widetilde{V}^T + (\Delta U)\widetilde{\Sigma}\widetilde{V}^T - U\Sigma\widetilde{V} + U\Sigma(\Delta V)^T$$

$$= U(\Delta\Sigma)\widetilde{V}^T + (\Delta U)\widetilde{\Sigma}\widetilde{V}^T + U\Sigma(\Delta V)^T. \tag{52}$$

Multiplying (52) with $U^T$ on the left and $\widetilde{V}$ on the right leads to:

$$U^T H \widetilde{V} = \Delta\Sigma + U^T(\Delta U)\widetilde{\Sigma} + \Sigma(\Delta V)^T \widetilde{V}. \tag{53}$$

Thus,

$$H = \widetilde{A} - A = \widetilde{U}\widetilde{\Sigma}\widetilde{V}^T - U\Sigma V^T$$

$$= \widetilde{U}\widetilde{\Sigma}(V + \Delta V)^T - (\widetilde{U} - \Delta U)\Sigma V^T$$

$$= \widetilde{U}\widetilde{\Sigma}V^T + \widetilde{U}\widetilde{\Sigma}(\Delta V)^T - \widetilde{U}\Sigma V^T + (\Delta U)\Sigma V^T$$

$$= \widetilde{U}(\Delta\Sigma)V^T + \widetilde{U}\widetilde{\Sigma}(\Delta V)^T + (\Delta U)\Sigma V^T. \tag{54}$$

Similarly, Multiplying (54) with $\widetilde{U}^T$ on the left and $V$ on the right leads to:

$$\widetilde{U}^T H V = \Delta\Sigma + \widetilde{\Sigma}(\Delta V)^T V + \widetilde{U}^T(\Delta U)\Sigma. \tag{55}$$

Denote $dP = \boldsymbol{U}^T \boldsymbol{H} \widetilde{\boldsymbol{V}}$, $d\bar{P} = \widetilde{\boldsymbol{U}}^T \boldsymbol{H} \boldsymbol{V}$, $\Delta \boldsymbol{\Omega}_U = \boldsymbol{U}^T (\Delta \boldsymbol{U})$, $\Delta \boldsymbol{\Omega}_V = \boldsymbol{V}^T (\Delta \boldsymbol{V})$. Notice that $\boldsymbol{I} = \widetilde{\boldsymbol{U}}^T \widetilde{\boldsymbol{U}} = \boldsymbol{U}^T \boldsymbol{U}$ gives $(\boldsymbol{U} + \Delta \boldsymbol{U})^T \widetilde{\boldsymbol{U}} = \boldsymbol{U}^T (\widetilde{\boldsymbol{U}} - \Delta \boldsymbol{U})$. Hence $\boldsymbol{U}^T (\Delta \boldsymbol{U}) = -(\Delta \boldsymbol{U})^T \widetilde{\boldsymbol{U}}$. Similarly, we also have $\boldsymbol{V}^T (\Delta \boldsymbol{V}) = -(\Delta \boldsymbol{V})^T \widetilde{\boldsymbol{V}}$. Plugging these into (52) and (54), we have

$$\begin{cases} dP = \boldsymbol{U}^T \boldsymbol{H} \widetilde{\boldsymbol{V}} = \Delta \boldsymbol{\Sigma} + \Delta \boldsymbol{\Omega}_U \widetilde{\boldsymbol{\Sigma}} - \boldsymbol{\Sigma} \Delta \boldsymbol{\Omega}_V, \\ d\bar{P} = \widetilde{\boldsymbol{U}}^T \boldsymbol{H} \boldsymbol{V} = \Delta \boldsymbol{\Sigma} + \widetilde{\boldsymbol{\Sigma}} \Delta \boldsymbol{\Omega}_V - \Delta \boldsymbol{\Omega}_U^T \boldsymbol{\Sigma}, \end{cases} \tag{56}$$

Next, from (56) we can cancel $\Delta \boldsymbol{\Omega}_V$ by

$$\mathbf{G}_U := dP \widetilde{\boldsymbol{\Sigma}}^{\mathrm{T}} + \boldsymbol{\Sigma} d\bar{P}^{\mathrm{T}}$$

$$= \widetilde{\boldsymbol{\Sigma}} \widetilde{\boldsymbol{\Sigma}}^T - \boldsymbol{\Sigma} \boldsymbol{\Sigma}^T + \Delta \boldsymbol{\Omega}_U \widetilde{\boldsymbol{\Sigma}} \widetilde{\boldsymbol{\Sigma}}^T - \boldsymbol{\Sigma} \boldsymbol{\Sigma}^T \Delta \boldsymbol{\Omega}_U. \tag{57}$$

Let $\Delta \boldsymbol{\Omega}_U = \{w_{ij}\}_{i,j=1}^n$. Then for all $1 \leq i, j \leq n$, the following formulas hold

$$(\mathbf{G}_U)_{ij} = \begin{cases} (\tilde{\sigma}_j^2 - \sigma_i^2) w_{ij}, & i \neq j, \\ (\tilde{\sigma}_j^2 - \sigma_i^2)(w_{ij} + 1), & i = j. \end{cases} \tag{58}$$

Here if $i > \min\{n, m\}$, we define $\sigma_i$ or $\tilde{\sigma}_i$ to be 0. Also, define $\boldsymbol{F}_U^{21}, \boldsymbol{F}_U^{12}, \boldsymbol{F}_V^{21}, \boldsymbol{F}_V^{12}$ as in the statement of **Lemma 2**. By assumption, $\sigma_{k-1} - \tilde{\sigma}_k > 0$ and $\tilde{\sigma}_{k-1} - \sigma_k > 0$, we can directly check that the denominators in these four matrices only have nonzero entries, thus are well defined. Consider the upper right part in $\Delta \boldsymbol{\Omega}_U = \boldsymbol{U}^T \boldsymbol{H}$, that is, $1 \leq i \leq k-1, k \leq j \leq n$, from (58) we have

$$w_{ij} = \frac{1}{\tilde{\sigma}_j^2 - \sigma_i^2} (\mathbf{G}_U)_{ij}, 1 \leq i \leq k-1, k \leq j \leq n. \tag{59}$$

Therefore,

$$\boldsymbol{U}_k^T \widetilde{\boldsymbol{U}}_\perp = \boldsymbol{U}_k^T (\Delta \boldsymbol{U}_\perp) = \boldsymbol{F}_U^{12} \circ \boldsymbol{G}_U^{12} = \boldsymbol{F}_U^{12} \circ \left( \boldsymbol{U}_k^T \boldsymbol{H} \widetilde{\boldsymbol{V}}_\perp \widetilde{\boldsymbol{\Sigma}}_\perp^T + \boldsymbol{\Sigma}_k \boldsymbol{V}_k^T \boldsymbol{H}^T \widetilde{\boldsymbol{U}}_\perp \right) \tag{60}$$

Following the same reasoning, we also obtain

$$\boldsymbol{U}_\perp^T \widetilde{\boldsymbol{U}}_k = \boldsymbol{U}_\perp^T (\Delta \boldsymbol{U}) = \boldsymbol{F}_U^{21} \circ \left( \boldsymbol{U}_\perp^T \boldsymbol{H} \widetilde{\boldsymbol{V}}_k \widetilde{\boldsymbol{\Sigma}}_k^T + \boldsymbol{\Sigma}_\perp \boldsymbol{V}_\perp^T \boldsymbol{H}^T \widetilde{\boldsymbol{U}}_k \right), \tag{61}$$

$$\boldsymbol{V}_\perp^T \widetilde{\boldsymbol{V}}_k = \boldsymbol{V}_\perp^T (\Delta \boldsymbol{V}_\perp) = \boldsymbol{F}_V^{21} \circ \left( \boldsymbol{\Sigma}_\perp^T \widetilde{\boldsymbol{U}}_\perp \boldsymbol{H} \widetilde{\boldsymbol{V}}_k + \boldsymbol{V}_\perp^T \boldsymbol{H}^T \widetilde{\boldsymbol{U}}_k \widetilde{\boldsymbol{\Sigma}}_k \right), \tag{62}$$

$$\boldsymbol{V}_k^T \widetilde{\boldsymbol{V}}_\perp = \boldsymbol{V}_k^T (\Delta \boldsymbol{V}_\perp) = \boldsymbol{F}_V^{12} \circ \left( \boldsymbol{\Sigma}_k^T \boldsymbol{U}_k^T \boldsymbol{H} \widetilde{\boldsymbol{V}}_\perp + \boldsymbol{V}_k^T \boldsymbol{H}^T \widetilde{\boldsymbol{U}}_\perp \widetilde{\boldsymbol{\Sigma}}_\perp \right). \tag{63}$$

**Lemma 4.** Assume $\sigma_{k-1} - \tilde{\sigma}_k > 0$ and $\tilde{\sigma}_{k-1} - \sigma_k > 0$, and let $\mathbf{B}_1 \in \mathbb{R}^{n-k+1,k-1}$, $\mathbf{B}_2 \in \mathbb{R}^{m-k+1,k-1}$, $\mathbf{B}_3 \in \mathbb{R}^{k-1,m-k+1}$, $\mathbf{B}_4 \in \mathbb{R}^{k-1,n-k+1}$ be some arbitrary matrices.

Then, we have

$$\left\| \boldsymbol{F}_U^{21} \circ (\boldsymbol{B}_1 \widetilde{\boldsymbol{\Sigma}}_k) \right\|_F \leq \frac{\tilde{\sigma}_{k-1}}{\tilde{\sigma}_{k-1}^2 - \sigma_k^2} \|\boldsymbol{B}_1\|_F, \left\| \boldsymbol{F}_U^{21} \circ (\boldsymbol{\Sigma}_\perp \boldsymbol{B}_2) \right\|_F \leq \frac{\sigma_k}{\tilde{\sigma}_{k-1}^2 - \sigma_k^2} \|\boldsymbol{B}_2\|_F, \tag{64}$$

$$\left\| \boldsymbol{F}_U^{12} \circ (\boldsymbol{B}_3 \widetilde{\boldsymbol{\Sigma}}_\perp^{\ T}) \right\|_F \leq \frac{\tilde{\sigma}_k}{\sigma_{k-1}^2 - \tilde{\sigma}_k^2} \|\boldsymbol{B}_3\|_F, \left\| \boldsymbol{F}_U^{12} \circ (\boldsymbol{\Sigma}_k \boldsymbol{B}_4) \right\|_F \leq \frac{\sigma_{k-1}}{\sigma_{k-1}^2 - \tilde{\sigma}_k^2} \|\boldsymbol{B}_4\|_F. \tag{65}$$

where $\|\cdot\|_F$ is the Frobenius norm. The same inequality holds for the spectral norm. Similar results also hold for $\boldsymbol{F}_V^{12}$ and $\boldsymbol{F}_V^{21}$.

***Proof.*** Here we only prove the first inequality in the above formulas, and the other three inequalities can be proved similarly. Recall that the definition of $\boldsymbol{F}_U^{21}$ is $(\boldsymbol{F}_U^{21})_{i-k+1,j} = \frac{1}{\tilde{\sigma}_j^2 - \sigma_{i+k-1}^2}$, $1 \leq i \leq n$, $1 \leq j \leq k-1$. We directly have

$$\boldsymbol{F}_U^{21} \circ (\boldsymbol{B}_1 \tilde{\boldsymbol{\Sigma}}_k) = \overline{\boldsymbol{F}}_U^{21} \circ \boldsymbol{B}_1 \tag{66}$$

where

$$(\overline{\boldsymbol{F}}_U^{21})_{i-k+1,j} = \frac{\tilde{\sigma}_j}{\tilde{\sigma}_j^2 - \sigma_i^2}, k \leq i \leq n, 1 \leq j \leq k-1 \tag{67}$$

Let $\boldsymbol{D}_1 = \boldsymbol{F}_U^{21} \circ (\boldsymbol{B}_1 \tilde{\boldsymbol{\Sigma}}_k)$. Then $\boldsymbol{B}_1 = \widetilde{\boldsymbol{F}}_U^{21} \circ \boldsymbol{D}_1$, where

$$(\widetilde{\boldsymbol{F}}_U^{21})_{i-k+1,j} = \frac{\tilde{\sigma}_j^2 - \sigma_i^2}{\tilde{\sigma}_j} = \tilde{\sigma}_j - \frac{\sigma_i^2}{\tilde{\sigma}_j}, k \leq i \leq n, 1 \leq j \leq k-1 \tag{68}$$

Inserting the above expression of $\widetilde{\boldsymbol{F}}$ into $\boldsymbol{B}_1 = \widetilde{\boldsymbol{F}}_U^{21} \circ \boldsymbol{D}_1$, we have

$$\boldsymbol{B}_1 = \boldsymbol{D}_1 \begin{pmatrix} \tilde{\sigma}_1 & & & \\ & \tilde{\sigma}_2 & & \\ & & \ddots & \\ & & & \tilde{\sigma}_{k-1} \end{pmatrix} - \begin{pmatrix} \tilde{\sigma}_k^2 & & & \\ & \tilde{\sigma}_{k+1}^2 & & \\ & & \ddots & \\ & & & \tilde{\sigma}_n^2 \end{pmatrix} \boldsymbol{D}_1 \begin{pmatrix} \frac{1}{\tilde{\sigma}_1} & & & \\ & \frac{1}{\tilde{\sigma}_2} & & \\ & & \ddots & \\ & & & \frac{1}{\tilde{\sigma}_{k-1}} \end{pmatrix} \tag{69}$$

Taking norm on both sides, we obtain

$$\||\boldsymbol{B}_1\|| \leq \tilde{\sigma}_{k-1} \||\boldsymbol{D}_1\|| - \frac{\sigma_k^2}{\tilde{\sigma}_{k-1}} \||\boldsymbol{D}_1\|| = \frac{\tilde{\sigma}_{k-1}^2 - \sigma_k^2}{\tilde{\sigma}_{k-1}} \||\boldsymbol{D}_1\||, \tag{70}$$

which further gives

$$\||\boldsymbol{D}_1\|| \leq \frac{\tilde{\sigma}_{k-1}}{\tilde{\sigma}_{k-1}^2 - \sigma_k^2} \||\boldsymbol{B}_1\|| \tag{71}$$

Similarly, other formulas can be obtained.

## 3  Comparing the two energy distance measures

Energy distance, initially proposed by Krioukov, enables better network properties identification[6]. However, a precise method for quantifying energy distance has yet to be established. Nevertheless, Krioukov demonstrated that |CN|, the number of common neighbors, provides an approximation of the complement of the energy distance.

Previous research has demonstrated that a higher value of the logarithm of thermodynamic activity facilitates more precise measurements of energy distance[6]. We further proved that $\alpha'$, representing the logarithm of thermodynamic activity for $d_{topo}$, exceeds $\alpha$ associated with |CN| (**Supplementary text 2.1**, **Theorem 1**). The precision of energy distance measures is determined by integrating graphons, based on the principle of maximum entropy[6]. Hence, it is difficult to directly compare the precision of two energy distance measures using real-world examples. We employed $\alpha' = 15$ and $\alpha = 5$ as an example to visually illustrate how enhancements in the logarithm of thermodynamic activity affect measurement precision (**Supplementary Fig. 44a**). As shown in the figure, $C_n(\alpha', r)$ is closer to $1 - r$ compared to $C_n(\alpha, r)$, indicating that $C_n(\alpha', r)$ provides a more precise measurement of $r$, where $C_n(\cdot)$ denotes the integration of all 2-hop paths between the two involving nodes[6]. $C_n(\alpha', r)$ represents a function of the integration of common neighbors and the energy distance $r$, where $\alpha'$ is a fixed value. Therefore, $d_{topo}$ offers superior precision compared to |CN|.

We demonstrated the difference in precision between |CN| and $d_{topo}$ using the logarithm of the thermodynamic activity. Next, we use a simple example to analyze the difference in their physical properties. Specifically, we generated an undirected graph with 500 nodes and 10,000 edges. The topological similarities between nodes are $\frac{|CN|}{\kappa_i + \kappa_j - |CN|}$. As shown in **Supplementary Fig. 44b**, the topological similarities in this network mostly range from 5% to 15%.

Firstly, we explored the relationship between the union of neighbors ($\kappa_i + \kappa_j - |CN|$) and different measures. Specifically, we analyzed node pairs with topological similarities around 5%, 10%, and 15%, within a range of ±0.5%. As shown in **Supplementary Figs. 44c, d**, when using |CN| as a measure, the same value can correspond to multiple topological similarities. For instance, |CN| = 14 can correspond to node pairs with topological similarities of 10% and 15%. In contrast, $d_{topo}$ more accurately distinguishes between node pairs with different topological similarities. Additionally, $d_{topo}$ exhibits a negative correlation with the union of neighbors. Previous research has shown that high-degree nodes negatively impact the detection of geometry induced by latent hyperbolic spaces[18]. Therefore, by assigning lower measurement values to high-degree node pairs, $d_{topo}$ can effectively align with the properties of latent hyperbolic geometry.

Secondly, we analyzed the relationship between topological similarities and measures (**Supplementary Figs. 44e, f**). Our findings indicate that a single |CN| value corresponds to a wide range of topological similarities (wide horizontal slices), while $d_{topo}$ provides a more detailed delineation of topological similarity.

These experiments demonstrate the physical properties of $d_{topo}$ in terms of both degree and topological similarity, confirming the conclusions of **Theorem 1**.

# 4    Analysis of $n$-hop path statistics

In the statistical analysis of n-hop paths, paths with loops are also included. By removing loops from these paths, $n$-hop paths can be categorized into $(n-1)$ types. Based on the number of hops after loop removal, these paths can be abstracted as polygons. For instance, 2-hop paths can be represented as triangles ($P_2$), while 4-hop paths correspond to pentagons ($P_4$). We observed a direct correlation between $|b_n(l)|$ and the degrees $\kappa_i$ and $\kappa_j$. Consequently, all n-hop paths are categorized into three types: $a_n$, $b_n$, and $c_n$. In this section, our focus is on $c_n$, primarily through the analysis of 2-hop, 4-hop, and 6-hop paths between nodes D and E (**Supplementary Fig. 45**). Paths without loops in the network are:

$a_2$: *D-C-E*.

$a_3$: *D-A-B-E*; *D-A-C-E*.

$a_4$: *D-A-B-G-E*; *D-C-A-B-E*; *D-C-H-G-E*.

$a_6$: *D-A-B-G-H-C-E*; *D-A-C-H-G-B-E*.

According to our proposed classification system, the statistics of n-hop paths can be represented as:

$$|n - \text{hop}| = \left( \sum_{t=2}^{n-2} |P_t| \right) + |P_n| \tag{72}$$

Consequently, we can determine the quantity of 2-hop paths:

$$|2 - \text{hop}| = |P_2| = |a_2| + |b_2| + |c_2| = |a_2| + 0 + 0 = 1 \tag{73}$$

where $|P_2|$ represents the quantity of $P_2$-type paths. The quantity of 4-hop paths is:

$$|4 - \text{hop}| = |P_1| + |P_2| + |P_4| = 0 + \left[ (\kappa_D + \kappa_E)|a_2| + \left( \sum_{t \in \mathcal{N}_2, t \notin \{D,E\}} \kappa_t \right) - 2|a_3| \right] + |a_4|$$

$$= |P_4| + (\kappa_D + \kappa_E)|a_2| + \left[ \left( \sum_{t \in \mathcal{N}_2, t \notin \{D,E\}} \kappa_t \right) - 2|a_3| \right]$$

$$= |a_4| + |b_4| + |c_4| = 3 + (5 \times 1) + (4 - 2) = 10 \tag{74}$$

where $\mathcal{N}_2$ represents the set of nodes involved in all 2-hop paths, while $\kappa_t$ denotes the degree of node $t$. Additionally, $\left[ \left( \sum_{t \in \mathcal{N}_2, t \notin \{D,E\}} \kappa_t \right) - 2|P_3| \right]$ represents the count of $P_3$-type paths related to the degrees of other nodes. The quantity of 6-hop paths is:

$$|6 - \text{hop}| = |P_1| + |P_2| + |P_3| + |P_4| + |P_6|$$

$$= 0 + |P_2| + |P_3| + \left[ \left( \sum_{t \in \mathcal{N}_4, t \notin \{D,E\}} \kappa_t \right) - 2|a_4| - \eta_1 \right] + (\kappa_D + \kappa_E)|a_4| + |a_6|$$

$$= 0 + |P_2| + |P_3| + (27 - 6 - 8) + 5|a_4| + |a_6|$$

$$= |a_6| + |P_2| + |P_3| + 5|a_4| + 13$$

$$= |a_6| + |P_2| + 2[l_3(D) + l_3(E)]|a_3| + \left( \sum_{t \in \mathcal{N}_3, t \notin \{D,E\}} l_3(t) \right) + 5|a_4| + 13$$

$$= |a_6| + |P_2| + 8 + 0 + 5|a_4| + 13$$

$$= |a_6| + |P_2| + 5|a_4| + 21$$

633 $$= 2 + 15 + |P_2| + 21$$

634 $$= 2 + 15 + p_1 + p_2 + p_3 + p_4 + 21$$

635 $$= 2 + 15 + (\kappa_D^2 + \kappa_E^2 + \kappa_D \kappa_E)|a_2| + p_2 + p_3 + p_4 + 21$$

636 $$= 2 + 15 + 19 + p_2 + p_3 + p_4 + 21$$

637 $$= |a_6| + |b_6| + p_2 + p_3 + p_4 + 21$$

638 $$= |a_6| + |b_6| + \sum_{t \in \mathcal{N}_2, t \notin \{B,D\}} (\kappa_t^2 - 3) + p_3 + p_4 + 21$$

639 $$= |a_6| + |b_6| + 13 + p_3 + p_4 + 21$$

640 $$= |a_6| + |b_6| + p_3 + p_4 + 34$$

641 $$= |a_6| + |b_6| + \left( \sum_{t \in \mathcal{N}_{D,n}, t \notin \{D,E\}} \kappa_t + \sum_{t \in \mathcal{N}_{E,n}, t \notin \{D,E\}} \kappa_t \right)|a_2| - \eta_3 + p_4 + 34$$

642 $$= |a_6| + |b_6| + 9 - 3 + p_4 + 34$$

643 $$= |a_6| + |b_6| + 2[l_4(D) + l_4(E)]|a_2| + \sum_{t \in \mathcal{N}_2, t \notin \{B,D\}} l_4(t) - \eta_4 + 40$$

644 $$= |a_6| + |b_6| + 2(1+3) + 27 - 22 + 40$$

645 $$= |a_6| + |b_6| + 53$$

646 $$= |a_6| + |b_6| + |c_6| = 89 \tag{75}$$

647  where $\mathcal{N}_4$ denotes the list of nodes, which may be repeated, involved in all 4-hop paths. $l_3(\cdot)$
648  measures the quantity of 3-hop paths returning to itself. $p_1$ represents $|b_n|$. $p_2$ denotes the number
649  of paths starting from common neighbors, jumping to non-D, non-E nodes, and then returning to
650  themselves. $p_3$ accounts for paths initiating from D or E, jumping to non-common neighbors, and
651  then looping back. $p_4$ counts 4-hop paths looping back to themselves via D, E, or common
652  neighbors. $\mathcal{N}_{D,n}$ denotes the list of neighbor nodes of D. $\mathcal{N}_{E,n}$ denotes the list of neighbor nodes
653  of E. $\eta_1$, $\eta_2$, $\eta_3$, $\eta_4$ represent the statistical redundancies of paths that satisfy multiple
654  classification criteria within their respective terms. The analysis reveals that the calculation of $|c_n|$
655  is predominantly related to the degrees of nodes other than $\kappa_i$ and $\kappa_j$. Hence, the study does not
656  investigate the influence of $|c_n|$ on the topological structure and degree information captured by
657  $\boldsymbol{D}_{topo}$.
658